\documentclass[english]{svmult}
\usepackage{todonotes,ulem}
\usepackage{amssymb,amsmath,mathrsfs}
\usepackage{graphicx,psfrag,color,rotate}
\usepackage{tikz}
\renewcommand{\vec}{\boldsymbol}
\renewcommand{\d}[1][]{\ensuremath{\,\mathrm{d}#1}}

\newcommand{\indi}[1]{\ensuremath{\operatorname{#1}}}
\newcommand{\op}[1]{\ensuremath{\operatorname{#1}}}

\makeindex 
\begin{document}
\title*{Frontiers in Mortar Methods for Isogeometric Analysis}
\titlerunning{Frontiers in Mortar Methods for IgA}
\author{Christian Hesch \and Ustim Khristenko \and Rolf Krause \and
Alexander Popp \and Alexander Seitz \and Wolfgang Wall \and
Barbara Wohlmuth}
\authorrunning{Hesch, Khristenko, Krause, Popp, Seitz, Wall, Wohlmuth}

\institute{
Christian Hesch \at Universit\"at Siegen, Chair of Computational Mechanics, Paul-Bonatz Str. 9-11, 57068 Siegen, \at \email{christian.hesch@uni-siegen.de}
\and 
Alexander Popp \at Universit\"at der Bundeswehr M\"unchen, Institute for Mathematics and Computer-Based Simulation, Werner-Heisenberg-Weg 39, 85577 Neubiberg, \at \email{alexander.popp@unibw.de}
\and
Alexander Seitz, Wolfgang Wall \at Technische Universit\"at M\"unchen, Institute for Computational Mechanics, Boltzmannstr. 15, 85748 Garching, \at \email{wall,seitz@lnm.mw.tum.de}
\and
Rolf Krause \at Universita della Svizzera italiana, Chair for Advanced Scientific Computing, Via Giuseppe Buffi 13, 6900 Lugano, Switzerland, \at \email{rolf.krause@usi.ch}
\and
Ustim Khristenko, Barbara Wohlmuth \at Technische Universit\"at M\"unchen, Institute for Numerical Mathematics, Boltzmannstr. 3, 85748 Garching, \at \email{khristen,wohlmuth@ma.tum.de}
}

%
% =================================================================================================

\maketitle
%
%
% -----------------------------------------------------------------------------
\abstract{Complex geometries as common in industrial applications consist of multiple patches, if spline based parametrizations are used. The requirements for the generation of analysis-suitable models are increasing dramatically since isogeometric analysis is directly based on the spline parametrization and nowadays used for the calculation of higher-order partial differential equations. The computational, or more general, the engineering analysis necessitates suitable coupling techniques between the different patches. Mortar methods have been successfully applied for coupling of patches and for contact mechanics in recent years to resolve the arising issues within the interface. We present here current achievements in the design of mortar technologies in isogeometric analysis within the  Priority Program SPP 1748, ”Reliable Simulation Techniques in Solid Mechanics.  Development of Non-standard Discretisation Methods, Mechanical and Mathematical Analysis”.}

\section{Introduction}
% Mortar allgemein.
Mortar methods have been developed in the early 1990s of the past century \cite{bernadi1994}, see also in \cite{belgacem1999} in the context of domain decomposition problems, originally applied to spectral and finite element methods, see, among many other \cite{wohlmuth2000b,wohlmuth2001,flemisch2005b}. Domain decomposition techniques provide powerful tools for the coupling of different, in general nonconforming meshes. A wide range of reason exists to create such interfaces; the characteristic idea of mortar methods rely on the weak, integral condition in contrast to strong point-wise couplings. Within the principle of virtual work and far beyond this mechanical concept, mortar methods enter the corresponding balance equations in a variational consistent manner.

Lagrange multipliers in a dual form have been proposed in \cite{wohlmuth2000a}. This allows for a cost efficient and effective way for the interface coupling. Several interpretation exist for the condensation, e.g. as elimination of the Lagrange multipliers via the Schur decomposition as presented in \cite{flemisch2005} or as null-space reduction scheme as shown in \cite{hesch2014b}. In the latter citation, mortar methods are used for overlapping domain decomposition methods in the context of fluid-structure interaction problems (FSI), also known as immersed techniques. Moreover, \cite{kloeppel2011} applied mortar methods on boundary fitted FSI, demonstrating the wide range of applicability of this methodology.

For contact mechanics, nodal wise enforcement of the non-penetration condition are used since the 1980s, see, e.g., \cite{hallquist1979,hallquist1985}. As shown in detail in \cite{elabbasi2001}, nodal wise formulations do not pass the patch test and do not converge correctly, which is a major drawback of these methods. In contrast, mortar methods used as variationally consistent contact interface conditions as considered in  \cite{puso2004a,puso2004b,hueber2005b,hueber2009,hesch2009,popp2010,popp2012,popp2013} pass the patch test. 

Isogeometric Analysis (IgA) as introduced in \cite{hughes2005} has become a widely used methodology, see \cite{cottrell2009, cottrell2007,cottrell2006, hoellig2003}. This framework facilitates the usage of NURBS basis functions, emanating from the field of computer aided design (CAD). Moreover, it allows for the construction of finite element basis functions with adjustable continuity across the element boundaries, in contrast to classical Lagrangian basis functions. This enables the numerical treatment of higher-order partial differential equations (PDE's), e.g.\ for Cahn-Hilliard or Cahn-Hilliard like formulations \cite{gomez2008}, in fracture mechanics \cite{borden2014,dittmann2017b,dittmann2018}, in structural mechanics, e.g.\ in \cite{benson2010,benson2011,dornisch2013,dornisch2016,Echter2013,kiendl2009,reali2015} and for generalized continua \cite{Fischer2011}.

A major drawback of IgA is the decomposition of the whole domain in patches. In standard industrial geometries, the domain is decomposed in hundreds or even thousands of patches, which necessitates the application of suitable interface conditions. Therefore, the combination of mortar methods and IgA has been proposed in \cite{hesch2012a,brivadis2015}. A wide range of issues arise, mostly related with the suitable choice of the Lagrange multiplier space. In a series of actual contributions \cite{schuss2018,horger2019,dittmann2018c,dittmann2020}, higher-order domain decomposition has been addressed as well, which allows for the application of higher-order PDE's in multi-patch geometries.

Finally, several contributions deal with IgA and mortar contact methods, see \cite{lorenzis2011,lorenzis2012,dittmann2014,hesch2016a,Seitz2016}. Biorthogonal splines for the effective condensation of the Lagrange multipliers have been developed in \cite{wunderlich2018}. Multidimensional coupling has been addressed recently in \cite{steinbrecher2019c} and \cite{hesch2020}.

The paper is structured as follows. In Section \ref{sec:basics} basic notations and IgA concepts are introduced within a most general framework on unconstrained and constrained elasticity. In Section \ref{sec:domain}, recent trends in IgA based mortar domain decomposition methods are shown. This is followed in Section \ref{sec:contact} by recent contributions for IgA mortar contact techniques, along with multidimensional coupling conditions in Section \ref{sec:multi}. Eventually, conclusions are drawn in Section \ref{sec:conclusions}.

%\todoB{ allgemeine Bemerkungen}
%\begin{itemize}
%\item moeglichst weit vorne die abkuerzung von partial differential equation (PDE) einfuehren
%\item bei mortar einheiltlich von sub-patches reden und die mit $\Omega_m$ bezeichnen $m=1, \ldots M$
%\item IgA statt IGA verwenden und auch als Abkuerzung einfuehren
%\item die Interfaces mit $\Gamma_l $ , $l= 1, \ldots L$ bezeichnen
%\item von slave und master seite sprechen
%\item bitte ergaenzen - damit jeder schnell sieht was gebraucht bzw schon eingefuehrt ist
%\end{itemize}
%\todoE{  }

\section{Coupled simulations with Mortar Methods in HPC}
Although mortar methods have been originally developed for coupling on non-overlapping subdomains, the idea of variational transfer has been applied to a much wider class of problems. A principal advantage of mortar methods is their ability to couple different discretizations, either on interfaces, surfaces, in the volume, and even across dimensions, i.e. between 1D and 3D models or 3D and 2D models. Using variants or derivatives of the mortar methods, coupled multi-physics problems, such as fluid structure interaction, can be realized using immersed methods with volume coupling~\cite{Nestola2018f}. For multi-scale simulations in mechanics, mortar methods can be used to couple molecular dynamics simulations with finite element approximations using an overlapping decomposition approach. Contact problems in mechanics can efficiently be dealt with using mortar methods for the coupling between surfaces \cite{TDickopf_RKrause_2008, Planta2020b}, and coupling across dimension or non-conforming meshes can be used for the simulation of flow in fracture networks in geo-sciences~\cite{schadle2019,Planta2019b,Planta2020a,berre2020verification}. Finally, mortar methods can also be employed to build multilevel approximation spaces for multigrid methods, thereby serving for the construction of linear or non-linear multigrid methods on complex geometries~\cite{dickopf_numerical_2013,zulian2017geometry}.

The flexibility mortar methods provide, however, is tightly connected to the capability of assembling the mortar transfer operator for volume coupling or surface coupling - or combinations thereof. This seemingly practical task turns out to play a pivotal role when more complex applications with possibly several non-connected or overlapping domains are considered. 
For large scale simulations, also the parallel assembly of the transfer operator has to be considered. The latter is  a challenging task in terms of efficiency and scalability, as in general no a priori information on the connectivity between two non-matching meshes is available, which then has to be generated and dealt with during the assembly.

In this section, we will present examples from cardiac simulation, computational mechanics, and fluid-structure interaction in cardiac simulation and geo-sciences, which illustrate the capabilities of the mortar method. Moreover, we will discuss the tedious and non-trivial assembly of the mortar operator in the context of massively parallel computations in HPC (e.g., \cite{Osborn2018}), which has been realized in the two libraries~\cite{moonolithgit}.

\begin{center}
\begin{figure}
 \centering{\includegraphics[width=0.8\linewidth]{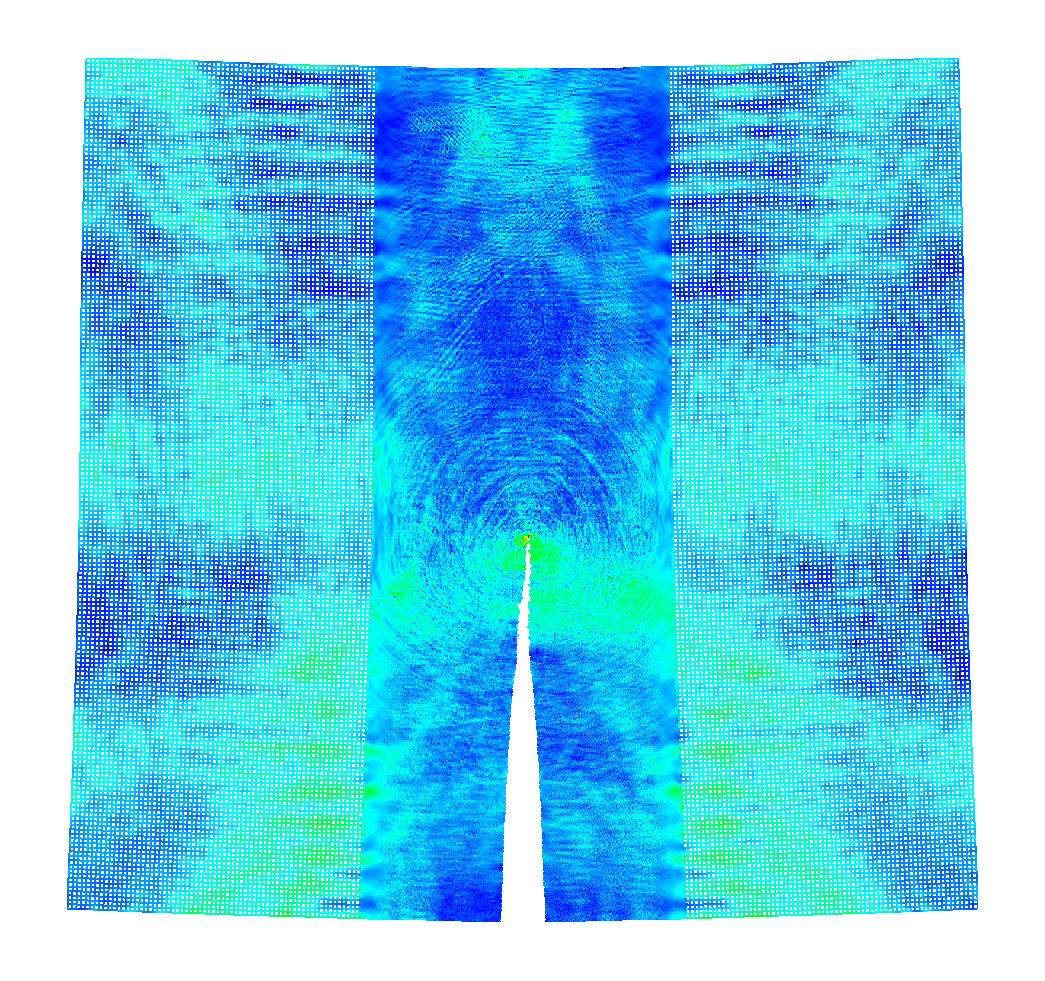}}
\caption{Multi-scale coupling of Molecular Dynamics and Finite Elements via  Partition of Unity and $L^2$-projection}
\label{fig:MD-FEM-coupling}
\end{figure}
\end{center}

One example for computationally demanding simulations are multi-scale simulations in mechanics.
The idea behind these multi-scale simulation is to resolve phenomena such as fracture locally by means of molecular dynamics and to use a continuum mechanics representation for the remaining body. Thus, on the discrete level, molecular dynamics simulation have to be coupled with finite element discretizations.
In molecular dynamics (MD), atoms are represented as point masses, which are subject to internal and external forces. For the positions of the atoms it is assumed that they follow Newton's equation of motion. The forces exerted on the atoms are modeled by means of the gradient of a potential, e.g. Lennard--Jones, which is describing the behavior of the material under consideration. This leads to a system of ordinary differential equations, which then is solved numerically. For the coupling of molecular dynamics simulations, we have to transfer quantities such as displacements, velocities, and forces between a finite element discretization, which is based on integral quantities, and the MD discretization, which is based on pointwise given information (atoms). In \cite{FackeldeyKrauseKrauseLenzen09} this coupling has been realized by attaching a partition of unity to the atoms and then using a mortar transfer operator for coupling between the MD and the FEM discretizations. As it turns out, the mortar method in this context can be shown to act as a frequency filter, which will effectively remove the high-frequency components of the MD displacements, which can not be represented on the FE mesh. As a consequence, mortar based multi-scale coupling eliminates a large part of the unphysical wave reflections at the coupling interface and gives rise to a stable coupling between MD and FEM, see Figure~\ref{fig:MD-FEM-coupling} and~\cite{FackeldeyKrauseKrauseLenzen09}.
In the context of fast solution methods, mortar methods have also been used to derive adaptive space-time discretization methods, see \cite{DKrause_RKrause_2014,krauseLargescaleScalableAdaptive2015}, which combine the advantages of structured meshes in terms of simple data-structures with the advantages of adaptive discretizations.
% Figure \ref{fig:ShallowTrees} shows the mesh hierarchy for simulated electric potential in a human heart. As can be seen, a coarse tessellation is refined locally with arbitrary depth, leading to non-conforming interfaces. 
%
%\begin{center}
%\begin{figure}
% \includegraphics[width=\linewidth]{pictures/_p_shallow_trees_krause.png}
%\caption{Shallow Trees with}
%\label{fig:ShallowTrees}
%\end{figure}
%\end{center}
%
Here, for the coupling at the interfaces, mortar methods have been employed, allowing for a local (to a single processor or core) treatment of the unknowns. Clearly, the load balancing needs to be adapted to this situation. The resulting decomposition cannot only be used for the design of adaptive parallel methods, but also for the construction of additive Schwarz-preconditioners in space and time based on non-conforming space-time decomposition, see \cite{DKrause_RKrause_2014,krauseLargescaleScalableAdaptive2015}.
\begin{figure}[!htb]
\minipage{0.49\textwidth}
  \includegraphics[width=0.8\linewidth]{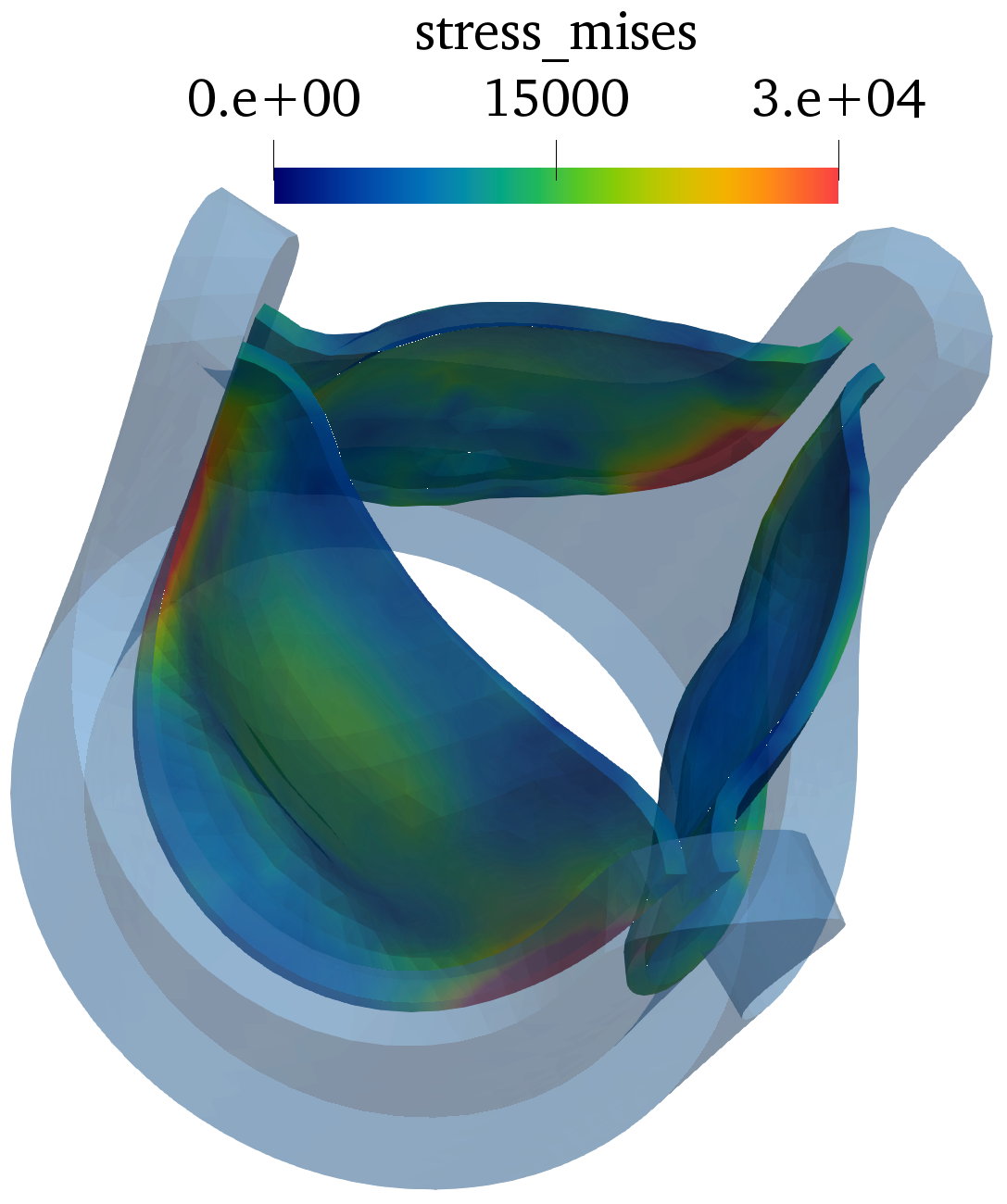}
  \caption{Spatial distribution of the von Mises stresses in the bio-prosthetic aortic valve}\label{fig:HeartValve-FSI}
\endminipage\hfill
% \end{figure}
% \begin{figure}[!htb]
\minipage{0.49\textwidth}
  \includegraphics[width=\linewidth]{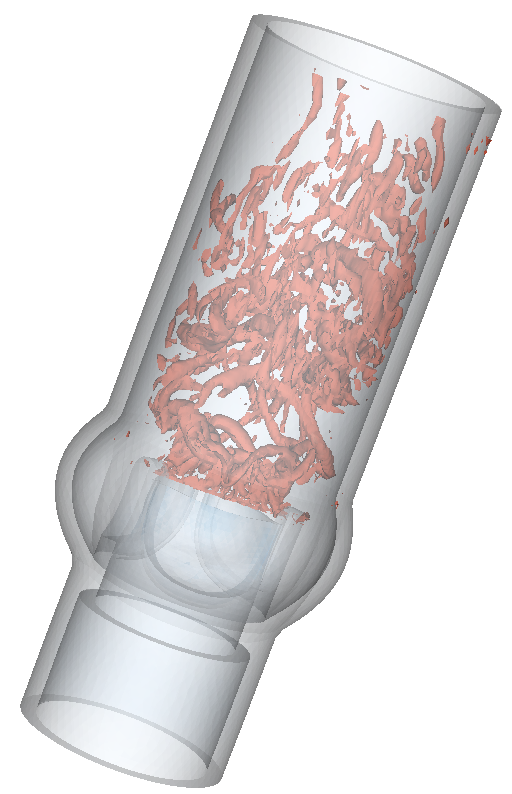}
  \caption{Vortical structures arising from the interaction of the bio-prosthetic aortic valve with the blood flow ~\cite{Nestola2018f}.}\label{fig:HeartValve-vonMises}
\endminipage\hfill
\end{figure}

\begin{figure}[!htb]
\minipage{0.49\textwidth}
  \includegraphics[width=0.7\linewidth]{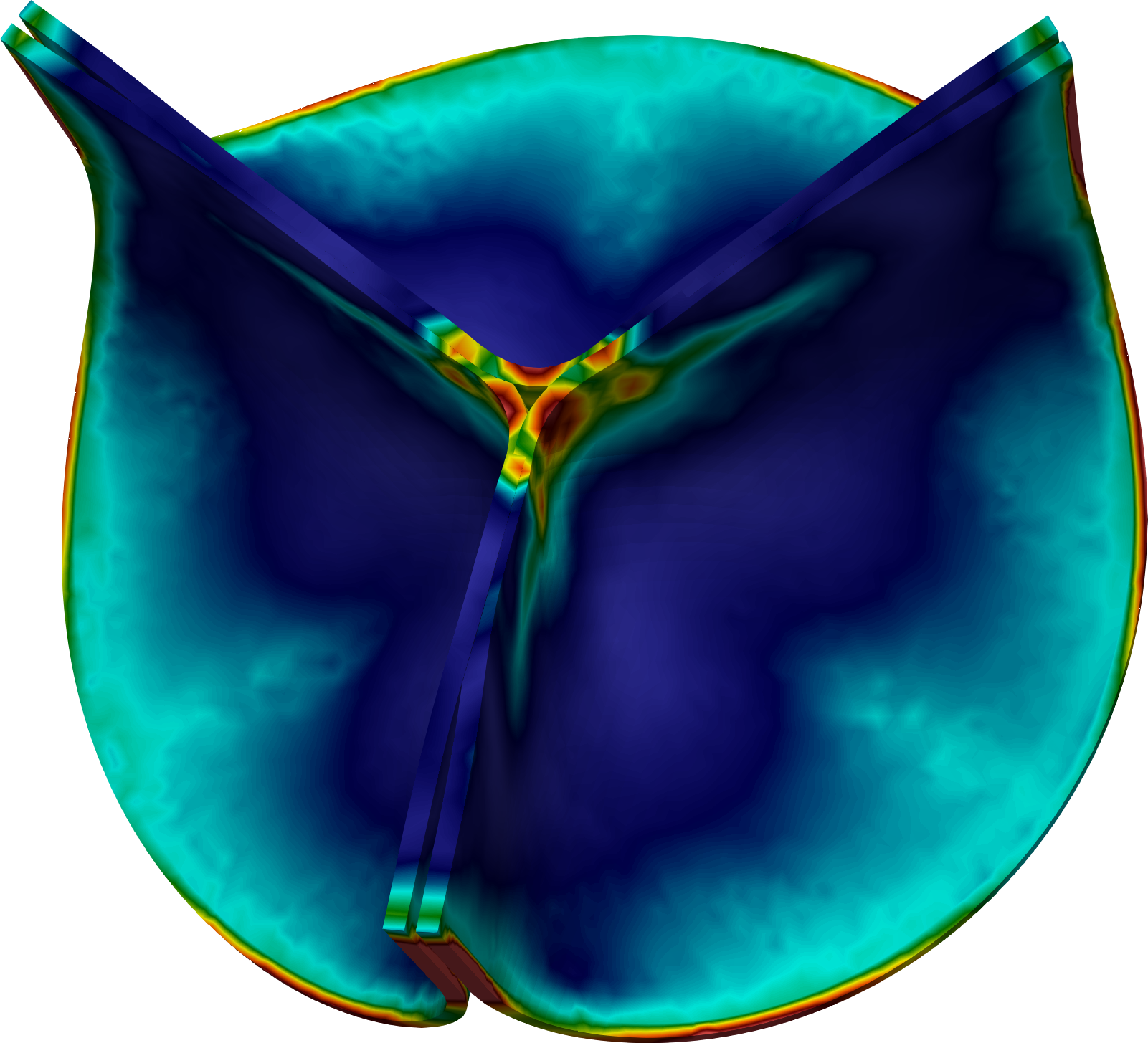}
  \caption{Spatial distribution of the von Mises stresses in the bio-prosthetic aortic valve during contact}\label{fig:HeartValve-FSI2}
\endminipage\hfill
% \end{figure}
% \begin{figure}[!htb]
\minipage{0.4\textwidth}
  \includegraphics[width=0.6\linewidth]{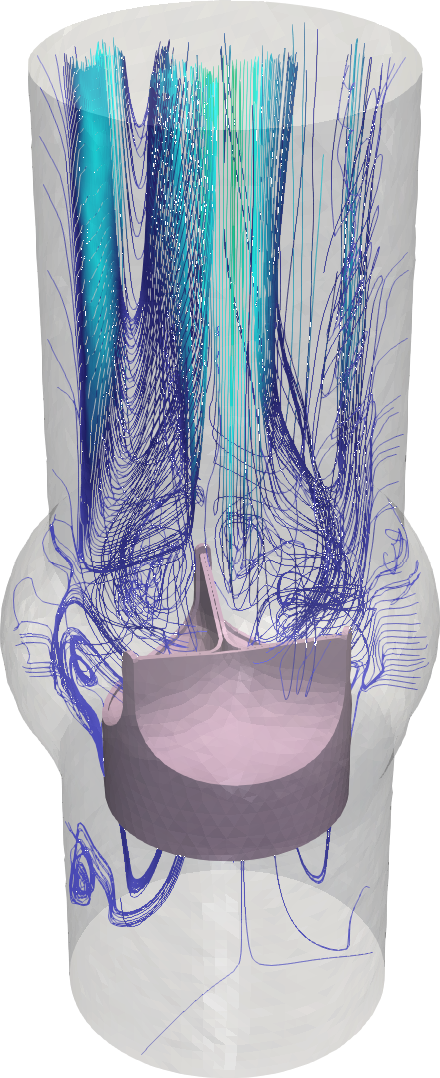}
  \caption{Streamlines of the velocity fluid field during the closure of the valve when contact occurs.}\label{fig:HeartValve-vonMises2}
\endminipage\hfill
\end{figure}
\begin{figure}
 \includegraphics[width=\linewidth]{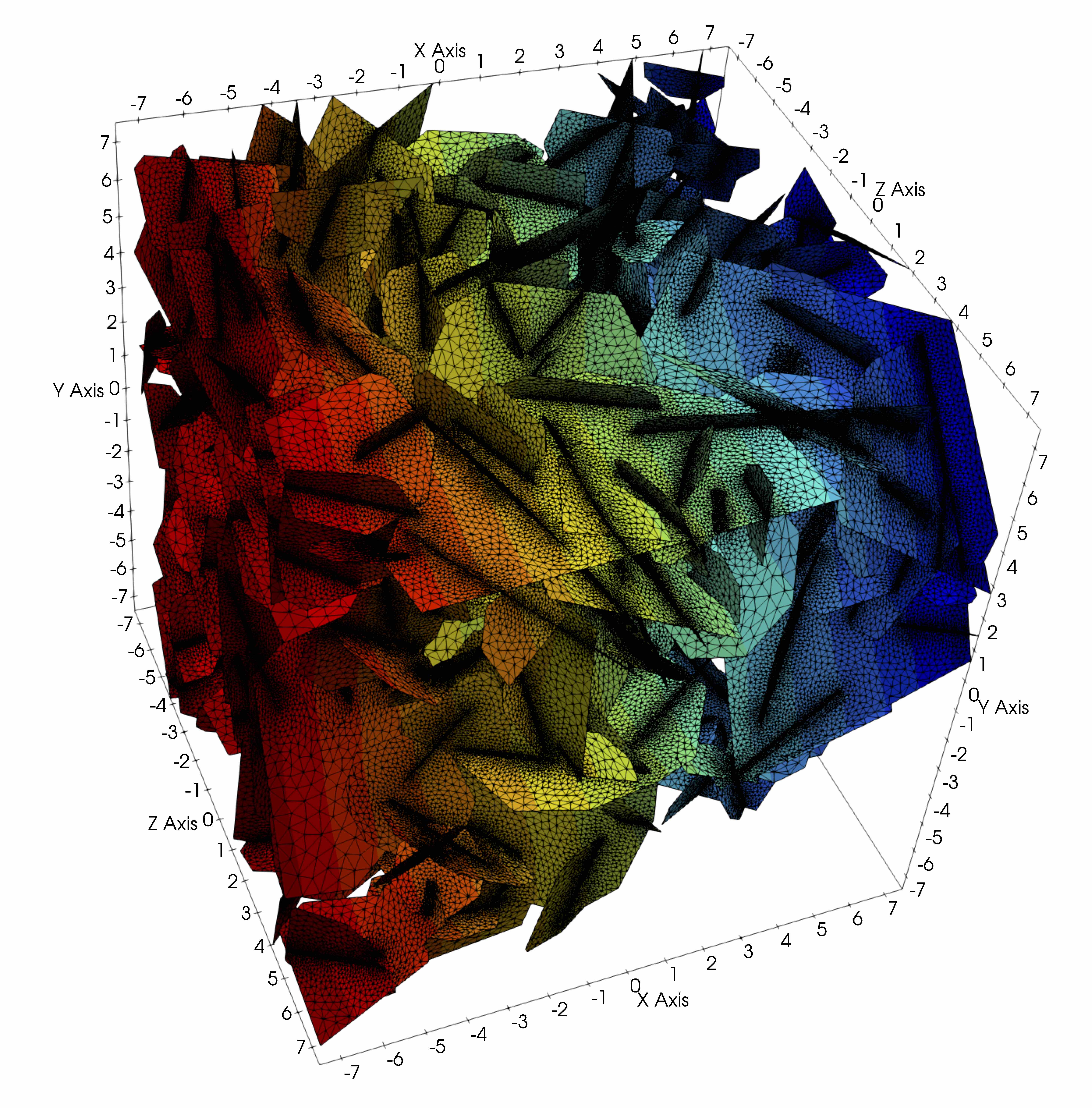}
\caption{Non conforming mesh method for flow in fracture porous media \cite{schadle2019}.}
\label{fig:FractureNetwork}
\end{figure}

Whereas in the above example the computation of the transfer operator at the domain interface is straight forward, it becomes more difficult and demanding in the case of coupling between non-matching or warped surfaces or in the case of volume coupling between bodies represented with unstructured meshes. By definition, the assembly of the transfer operator requires the evaluation of integrals on the intersection of two non-matching meshes. We refer to Figure~\ref{fig:FractureNetwork}, which illustrates the complexity of the resulting intersections for a fracture network. In the case of surface coupling, meshes need to be projected from one surface onto another~\cite{TDickopf_RKrause_2008}. In a parallel setting, involving two or more domains or bodies, a major difficulty is that a priori we don't know which elements of which subdomain will have a non-empty intersections \cite{krause_parallel_2016}.
We note that in a parallel computation these might be on completely different processors, so that a global search has to be carried out. A global search however is not advisable due to the resulting quadratic complexity. Thus, more efficient strategies, i.e. hierarchic strategies based on $kd$-trees, are employed for detecting possibly intersecting elements. The resulting cut-candidates are then checked in detail for possible intersections, so that the quadratures on the intersections can be carried out, in order to compute the entries of the mass matrix.
From a technical point of view, detection and computation of the intersections has to be handled very carefully, as small cuts or ill-conditioned sub-problems will show up. Additionally, in order to guarantee scalability, the computation of the intersections and the computation of the local integrals are distributed globally to ensure equal load balancing. For surface related coupling, e.g. contact problems, this imbalance is quite obvious. It will, however, also show up for volume coupling. 
One possibility to ensure a good load balancing, is to use space-filling curves. We refer to \cite{krause_parallel_2016}, where this approach is described in detail as well as to the library MoonoLith \cite{moonolithgit}, which to our knowledge is the only currently available library implementing variational transfer on arbitrary meshes in parallel for surfaces and volumes.  As a consequence, the flexibility of mortar methods is counterbalanced by the complex assembly of the transfer operator for complex geometries.

\begin{figure}[htb]
 \includegraphics[width=\linewidth]{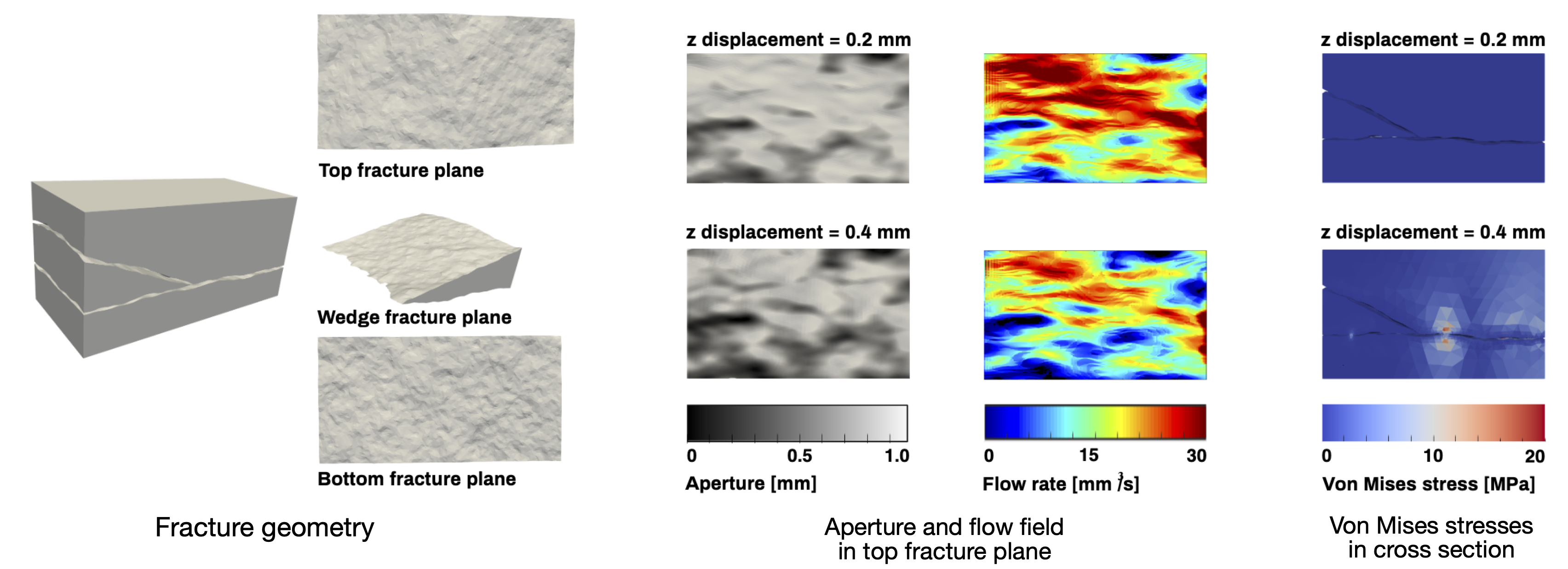}
\caption{Fluid Structure Interaction with contact for geo-sciences (from \cite{Planta2020a}).}
\label{fig:geo_fsi_contact}
\end{figure}
Once available, however, variational transfer can be used to design new numerical methods for coupled multiphysics simulations. Here, as an example, we consider immersed methods for fluid structure interaction in cardiac simulation. Figures~\ref{fig:HeartValve-FSI} and \ref{fig:HeartValve-vonMises} show the fluid-structure interaction interaction
between the turbulent systolic jet and the leaflets of a bio-prosthetic aortic valve. The fluid-structure interaction formulation relies on a mortar approach to couple a finite difference discretization of the Navier-Stokes equations, with an finite-element formulation of an anisotropic fiber-reinforced model~(\cite{Nestola2018f,henniger2010high,nestola2019b}). In contrast to other immersed approaches, the coupling strategy allows for an implicit treatment of the equations describing the solid dynamics. Moreover, it effectively prevents leakage at the interface, as the basis functions of the multiplier space form a partition of unity.
Figures~\ref{fig:HeartValve-FSI2} and \ref{fig:HeartValve-vonMises2} show the combination of volume coupling with surface coupling, i.e. of FSI with contact for the considered bio-prosthetic aortic valve. Here, transfer  on the entire volume as well as on the surface have to be carried out in order to satisfy the  equality (FSI) constraint and inequality (contact) constraints.

\section{Basic equations and Isogeometric Analysis}\label{sec:basics}
We start with a short summary of elasticity with constraints to introduce the basic concepts to be dealt with in the following. Therefore, we consider a Lipschitz bounded domain with reference configuration $\mathcal{B}_{0}\subset\mathbb{R}^d$, $d\in\{2,3\}$, undergoing a motion characterized by a time dependent deformation mapping $\vec{\varphi}: \mathcal{B}_{0} \times \mathcal{I}\rightarrow\mathbb{R}^d$, where $\mathcal{I}=[0,T]$ is the time interval elapsed during the motion. The current configuration is denoted by $\mathcal{B}_{t} = \vec{\varphi}_t(\mathcal{B}_{0})$, material points are labeled by $\vec{X}\in\mathcal{B}_{0}\rightarrow\mathbb{R}^d$. 

%The material velocity is given by $\vec{v}:\mathcal{B}_{0}\times \mathcal{I}\rightarrow\mathbb{R}^n,\,\vec{v}=\partial\vec{\varphi}/\partial t$.

\textbf{Unconstrained elasticity.}
For the most basic setting in elasticity, we introduce the virtual work of the internal and external contribution and postulate that the principle of virtual work is valid. In a first step, the spaces of solution and weighting functions
\begin{align}                     
\mathcal{S} &= \left\{\phantom{\delta}\vec{\varphi}\in \text{H}^1(\mathcal{B}_0)\,|\,\phantom{\delta}\vec{\varphi} = \bar{\vec{\varphi}}\;\text{on}\;\Gamma^{u}\right\},\\
\mathcal{V} &= \left\{\delta\vec{\varphi}\in \text{H}^1(\mathcal{B}_0)\,|\,\delta\vec{\varphi} = 0\;\;\text{on}\;\Gamma^{u}\right\},
\end{align}
are defined. Here, we make use of the standard notation for the Sobolev space $\text{H}^s(\mathcal{B}_0)$, $s\geq 1$, of square integrable functions $\vec{\varphi}$ with square integrable weak derivatives of the given order. Note that second and third gradient materials require $\vec{\varphi}\in \text{H}^2(\mathcal{B}_0)$ and $\vec{\varphi}\in\text{H}^3(\mathcal{B}_0)$, respectively. In accordance with common nomenclature, we denote the Dirichlet boundary conditions $\Gamma^{u}$ and Neumann conditions $\Gamma^{n}$, satisfying $\partial \mathcal{B}_0 =\Gamma^{u}\cup \Gamma^{n}$ and $\Gamma^{u} \cap \Gamma^{n}=\emptyset$ throughout the time interval $\mathcal{I}$. The principle of virtual work now reads: find $\vec{\varphi}\in \mathcal{S}$ such that 
\begin{equation}
a(\vec{\varphi},\delta\vec{\varphi}) = l(\vec{\varphi},\delta\vec{\varphi}) \quad\forall\;\delta\vec{\varphi}\in\mathcal{V},
\end{equation}
where
\begin{align}
a(\vec{\varphi},\delta\vec{\varphi}) &:= \int\limits_{\mathcal{B}_0}\nabla_{\vec{X}}(\delta\vec{\varphi}):\vec{P} \d V,  \label{weak1} \\   
l(\vec{\varphi},\delta\vec{\varphi}) &:= \int\limits_{\mathcal{B}_0}\delta\vec{\varphi}\cdot\vec{B}(\vec{\varphi}) \d V +
    \int\limits_{\Gamma^n}\delta\vec{\varphi}\cdot\vec{T}(\vec{\varphi}) \d A,
\end{align}
are the internal and external contributions to the virtual work. Here, $\vec{P}$ denotes the first Piola-Kirchhoff stress tensor, $\vec{B}$ body forces and $\vec{T}$ surface loads, acting on $\Gamma^n$. For linear elasticity, the second term on the right hand side of \eqref{weak1} has to be linearized, hence, $a(\vec{\varphi},\delta\vec{\varphi})$ is a bi-linear form.

\textbf{Constrained elasticity.}
Elastic systems can be subject to a wide range of different constraints. For incompressible systems, constraints are defined throughout the whole domain. For overlapping domain decomposition methods as used for fluid-structure interaction (immersed techniques) as well as for solid-solid interaction problems, the constraints are defined on at least parts of the domain. Plasticity takes a special role, as the corresponding constraints are defined as Karush-Kuhn Tucker inequality conditions within the whole domain, almost always locally condensed. Many formulations focus on conditions at certain internal and external interfaces; classical domain decomposition problems act on fixed interior interfaces, whereas boundary fitted fluid-structure interaction formulations rest on moving internal interfaces. On the other hand, constraints acting on external interfaces like Dirichlet and control conditions can be applied as well as contact problems, which, similar to plasticity, are given as a set of Karush-Kuhn Tucker inequality conditions. 

This principle of virtual work reads now
\begin{align}
a(\vec{\varphi},\delta\vec{\varphi}) +  b(\vec{\lambda},\delta\vec{\varphi}) &= l(\vec{\varphi},\delta\vec{\varphi}) &&\forall\;\delta\vec{\varphi}\in\mathcal{V},\\
b(\vec{\mu},\vec{\varphi}) &= 0&&\forall\;\phantom{\delta}\vec{\mu}\in\mathcal{N},
\end{align}
where the form $b(\vec{\mu},\vec{\varphi})$ is to be defined corresponding to the considered constraints. Detailed formulations in the context of the mortar methods under investigations in this article are presented in subsequent sections. All formulations require appropriate definitions for the spaces of solution functions $\mathcal{M}$ and weighting functions $\mathcal{N}$ of the Lagrange multipliers $\vec{\lambda}$, depending on the chosen problem to be taken into account. Note that this kind of problems leads in general to a saddle point structure, such that the chosen spaces have to obey the inf-sup conditions.

\textbf{B-spline and NURBS spaces.}
Next, we introduce in a nutshell suitable B-spline and NURBS approximations. We refer to Hesch et al. \cite{hesch2016a} for more details on the construction of the spaces including hierarchical refinement procedures. A multivariate B-spline basis of degree \(\vec{p}=[p_1,\hdots,p_d]\) is defined by the dyadic product \(\vec{\Theta} = \Theta_1\otimes\hdots\otimes\Theta_d\) of univariate knot vectors, built by a sequence of knots \(\Theta_l= [\xi^l_1\leq\xi^l_2\leq\hdots\leq\xi^{l}_{\mathfrak{n}+p_l+1}]\), \(l\in\{1,\hdots,d\}\) and $\mathfrak{n}$ the number of basis functions. In the absence of
repeated knots, the partition \([\xi^1_{i_1},\xi^1_{i_1+1}]\times\hdots\times[\xi^d_{i_d},\xi^d_{i_d+1}]\) form an element of the mesh in the parametric domain\footnote{We define the
number of repetitions in \(\vec{\Theta}\) at node \(\vec{i}\) as multiplicity \(m_{\vec{i}}\).}. A single multivariate B-spline \(B^A\) is then defined by
\begin{equation}\label{eq:tensor}
B^{A} = B^{\vec{i}}_{\vec{p}}(\vec{\xi}) =  B^{\vec{i}}_{\vec{p}}(\xi^1,...,\xi^d) = \prod\limits_{l = 1}^{d}N_{i_l,p_l}(\xi^l),
\end{equation}
with multi-index \(\vec{i} = [i_1,\hdots,i_d]\) and \(\op{supp}(B^A) = [\xi^1_{i_1},\xi^1_{i_1+p_1+1}]\times\hdots\times[\xi^d_{i_d},\xi^d_{i_d+p_d+1}]\), providing the necessary support for the required continuity. The recursive definition of a univariate B-spline is given as follows
\begin{equation}
 N_{i_l,p_l} = \frac{\xi-\xi^l_{i_l}}{\xi^l_{i_l+p_l}-\xi^l_{i_l}}\,N_{i_l,p_l-1}(\xi)+\frac{\xi^l_{i_l+p_l+1}-\xi}{\xi^l_{i_l+p_l+1}-\xi^l_{i_l+1}}\,N_{i_l+1,p_l-1}(\xi),
\label{eq:BasisDef1}\end{equation}
starting with
\begin{equation}
 N_{i_l,0}(\xi) = \left\{\begin{array}{l}1\quad\text{if}\;\xi^l_{i_l}\leq\xi<\xi^l_{i_l+1} \\ 0\quad\text{otherwise}\end{array}\right..
\label{eq:BasisDef2}\end{equation}
The collection of B-splines \(B^A\), \(A\in[1,\hdots,\mathfrak{n}]\) is defined on \(\vec{\Theta}\) and the corresponding spline space, defined as \(S(\vec{\Theta}) = \op{span}(B^A)\). Moreover, the extension to the NURBS space is given by
\begin{equation}\label{eq:nurbs}
R^A = R^{\vec{i}}_{\vec{p}}(\vec{\xi}) = \frac{\prod\limits_{l = 1}^{d}N_{i_l,p_l}(\xi^l)\,w_{\vec{i}}}{\sum_{\hat{\vec{i}}}\prod\limits_{l = 1}^{d}N_{\hat{i}_l,p_l}(\xi^l)\,w_{\hat{\vec{i}}}},
\end{equation}
along with the corresponding NURBS weights \(w_{\vec{i}}\). The fundamental properties of a basis are typical for B-spline and NURBS spaces:
\begin{itemize}
\item Linear independence, \(\sum\limits_{A}c_{A}\,R^A(\vec{\xi}) \equiv 0 \Leftrightarrow c_{A} = 0\).
\item Partition of unity, \(\sum\limits_{A}R^A(\vec{\xi}) = 1\).
\item Local support of B-splines and of NURBS
\item Smoothness is related to knot multiplicity \(m_{\vec{i}}\)\footnote{$C^m$ continuity of the approximation spaces relates to solution functions $u$ which are at least $u\in \mathcal{H}^{m+1}(\mathcal{B})$, with $\mathcal{H}^{m+1}(\mathcal{B})$  being the Hilbert space of square integrable functions with $m+1$ square integrable derivatives.}
\item Nonnegativity, i.e. \(R^A(\vec{\xi}) \geq 0\) .
\end{itemize}
Please note that the Kronecker delta property \(R^A(\vec{\xi}_{\vec{i}} )= \delta^A_{\vec{i}}\), which is common for Lagrangian basis functions, is in general not fulfilled here. The parametric domain as defined in the knot vector \(\vec{\Theta}\) correlates to the domain \(\hat{\mathcal{B}}\), independent of the existence of local refinements. The shape functions \(R^A\) can be associated with a net of control points \(\vec{q}_A\in\mathbb{R}^{d}\), such that a geometrical map \(\vec{\mathfrak{F}}:\hat{\mathcal{B}}_0\to\mathcal{B}_0\) can be defined to link the parameter and the physical space
\begin{equation}
\vec{\varphi}^{h} := \vec{\mathfrak{F}}(\vec{\xi}) = R^{A}(\vec{\xi})\,\vec{q}_{A} = R^{\vec{i}}_{\vec{p}}(\vec{\xi})\,\vec{q}_{\vec{i}},
\end{equation}
cf. da Veiga et al. \cite{Veiga2012a}. The physical domain is the image of the parametric domain through \(\vec{\mathfrak{F}}\), and we note that both domains share the same regularity properties, see Brivadis et al. \cite{brivadis2015}.

\section{Mortar techniques for Isogeometric Analysis}\label{sec:domain}
To combine mortar with IgA techniques is of special interest for complex domains which require a multi-patch representation. The use of multi-patches can also help
to avoid singularities in the domain mapping for relatively simple domains. To allow for patch-wise independent mesh generation, it is a must to tear and interconnect
the discrete solution in a suitable way. Different alternatives such as discontinuous Galerkin (DG) based interior penalty approaches or mortar based Lagrange multiplier formulations exist.
Here we focus on mortar techniques and guarantee always a variational consistent weak $C^0$-coupling at the interior patch boundaries. We review two conceptual different strategies to realize higher-order continuity and address the cost of static elimination of the Lagrange multiplier. A rigorous mathematical analysis of standard mortar IgA techniques can be found in \cite{brivadis2015}.
\subsection{Biorthogonal splines for Isogeometric Analysis} \label{sub:biorth}
This subsection is based on \cite{wunderlich2018}. We provide the abstract construction framework of biorthogonal Lagrange multiplier basis functions. The handling of non-matching meshes in terms of Lagrange multipliers may result in a uniformly stable and variationally consistent discrete formulation, and thus from a theoretical point of view it is well-understood and attractive. However, it gives, in general, rise to a saddle point system, which is more challenging for iterative solvers due to its indefinite algebraic character. The analysis of such a system can be based on a mixed approach and requires continuity of the bilinear forms, approximation properties of the primal and dual space and a uniform inf-sup stability between the discrete spaces.

Alternatively the saddle point system can be formally condensed. Then we are in the setting of a positive definite system on a constrained primal space. The dual variable is eliminated, and at the same time constraints are incorporated in the primal space. Now we can apply the theory of nonconforming finite elements and have to analyze the consistency error. In the mortar case, the consistency error is directly related to the jump of the discrete solution across the interface. Due to the weak continuity, which is enforced by the discrete Lagrange multiplier, it can be shown, in case of an optimal mortar method, that it is at least of the same order as the best approximation error in the unconstrained primal space. 

Using a discrete Lagrange multiplier space, which is obtained, up to modifications at the crosspoints, as trace space of the primal space restricted to the slave side, there exists no basis of the discrete constrained space such that all basis functions have a local support. This is related to the fact that a typical mass matrix is sparse but has a dense inverse. Consequently, we obtain basis functions having a support on the slave side, which is  local in the direction normal to the interface but global along the interface. This also holds for classical low-order mortar finite elements but is even more pronounced in the higher-order IgA framework.

To obtain an optimal mortar IgA approach, which results in a sparse positive definite system on the constrained space, the discrete Lagrange multiplier space has to satisfy four elementary properties:\\
\begin{tabular}{c lcl}
$\bullet$&(BA) &-- & best approximation property with respect to the dual norm,\\[-2mm]
$\bullet$&(LS) &-- & local support of the basis functions,\\[-2mm]
$\bullet$&(BC) &-- & biorthogonality condition between the trace of the primal and the\\[-2mm]
& &&dual basis functions,\\[-2mm]
$\bullet$&(UC) &--& uniform inf-sup condition.
\end{tabular}
 
\begin{figure}[ht]
\begin{center}
\includegraphics[width=0.45\textwidth]{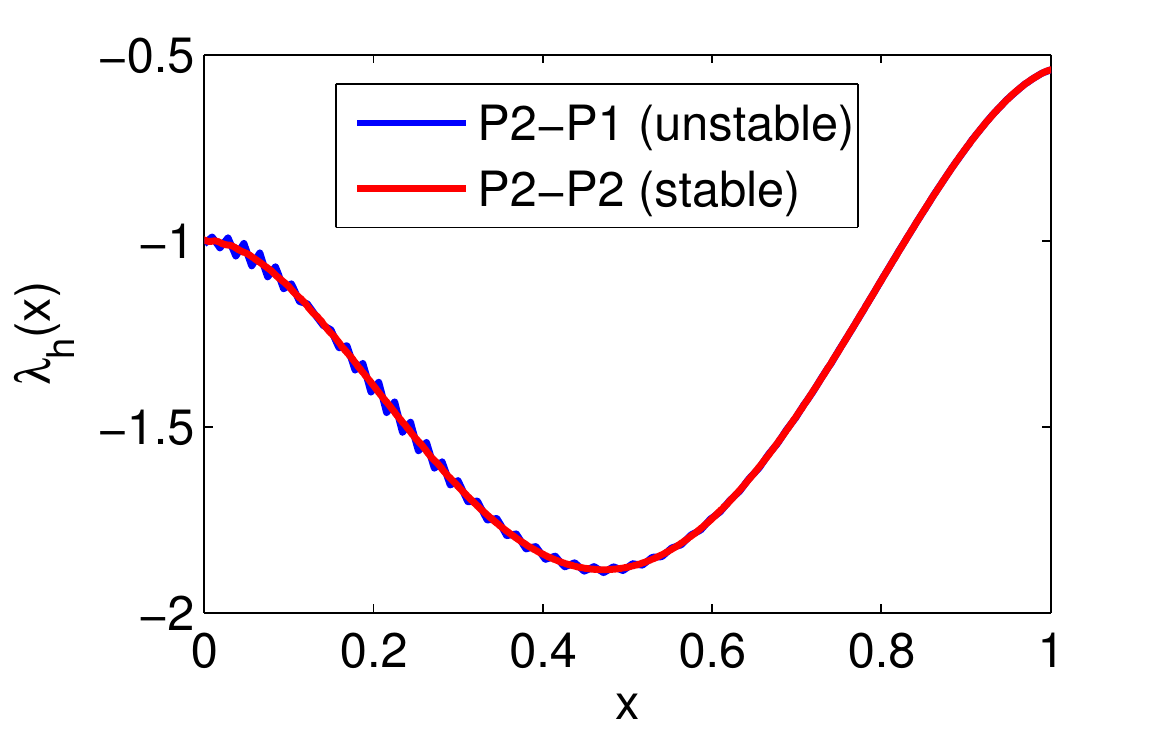}
\includegraphics[width=0.45\textwidth]{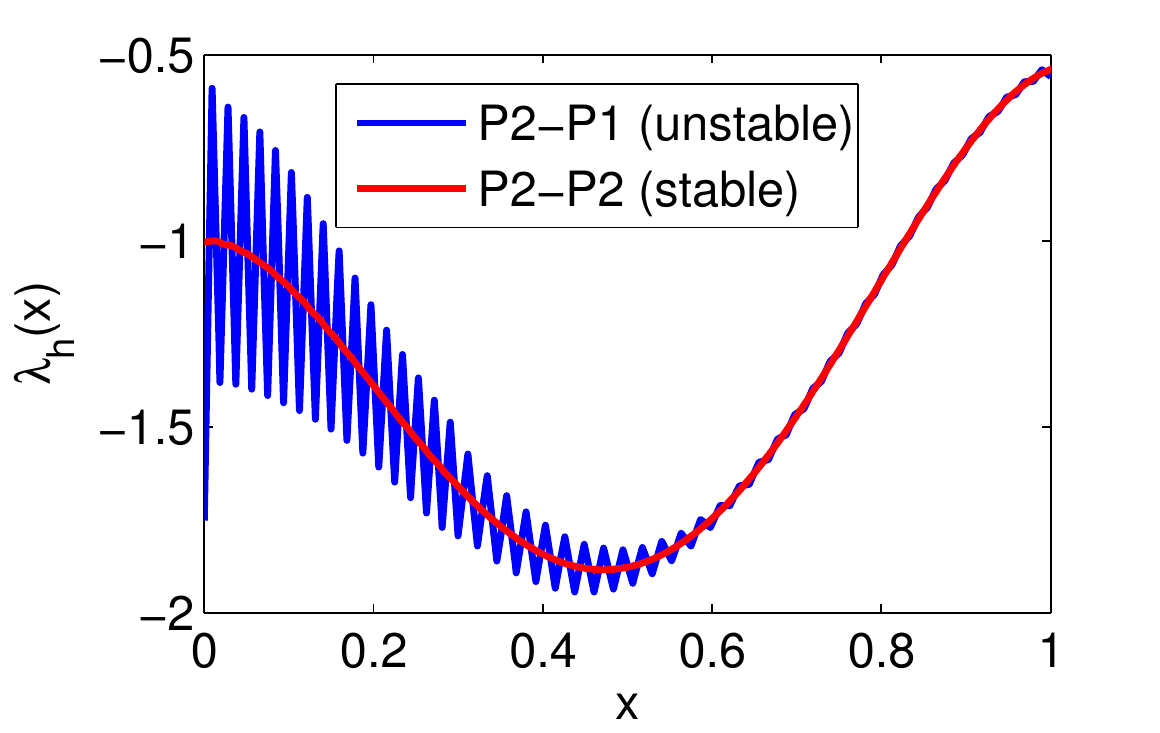}
\end{center}
\caption{Spurious oscillations in the Lagrange multiplier for $p=1$ in case of quadratic splines as primal space and uniformly stable results for $p=2$:
Dirichlet boundary conditions (right) and Neumann boundary conditions (left).}
\label{fig:counting}
\end{figure}

While (LS) and (BC) influence mainly the computational aspect, (BA) and (UC) are essential from the theoretical point of view. To see that (UC) is not only a counting argument of the dimensions of the involved spaces, we give a simple illustration of the effect of a mesh-dependent constant in the inf-sup condition. To do so, we consider $p=2$ for the primal space and two different pairings in the discrete Lagrange multiplier space. One is obtained by the trace space with $p=2$ and one with $p=1$. Here, a counting point argument would be very misleading. In this case, the choice  $p=1$ for the discrete Lagrange multiplier is unstable while $p=2$ is uniformly stable, as illustrated in Figure \ref{fig:counting}. The Dirichlet condition case is more pronounced to instabilities due to the strong constraint of the primal space at the boundary. For more details and a theoretical analysis of the uniformly stable pairing, we refer to \cite{brivadis2015}.

In the following, we briefly sketch the main steps in the construction of such a biorthogonal set of basis functions. The technical details can be found in \cite{wunderlich2018}. For simplicity of notation, we consider only one interface $\Gamma$.

\textbf{Step I:}
In a first step, we embed the trace space on the slave side of the interface $\Gamma$ into the product space of piecewise polynomials having no continuity constraints between the elements, i.e., the spline space of lowest regularity. Using for this higher dimensional product space an elementwise defined basis obtained by multiplying the original basis functions $\phi_i$, $i = 1 , \ldots I$,  with elementwise cut off functions $\chi_e$, $e \in {\mathcal E}$. Here $I$ is the dimension of the trace space and $\mathcal E$ the set of elements. We note that due to the locality of the support of $\phi_i $ many product terms $\phi_i \chi_e$ yield the zero function. The non-trivial ones can be reordered and denoted by  $\phi_{i,e}$, $i=1, \ldots , (p+1)(d-1)$ where $p$ is the polynomial order of the trace space and $d$ the dimension of the domain.  They form the new basis functions of the product space. It is now easy to construct a biorthogonal basis by inverting a local mass matrix of size $(p+1)(d-1) \times (p+1)(d-1)$.  More precisely, we require $\psi_{i,e} \in Q_p (e)$  such that
\begin{align}
\int_{e} \phi_{i,e}\, \psi_{j,e} \d s = \int_e \phi_{i,e} \d s \,  \delta_{i,j}, \quad i,j =1 , \ldots ,(p+1)(d-1) . 
\end{align}
Then by construction the biorthogonal basis defined as product space satisfies (BA), (LS) and (BC) but unfortunately not (UC), and thus we cannot guarantee unique solvability of the system.

\textbf{Step II:} In a second step, we reduce the dimension such that we have equality in the dimensions of the trace space and the one spanned by our modified biorthogonal basis functions. Each basis function $\phi_i$ of the trace space can be written uniquely as linear combination of the basis functions of the product space
\begin{align}
\phi_i  = \sum_{e \in {\mathcal E}, e \subset \text{supp } \phi_i } \phi_{g(i,e), e} , \quad i = 1, \ldots , I,
\end{align}
where $g(i,e) \leq (p+1)(d-1)$ such that $\phi_i \chi_e = \phi_{g(i,e), e} $. We note that if $i \not= j$ then for $e \subset \text{supp } \phi_i \cap \text{supp } \phi_j =: S_{ij}$ we get $g(i,e) \not= g(j,e)$. To define now a smaller set of biorthogonal basis functions, we glue the ones of Step I together by using the same coefficients in the linear combination, i.e.,
\begin{equation}
\psi_i  := \sum_{e \in {\mathcal E}, e \subset \text{supp } \psi_i } \psi_{g(i,e), e} , \quad i=1, \ldots , I .
\end{equation}
 By doing so, we obtain for each basis function of the trace space one basis function in the dual space and (UC) can be shown. By construction it satisfies (LS), i.e., supp $\phi_i =$ supp  $\psi_i $,  and (BC), i.e., 
\begin{equation}
\begin{aligned}
\int_{\Gamma} \phi_i \,\psi_j \d s & = \sum_{ e \in \mathcal E} \int_e \phi_i\, \psi_j \d s =
\sum_{ e \in \mathcal E, e \subset S_{ij}} \int_e \phi_{g(i,e),e} \,\psi_{g(j,e),e}  \d s \\ &= \sum_{ e \in \mathcal E, e \subset S_{ij}} \int_e \phi_{g(i,e),e}  \d s  \, \delta_{g(i,e), g(j,e)}  \\ &= \sum_{ e \in \mathcal E} \int_e \phi_{g(i,e),e}  \d s  \, \delta_{i, j}= \int_\Gamma \phi_i \d s \, \delta_{i,j} .
\end{aligned}
\end{equation}
Unfortunately by reducing the number of basis function, we have gained the property (UC) but lost for $p>1$ the property (BA).

\begin{figure}[ht]
\begin{center}
\includegraphics[width=0.85\textwidth]{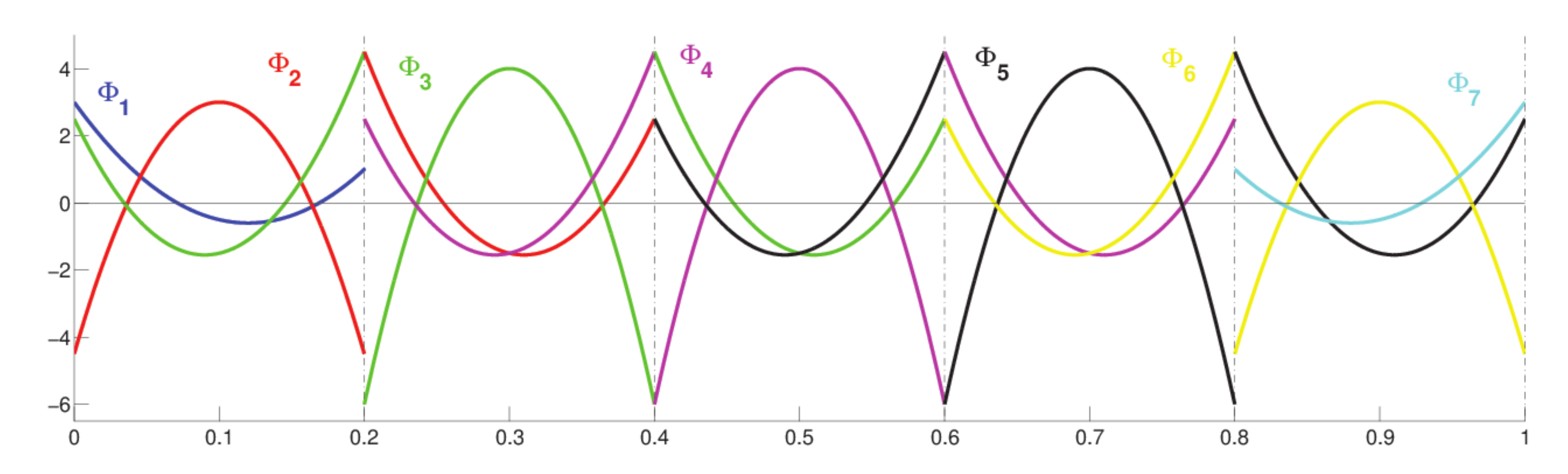}
\end{center}
\caption{Example of biorthogonal basis functions satisfying (BC), (UC) and (LS) but not (BA).}% \textcolor{red}{Shouldn't this be $\phi_i$???}}
\label{fig:localdual}
\end{figure}

In Figure \ref{fig:localdual}, we illustrate the result of Steps I-II for $p=2$ and a seven dimensional trace space. The basis functions have exactly the same support as the ones of the trace space, i.e., at most three elements, and are discontinuous across the elements.  We note that these basis functions cannot reproduce 
a linear function with mean value zero, and thus the required best approximation property is not satisfied. 
As can be easily observed, all interior basis functions $\phi_3, \phi_4, \phi_5$ have the same shape and can be obtained from each other by a simple coordinate shift.

\textbf{Step III:} In the third step, we modify the biorthogonal basis that we have obtained in Step II such that (LS), (BC) and (UC) will be preserved and the best approximation property (BA) will be restored.  We note that adding to a biorthogonal basis function a function which is orthogonal to the trace space does not destroy (BC).  As a preliminary step, we define out of the biorthogonal product space locally defined functions being orthogonal to the trace space. Having these basis functions at hand, we then define coefficients by solving systems of small size. We point out that the system size depends on the order $p$ but not on the meshsize. The small size of the system then guarantees that (LS) is preserved. Let us assume for the moment that we have calculated the coefficients, we then obtain the  modified dual basis function by adding a linear combination of globally orthogonal functions where the computed coefficients are used.  The system to be solved is given in such a way that the new basis satisfies by construction (BA). In other word the condition (BA) determines the small size system to be solved. Since on a uniform mesh all interior basis function have the same form, only a small number of different systems has to be solved, and this step is, as all other steps, computationally cheap. We point out that we have here to enlarge the support from at most $p+1$ elements to $2p+1$ elements. A biorthogonal basis having the same support as the trace basis and optimal reproduction property does not exist for $p > 1$. In the standard finite element case, we refer to \cite{wohlmuth2000b} for $p=1$ and to \cite{OW02} for $p> 1$.

\begin{figure}[ht]
\begin{center}
\includegraphics[width=0.8\textwidth]{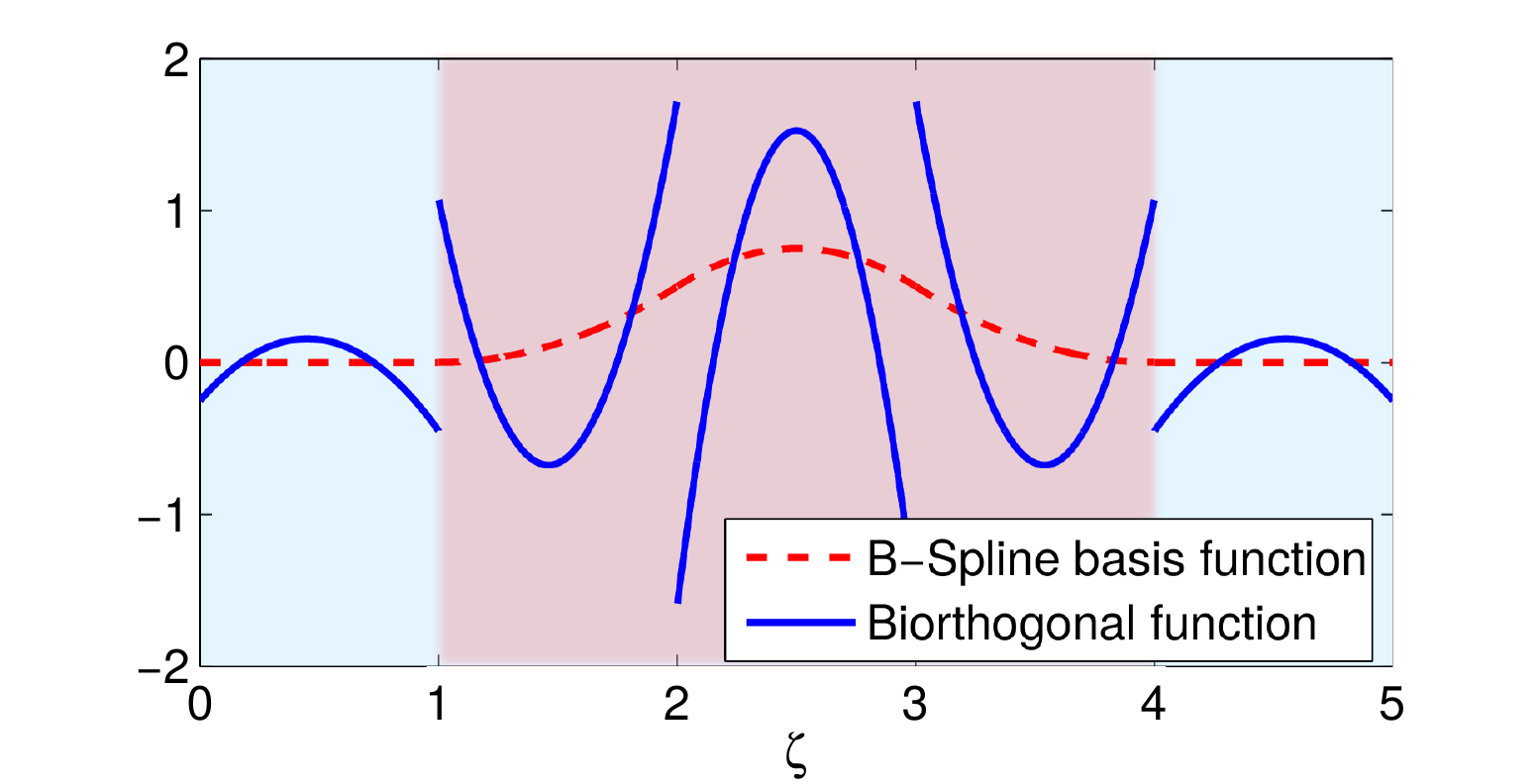}
\end{center}
\caption{Example of a interior biorthogonal basis function, $p=2$.}
\label{fig:localdualex}
\end{figure}

Figure \ref{fig:localdualex} illustrates one interior dual basis function for $p=2$. In this case the support is enlarged from 3 to 5 elements as indicated by the shadowed regions.

\textbf{Step IV:} In case of crosspoints/wirebaskets, we further reduce the dimension of the basis functions such that (UC) holds with respect to a smaller trace space.
It is of importance to note that this step has to be worked out carefully such that (BA) is not lost.

Although, Steps I-IV are quite technical and can be for $d=3$ on non-uniform meshes prone to coding bugs, all steps are local in the sense that the size of all involved systems to be solved does depend on $p$ and $d$ but not on the meshsize. 

\begin{figure}[ht]
\begin{center}
\includegraphics[width=0.88\textwidth]{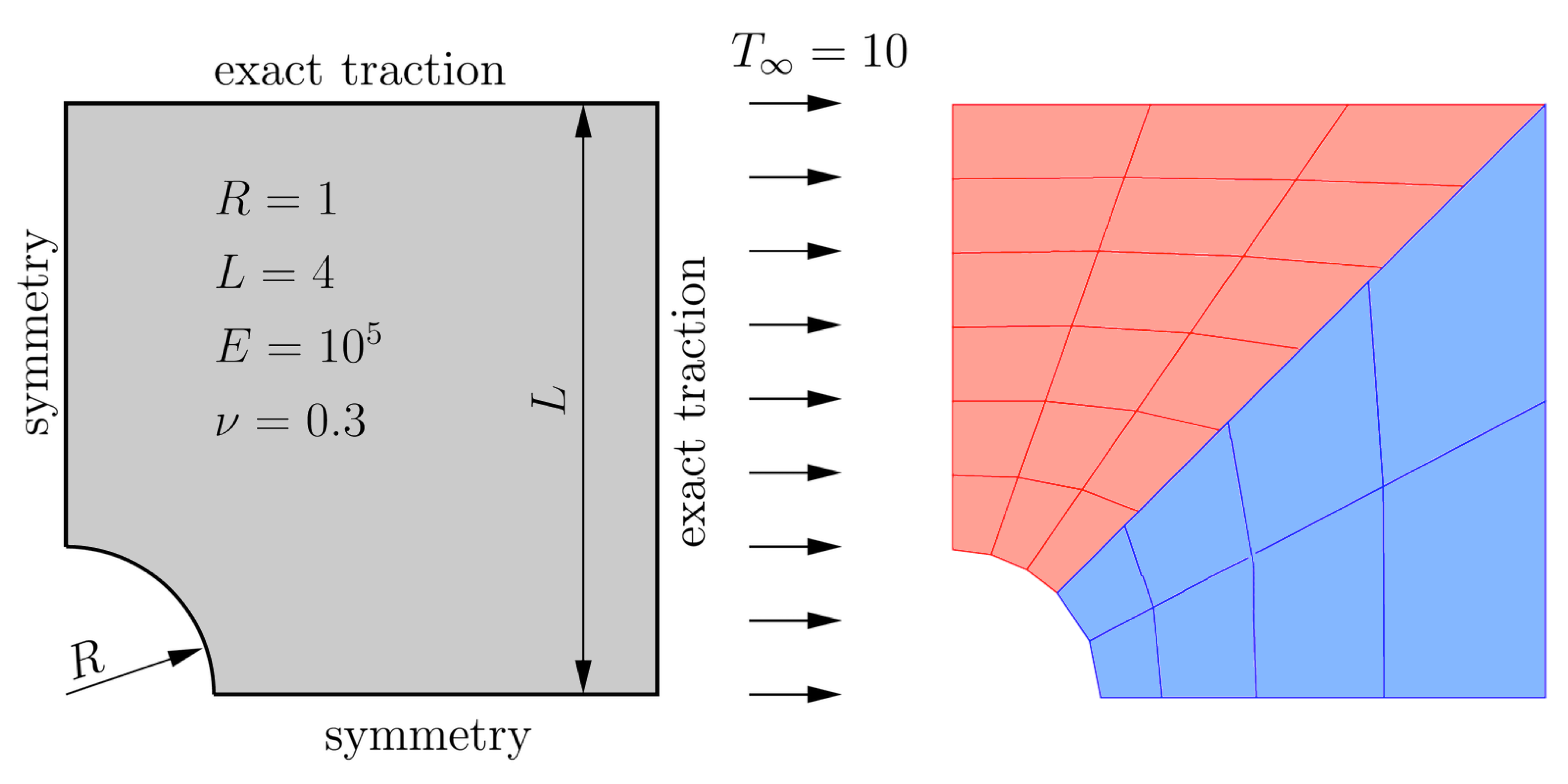}
\end{center}
\caption{2D plate with a hole - Geometry and setup (left) and two-patch parametrization (right) with an exemplary mesh ratio of 2:3. Reproduced and slightly modified from \cite{wunderlich2018}.}
\label{fig:platewithhole_setup}
\end{figure}

In the rest of this subsection, we illustrate the robustness and flexibility of the approach by a numerical example previously discussed in full detail in \cite{wunderlich2018}. As test case, we consider the well-known 2D benchmark of an infinite plate with a hole with the equations of linear elasticity. Due to symmetry, only a quarter of the plate is considered, and the infinite geometry is cut with the exact traction being applied as a boundary condition. As exemplary geometric setup, we choose two patches with a straight interface, but where the parametrization of the interface is different in the two patches. The entire setting is illustrated in Figure \ref{fig:platewithhole_setup} for a mesh ratio of 2:3.

Numerical investigations are outlined here for quadratic ($p=2$) and cubic ($p=3$) splines and three different choices for the Lagrange multiplier bases. Specifically, a so-called standard Lagrange multiplier basis ('std'), which is constructed as trace space of the primal space restricted to the slave side and therefore does not satisfy the (LS) condition, is compared with the two variants of biorthogonal Lagrange multiplier bases introduced above: the elementwise approach ('ele dual') from Figure \ref{fig:localdual}, which violates the (BA) property, and the approach with slightly enlarged support from Step III above ('optimal'), which satisfies all four conditions (BA), (LS), (BC) and (UC).

\begin{figure}[ht]
\begin{center}
\includegraphics[width=1.0\textwidth]{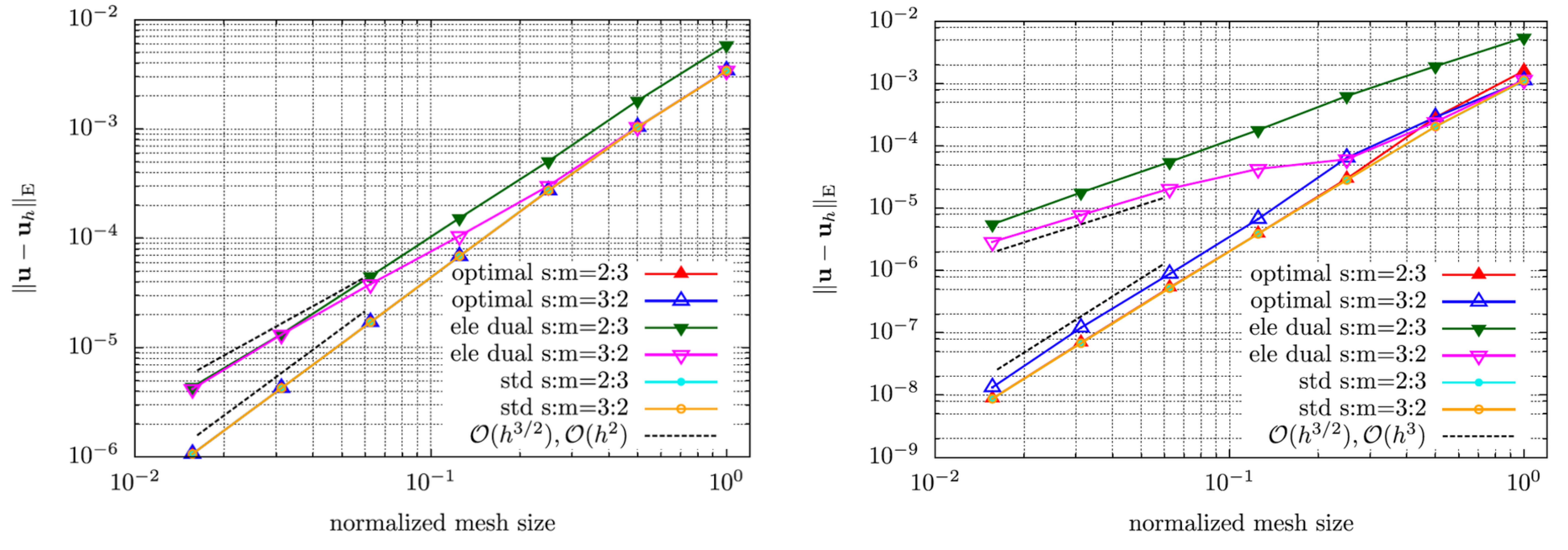}
\end{center}
\caption{2D plate with a hole - Convergence results for different Lagrange multiplier bases for $p=2$ (left) and $p=3$ (right) with exemplary mesh ratios of 2:3 and 3:2. Reproduced and slightly modified from \cite{wunderlich2018}.}
\label{fig:platewithhole_results}
\end{figure}

Convergence results under uniform mesh refinement, with the discretization error of the displacement field $\vec{u}$ measured in the energy norm $\Vert \vec{u} - \vec{u}_h \Vert_E$, are reproduced from \cite{wunderlich2018} in Figure \ref{fig:platewithhole_results}. Several key results can be identified: first and foremost, both the standard Lagrange multiplier basis as well as the optimal biorthogonal basis yield the expected convergence order of $\mathcal{O}(h^p)$ in all considered cases, and the absolute error levels are comparable. It should be kept in mind, however, that only the optimal biorthogonal basis from Step III above at the same time guarantees local support of the basis functions (LS). Second, the simple elementwise biorthogonal basis from Step II clearly cannot provide optimal results in all cases, but may exhibit a deteriorated convergence order of $\mathcal{O}(h^{3/2})$. This becomes particularly apparent if the slave mesh is coarser than the master mesh and for higher-order interpolations (here $p=3$). The results underline the importance of all three steps in the construction of biorthogonal Lagrange multiplier basis functions  highlighted above.

%\todoB{an Alexander: hier wuerde noch ein weiteres Bsp  aus der Diss von A. Seitz welches nicht im Paper ist gut dazu passen}
%\todoA{gute Idee aber geht leider nicht, weil wir in der Diss von A. Seitz diesen Fall mit den optimalen IgA-Basen nicht behandelt haben... ich habe stattdessen eine knappe Zusammenfassung der Platte mit Loch reingenommen. Neue Beispiele aus der Diss A. Seitz, die nicht in den Papers enthalten sind, gibt es dann eher in Kap. 4.1 / 4.2.}

\subsection{Multi-patch analysis for Kirchhoff–Love shells}\label{sec:higher}

As a first prototypical example for the use of IgA concepts and advanced mortar methods in elasticity, we focus on Kirchhoff–Love (KL) shell elements. The developments here are based on \cite{schuss2018}. The kinematical assumptions of KL shells rely on the out of plane curvature terms, used to describe the bending of the shell. This approach requires a general $G^1$ continuity across the whole domain (see \cite{kiendl2009} for details on $G^1$ continuity), in contrast to Hellinger-Reissner (HR) beams developed throughout the past decades. As IgA naturally allow us to deal with equations of higher-order, the central drawback of KL shell elements is removed. 

For complex geometries, the domain is always divided into sub-patches $\Omega_m$, $m = 1,\hdots,M$ with interfaces $\Gamma_l$, $l = 1,\hdots,L$. In particular, we require  $G^1$ continuous patch connection of the in general non-conform discretized patches. Within a classical mortar method to enforce $C^0$ continuity across the interface, a Lagrange multiplier space is introduced by the trace space of the displacements restricted to the slave side $\Gamma_{1}$. Now we can state that for a given $\vec{\varphi}_\mathrm{h}^{(2)}$ at the interface $\Gamma_{2}$ on the mortar side we assume now that a $\vec{\varphi}_\mathrm{h}^{(1)}$ at the interface $\Gamma_{1}$ on the slave side can be found, such that the minimization problem
\begin{equation}\label{eq:mortarMin}
\|\vec{\varphi}^{(1)}_{\indi{h}}-\vec{\varphi}^{(2)}_{\indi{h}}\|_{L^2(\Gamma_{1})}^2 = 
\inf_{\vec{w}\in W_{\indi{h}}^{(1)}}\|\vec{w}-\vec{\varphi}^{(2)}_{\indi{h}}\|_{L^2(\Gamma_{1})}^2,
\end{equation}
is satisfied. Here, \(W_{\mathrm{h}}^{(1)}=\op{span}\{\vec{N}^r_{(1)}\}\), where $\vec{N}^{r}_{(j)}$ are B-Spline shape functions on side $j$, restricted to the subset $r$, see \cite{schuss2018}, Section 3.2 for details. This leads to the classical mortar formulation of the constraints
\begin{equation}\label{eq:mortar1}
 \Phi^{0} := 
 \int\limits_{\Gamma_{1}}\left(\vec{N}_{(1)}^{r_1} \cdot\vec{N}_{(1)}^{r_2}\,q^{(1)}_{r_2}-\vec{N}_{(1)}^{r_1}\cdot\vec{N}_{(2)}^{r_3}\,q^{(2)}_{r_3}\right)\d \Gamma,
\end{equation}
where we have made use of
\begin{equation}\label{eq:basemod23}
\vec{\varphi}_{\mathrm{h}}^{(1)} = \sum_{r=1}^{3\,\mathfrak{n}^{(1)}} q_r^{(1)}\, \vec{N}^{r}_{(1)},\quad
\vec{\varphi}_{\mathrm{h}}^{(2)} = \sum_{r=1}^{3\,\mathfrak{n}^{(2)}}q_r^{(2)} \,\vec{N}^{r}_{(2)}.
\end{equation}
Note that we use the discrete Lagrange multiplier space with biorthogonality conditions between the primal and the dual basis functions, see \ref{sub:biorth} for details. The situation is different for a \(G^{1}\) continuous coupling. Therefore, we assume again that for a given $\vec{\varphi}_\mathrm{h}^{(2)}$ at the interface $\Gamma_{2}$ on the mortar side a $\vec{\varphi}_\mathrm{h}^{(1)}$ at the interface $\Gamma_{1}$ on the slave side can be found, such that
\begin{figure}[t]
\begin{center}
\begin{minipage}[t]{0.49\textwidth}
\includegraphics[width=\textwidth]{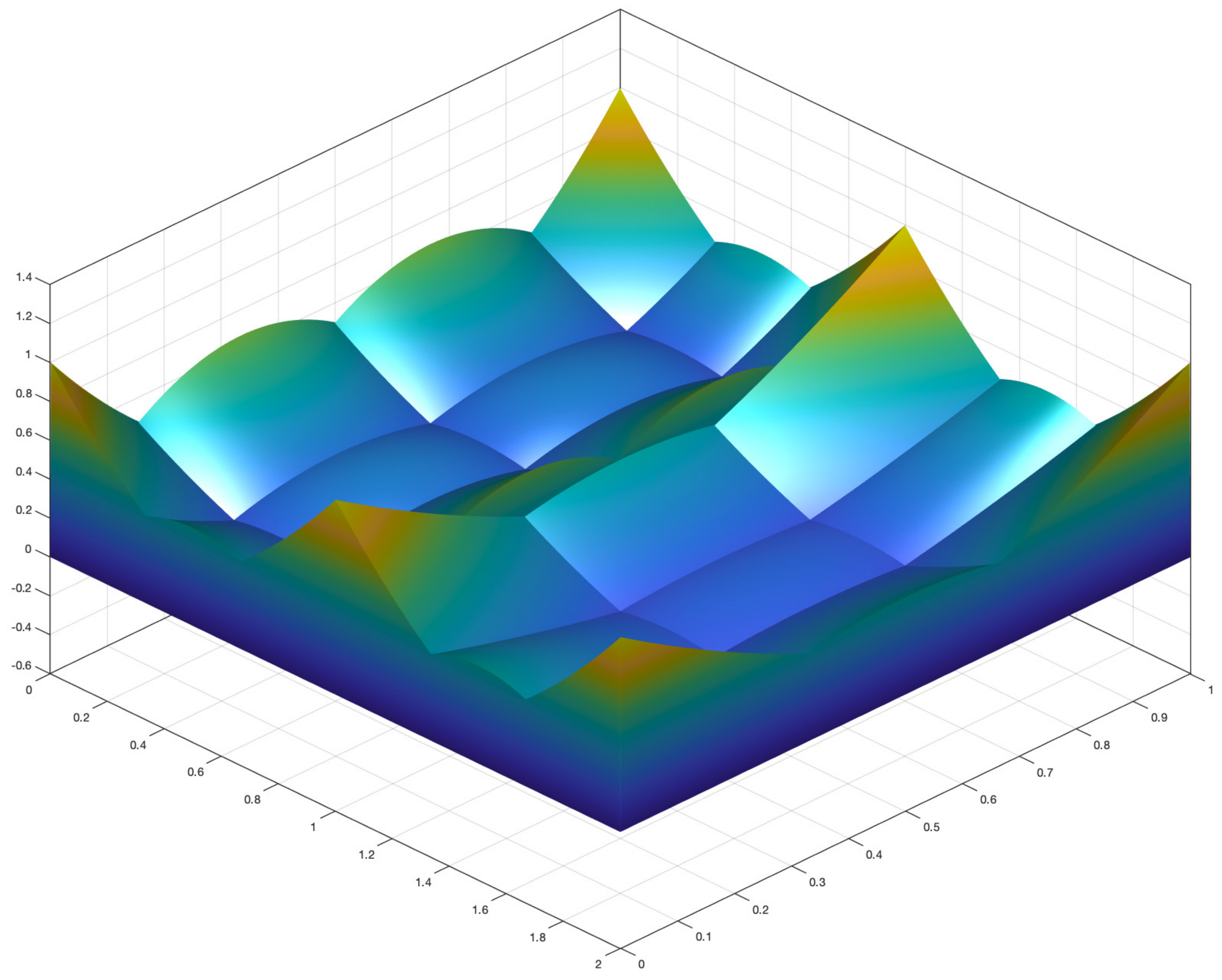}
\end{minipage} \hspace{0.0025\textwidth}
\begin{minipage}[t]{0.49\textwidth}
\includegraphics[width=\textwidth]{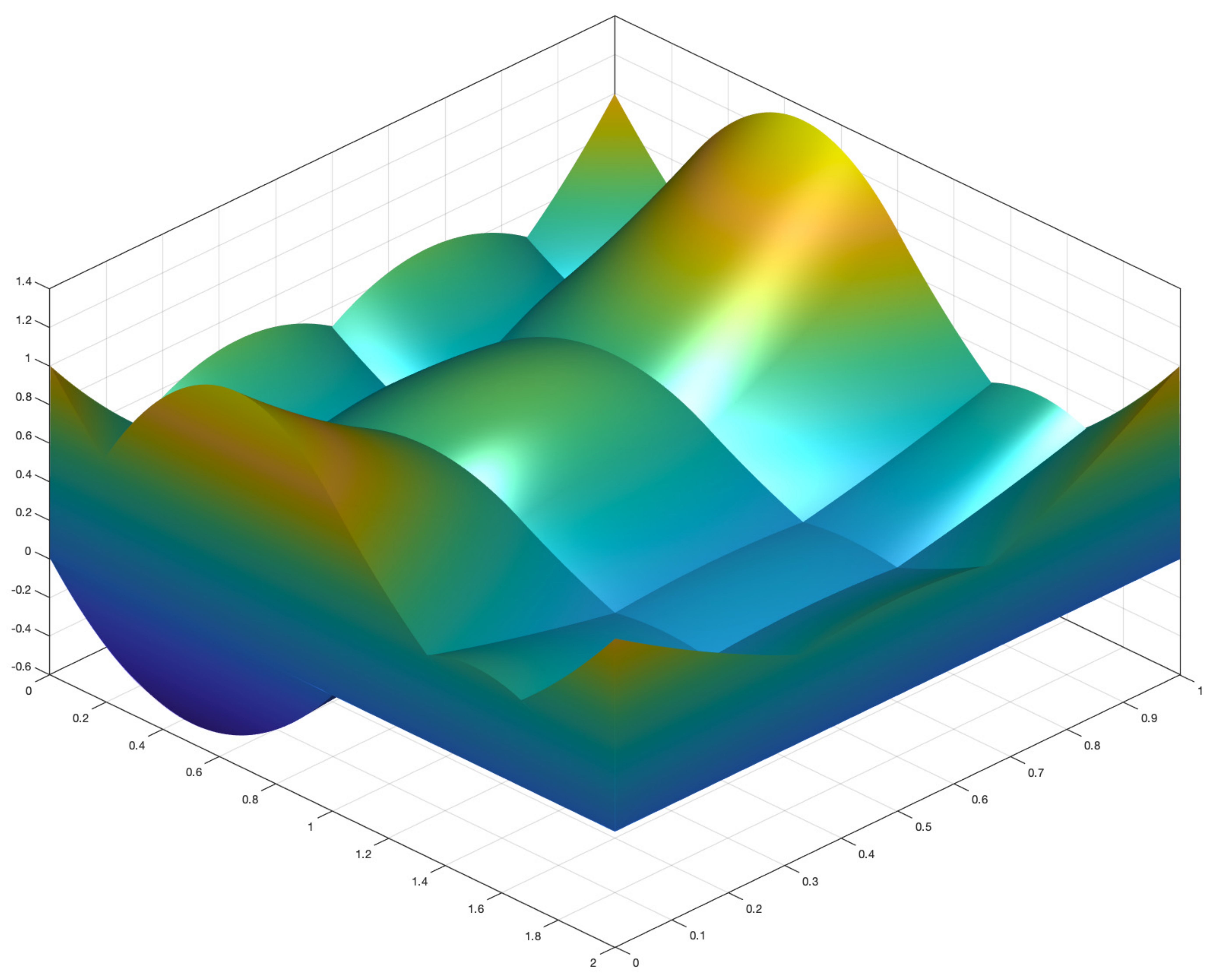}
\end{minipage} 
\end{center}
\vspace*{-4mm}
\caption{Initial bases (left) and modified bases functions (right).}
\label{fig:basemod1}
\end{figure}
\begin{equation}\label{eq:g1}
\begin{aligned}
\|\vec{\varphi}^{(1)}_{\indi{h}}-\vec{\varphi}^{(2)}_{\indi{h}}\|_{L^2(\Gamma_{1})}^2+
\sum\limits_{\alpha=1}^{2}\,\int\limits_{\Gamma_{1}}\left(\vec{\varphi}_{\mathrm{h},\alpha}^{(1)}-
\sum_{k=1}^2 \lambda_k^\alpha\,\vec{\varphi}_{\mathrm{h},k}^{(2)}\right)^2\,\d\Gamma = \\
\inf_{\vec{w}\in W_\mathrm{h}^{(1)}}
\left[  
\|\vec{w}-\vec{\varphi}^{(2)}_{\indi{h}}\|_{L^2(\Gamma_{1})}^2 +
\sum\limits_{\alpha=1}^{2}\,\int\limits_{\Gamma_{1}}\left(\vec{w}_{,\alpha}-
\sum_{k=1}^2 \lambda_k^\alpha\,\vec{\varphi}_{\mathrm{h},k}^{(2)}\right)^2\,\d\Gamma\right]
\end{aligned}
\end{equation}
is satisfied, where we make use of the notation $(\bullet)_{,\alpha}$ for the derivative with respect to the direction $\alpha$. This is equivalent to enforce
\begin{equation}\label{eq:mortar2} 
\begin{aligned}
 \Phi^{1} := \,&\Phi^{0} + \int\limits_{\Gamma_{1}}\vec{N}_{(1),1}^{r_1} \cdot\vec{N}_{(1),1}^{r_2}\,q^{(1)}_{r_2}-\vec{N}_{(1),1}^{r_1}\cdot
\left(\sum\limits_{k=1}^2\lambda_k^1\, \vec{N}_{(2),k}^{r_3}\right)q^{(2)}_{r_3}\d \Gamma+ \\
&\int\limits_{\Gamma_{1}}\vec{N}_{(1),2}^{r_1}\cdot\vec{N}_{(1),2}^{r_2}q^{(1)}_{r_2}-\vec{N}_{(1),2}^{r_1}\cdot
\left(\sum\limits_{k=1}^2\lambda_k^2\, \vec{N}_{(2),k}^{r_3}\right)q^{(2)}_{r_3}\d \Gamma.
\end{aligned}
\end{equation}
\begin{figure}[t]
\begin{center}
\begin{tabular}{cc}
\footnotesize
\psfrag{f}{\(\vec{t}\)}
\vspace{5mm}\includegraphics[width=0.27\textwidth]{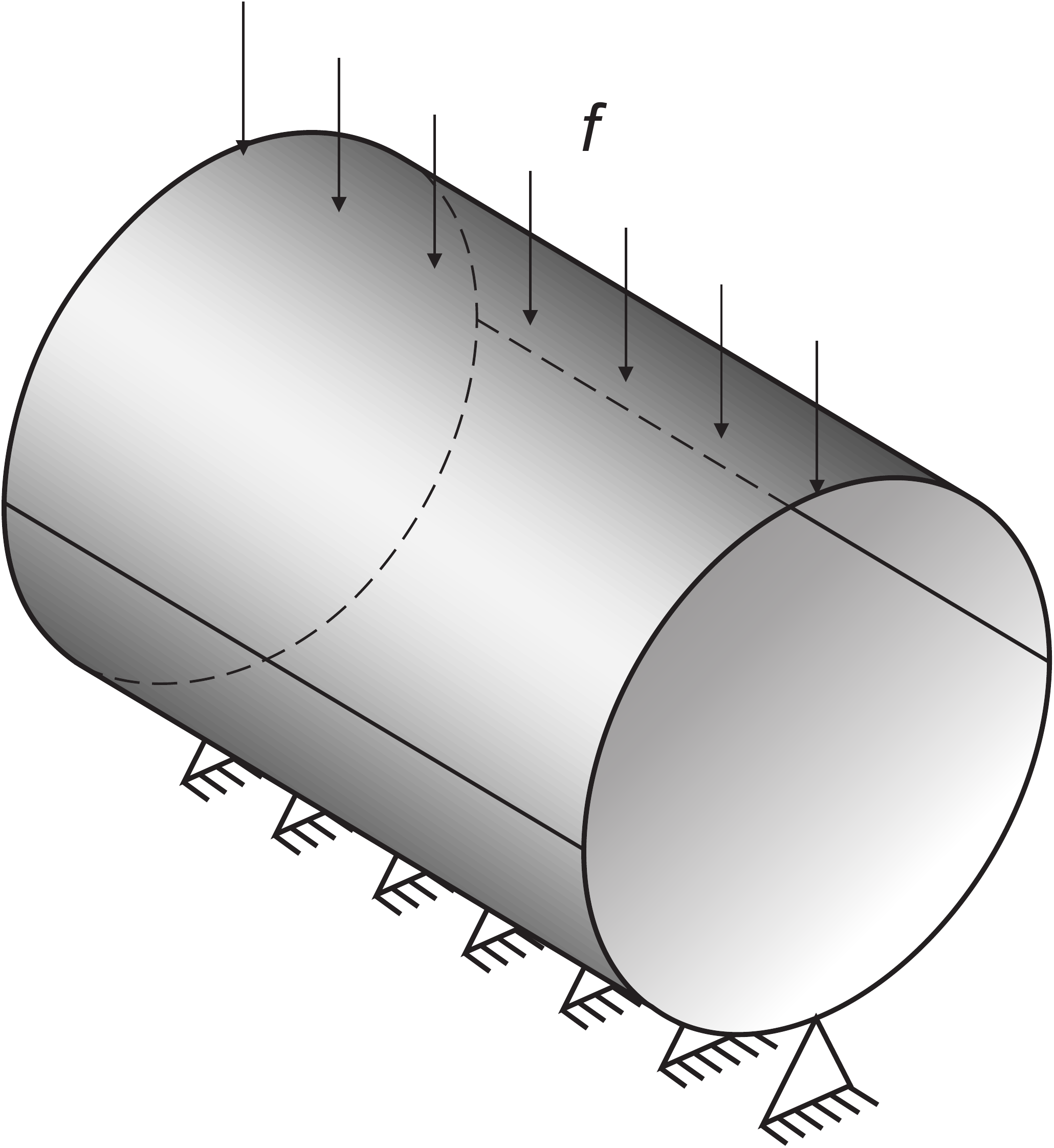}&
\includegraphics[width=0.6\textwidth]{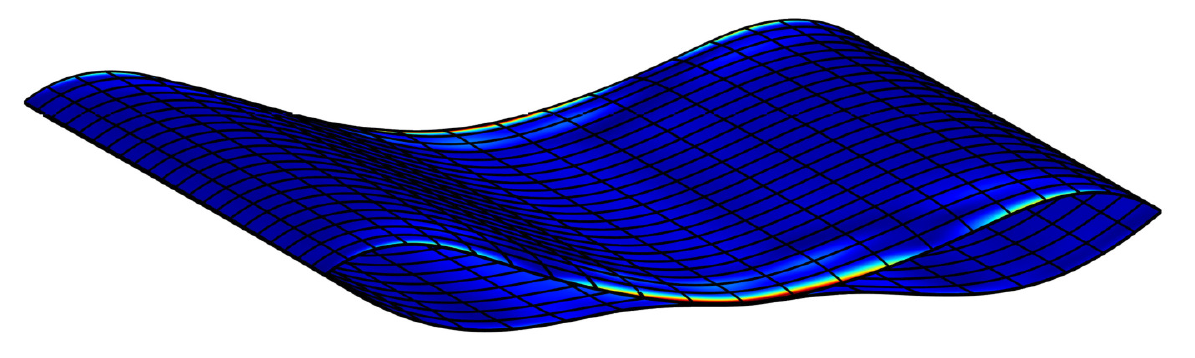}\\
&\hspace{-50mm}\includegraphics[width=0.6\textwidth]{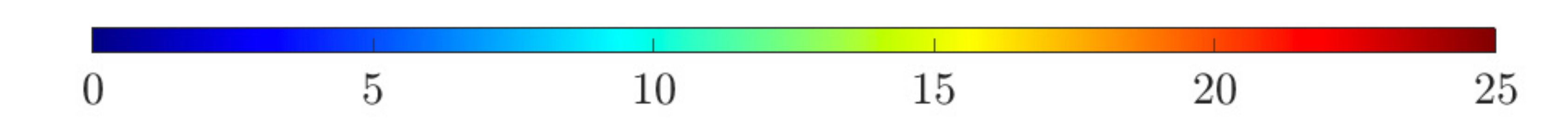}
\end{tabular}
\end{center}
\caption{Reference configuration and boundary conditions (left) and  von Mises stress distribution with $C^0$ continuous mortar coupling (right)}
\label{fig:cylinder}
\end{figure}
Here, $\lambda_i^j$, $i,j = 1,2$ are four real numbers, defined at each point of the surface to control the $G^1$ continuity. Note that for $\lambda_i^j = \delta_i^j$ we obtain $C^1$ continuity.  Certain ways exist to enforce the mortar constraints. Here, we provide some information for a local condensation procedure where we calculate modified basis functions to avoid the explicit usage of Lagrange multipliers. Therefore, we distribute points $\vec{\xi}_i$ on the parametric domain of the finer-meshed surface and determine the corresponding parameters $\vec{\xi}'_i$ with $\bar{\vec{\varphi}}_{\mathrm{h}}^{(2)}(\vec{\xi}'_i)=\bar{\vec{\varphi}}_{\mathrm{h}}^{(1)}(\vec{\xi}_i)$ using an orthogonal projection. Afterwards we use the information from \eqref{eq:g1} evaluated at the distributed points to calculate new bases functions, see Figure \ref{fig:basemod1} for a graphical representation of modified functions. This procedure can be considered as a local null-space reduction scheme, acting on the space of basis functions.

\begin{figure}[t]
\begin{center}
\includegraphics[width=0.65\textwidth]{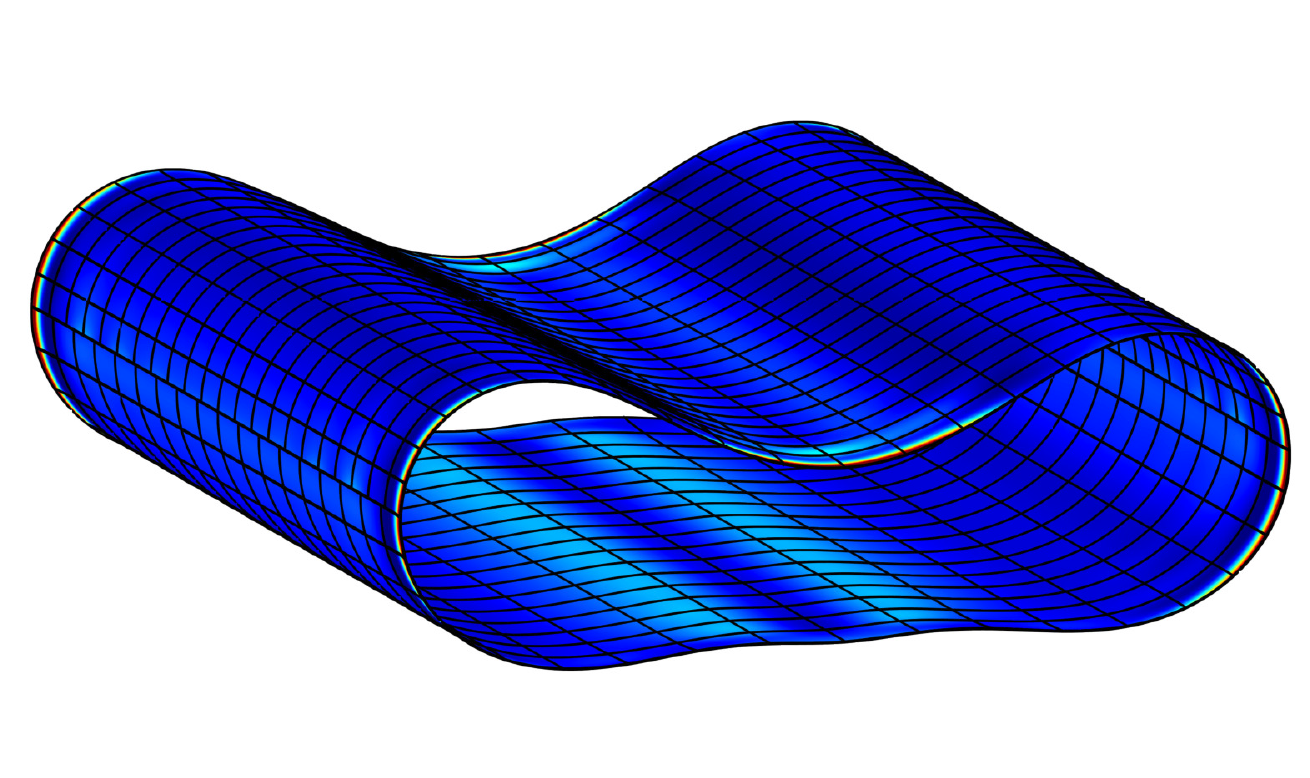}\\
\vspace{5mm}
\includegraphics[width=0.65\textwidth]{pictures/cylinder_colorbar_colormap}
\end{center}
\caption{Von Mises stress result for \(G^{1}\) coupling.}
\label{fig:cylinderStress}
\end{figure}

To demonstrate the applicability to KL shells, we investigate a cylinder composed of two, initially curved shell patches discretized by \(18\times 18\) and \(20\times 20\) cubic NURBS elements, respectively. Note that the NURBS weights are chosen such that two perfect half cylinders with a radius of \(1\,\mathrm{m}\), a length of \(3\,\mathrm{m}\) and a thickness of \(0.02\,\mathrm{m}\) are obtained. The cylinder is fixed along a bottom line and a line load of \(28\,\mathrm{N/m}\) is applied on the opposite side as shown in Figure \ref{fig:cylinder}, left side. On the right side, a classical $C^0$ continuous mortar method is applied, which does not allow for a transfer of bending moments across the interface. The effects on the deformed geometry displayed are obvious. In contrast, the results in Figure \ref{fig:cylinderStress} demonstrate that the coupling conditions satisfying \eqref{eq:g1} at the interface can counterbalance the non-matching meshes. Note that the $G^1$ coupling conditions are in general linear constraints, such that linear and angular momentum are conserved quantities throughout the interface, providing that the constraints are fulfilled in the reference configuration.

\subsection{Weak $C^n$ coupling for solids}

In \cite{dittmann2018c}, the previously introduced concept of high-order mortar coupling conditions is extended towards the application on Cauchy continua. Moreover, we investigate different evaluations of general $C^n$ continuous coupling conditions, written in terms of a saddle point system. The constraints for $C^0$ and $C^1$ continuous coupling, respectively, have been introduced in \eqref{eq:mortar1} and \eqref{eq:mortar2}, using $\lambda_i^j = \delta_i^j$. The extension for $C^2$ follows immediately via
\begin{equation}
 \Phi^{2} := \Phi^{1} + 
 \sum\limits_{\substack{j,l=1 \\ j\geq l}}^{3}\;
 \int\limits_{\Gamma_{1}}\left(\vec{N}_{(1),jl}^{r_1}\cdot\vec{N}_{(1),jl}^{r_2}\,q^{(1)}_{r_2}-
 \vec{N}_{(1),jl}^{r_1}\cdot\vec{N}_{(2),jl}^{r_3}\,q^{(2)}_{r_3}\right)\d \Gamma,
\end{equation}
which can be extended towards general $C^m$ continuity in a straight forward manner. 

One particular challenge in realization of a mortar method is the evaluation of the interface integral. This might be more a technical issue, but extremely important for an efficient implementation. Any quadrature rule based on the slave mesh does not respect the mesh lines of the master mesh and vice versa for a quadrature rule on the master mesh. Therefore it is common to use a quadrature rule based on a merged mesh, i.e.\ a mesh leads to an exact evaluation of the integral, if it respects the reduced smoothness of the master and slave functions at their respective lines. The standard mortar analysis assumes that the interface is resolved by the mesh on the master and the slave side. In that case, no projection of points on the discrete master side onto the discrete slave side and vice versa is required. Consequently the construction of the common mesh, named segmentation process, is still challenging but does not result in an additional variational crime. The situation is different for curved interfaces where the discrete interfaces do not match in general. Then for the segmentation process a mapping from the vertices of the master side onto the discrete slave interface is required. For the evaluation of the basis function on the master side the quadrature points of the merged mesh have to be mapped back onto the discrete master side, which results in an additional error contribution. 

\begin{figure}[t]
\begin{center}
\footnotesize
\begin{tabular}{cc}
\psfrag{loadp}[c][c]{\(\bar{\vec{T}}\)}
\includegraphics[width=0.24\textwidth]{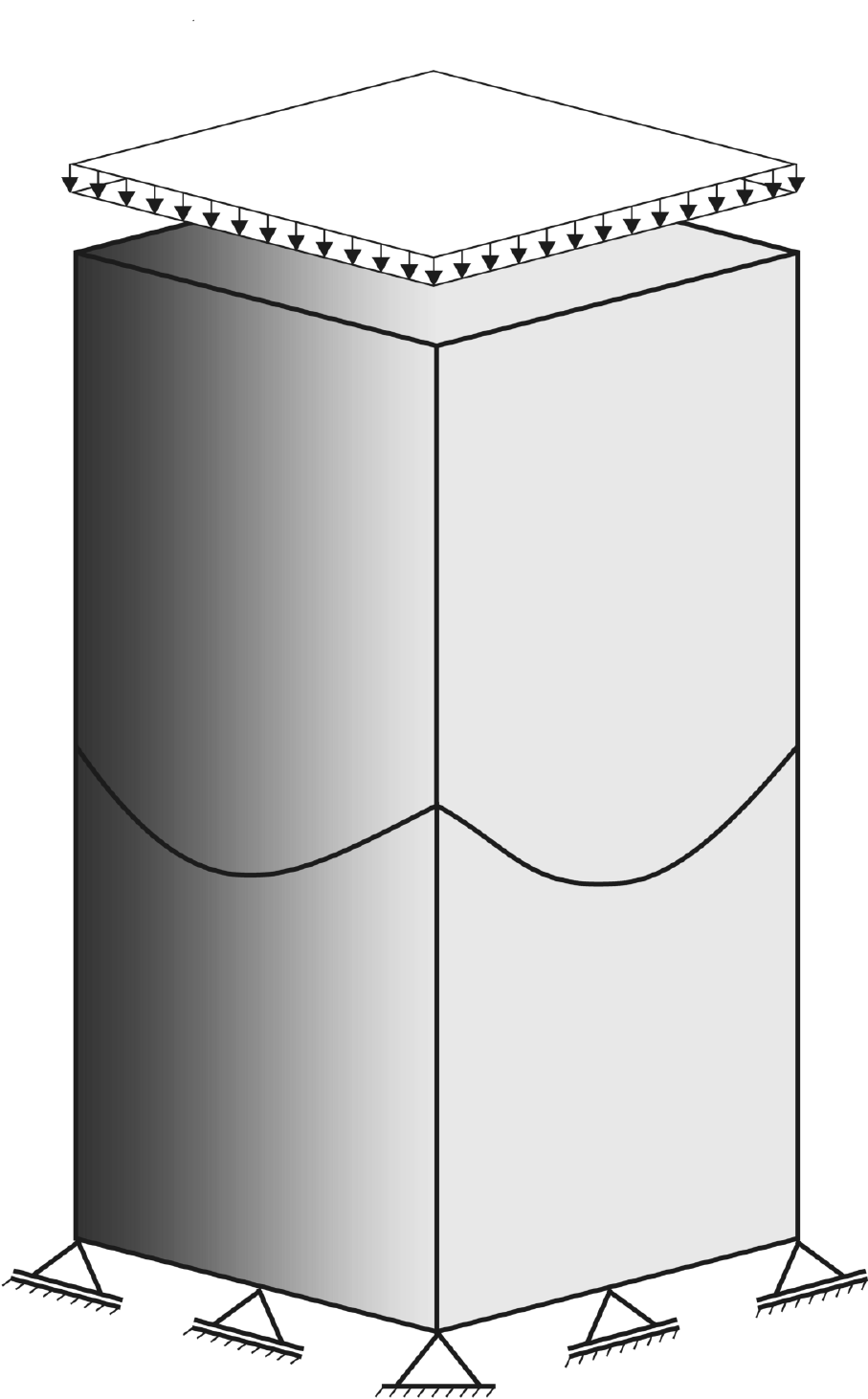}\hspace{15mm}
\includegraphics[width=0.34\textwidth]{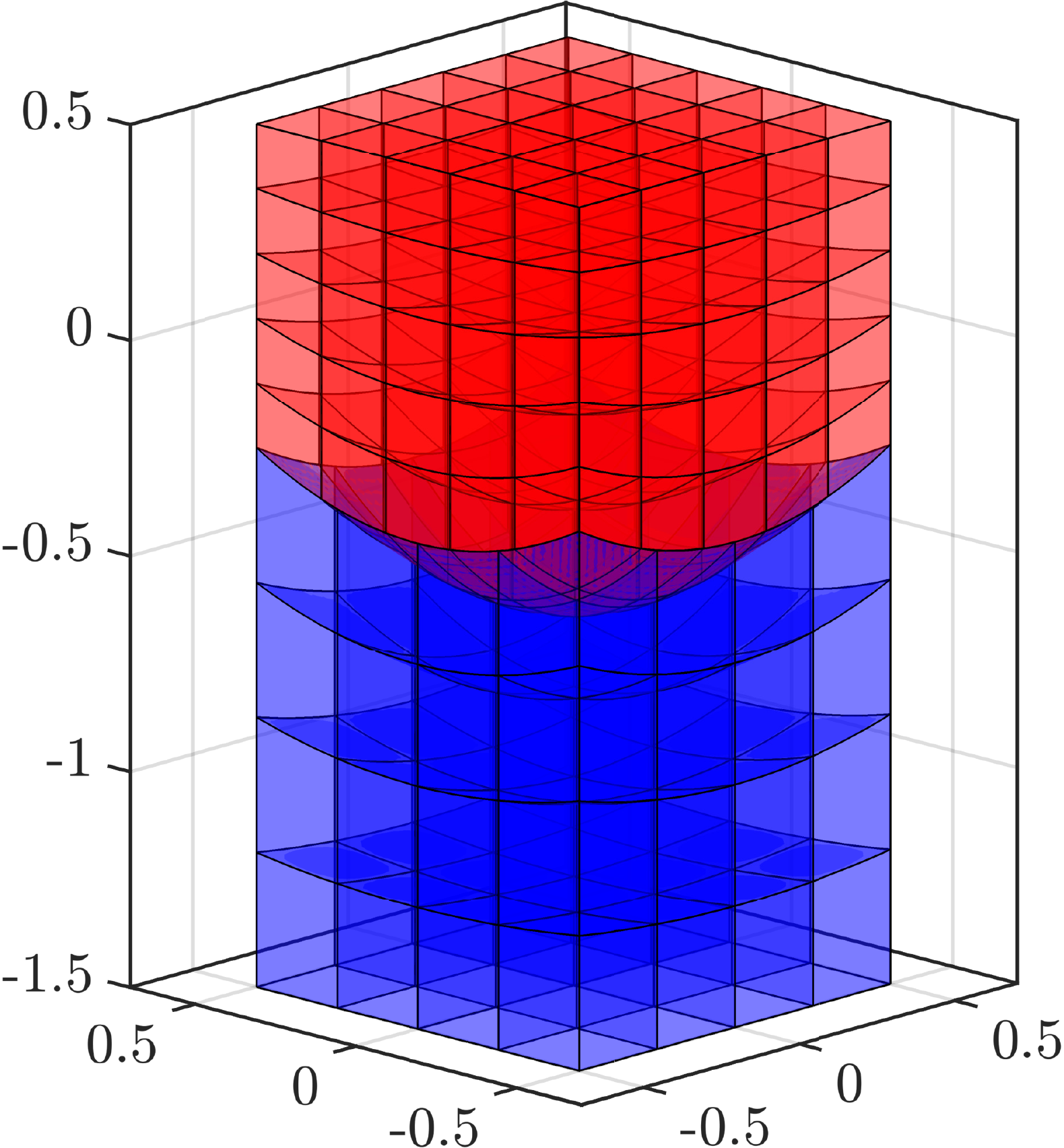}\scriptsize\([\mathrm{m}]\)
\end{tabular}
\end{center}
\caption{\textbf{Patch test.} Reference configuration (left) and computational mesh (right) for a patch test with curved interface.}
\label{fig:patchTestCurvedRef}
\end{figure}

In contrast, equidistant sample points on the parametric domain can be taken into account, which can be interpreted as a midpoint quadrature formula on a sub-mesh with respect to a weighted Lebesgue measure. More precisely, we use the Lebesgue measure on the parametric domain assuming a uniform decomposition. Then the quadrature weights are identically given by \(h^2\) and are just a constant scaling which does not alter the least-squares approach. To separate the effect of the approximation error we reconsider a patch test and recall, that the stress is a constant in the domain. Therefore, a fixed number of elements is applied with a curved interface in between, see Figure \ref{fig:patchTestCurvedRef}.
\begin{figure}[t]
\begin{center}
\footnotesize
\includegraphics[width=0.8\textwidth]{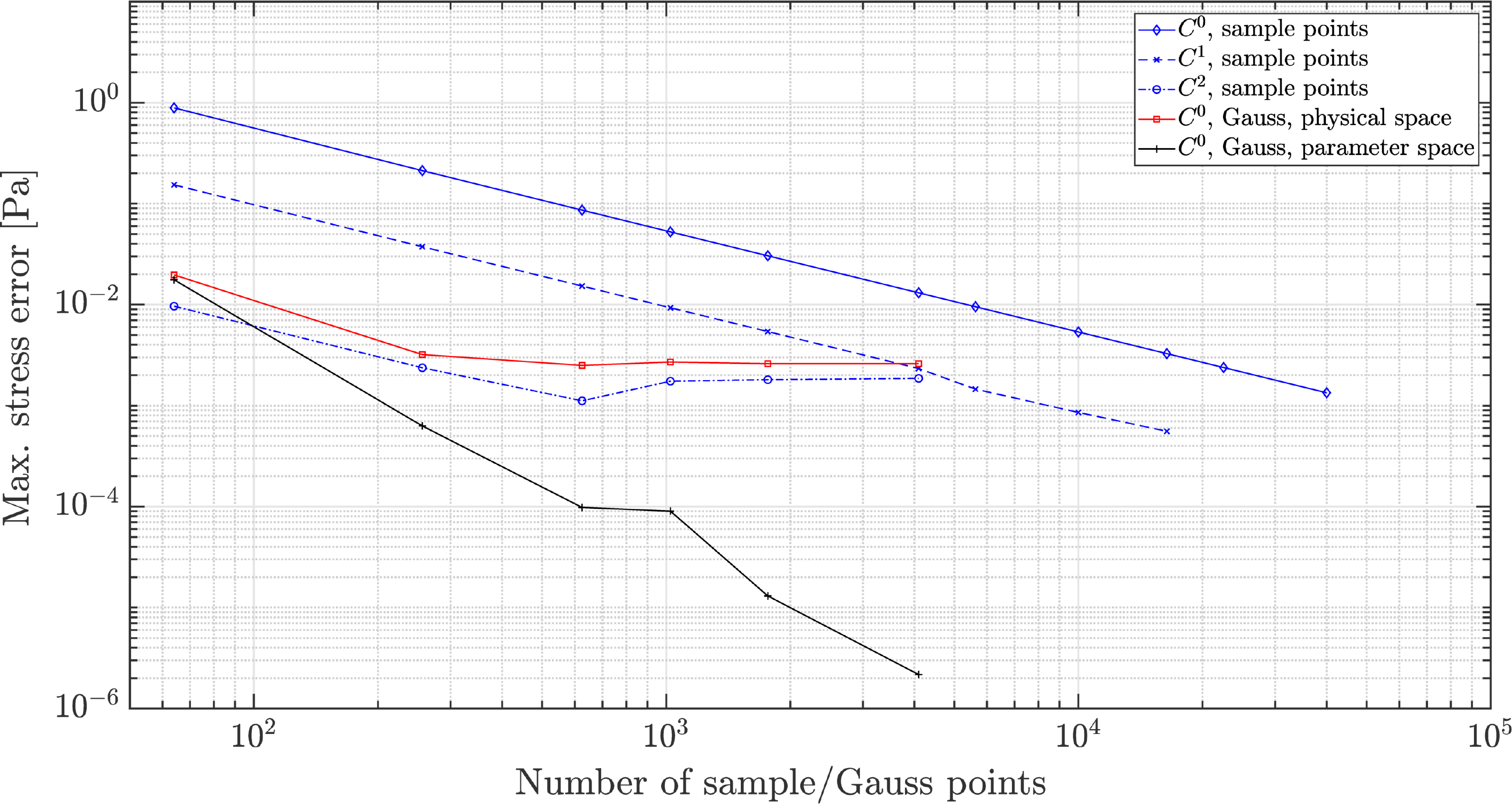}
\end{center}
\caption{Maximum error in the von Mises stress result of a patch test plotted over the total number of sample points per element.}
\label{fig:patchTestCnvergenceLog2}
\end{figure}
Figure \ref{fig:patchTestCnvergenceLog2} shows the results for a $C^0$, $C^1$ and $C^2$ coupling using different numbers of sample points. The error for the sample point evaluation decays with the same order independent of the enforced continuity across the interface, whereby the asymptotic limit for \(C^{2}\) coupling is already reached at 25 sample points per element and dimension. For comparison, a Gauss integration with rising number of Gauss points is presented as well, enforcing $C^0$ continuity across the interface. To be precise, a Gauss integration on the parameter space as well as on the physical space using the usual transformation rule evaluated at each Gauss point on the interface, i.e.\ \(\|\vec{\varphi}_{\indi{h},\xi_1}\times\vec{\varphi}_{\indi{h},\xi_2}\|\) for the area transformation, has been applied. As can be seen, the asymptotic limit is reached for a small number of Gauss points on the physical space, whereas on the parameter space the Gauss integration converges. 

\begin{figure}[t]
\begin{center}
\footnotesize
\begin{tabular}{ccccc}
\includegraphics[height=0.3\textheight]{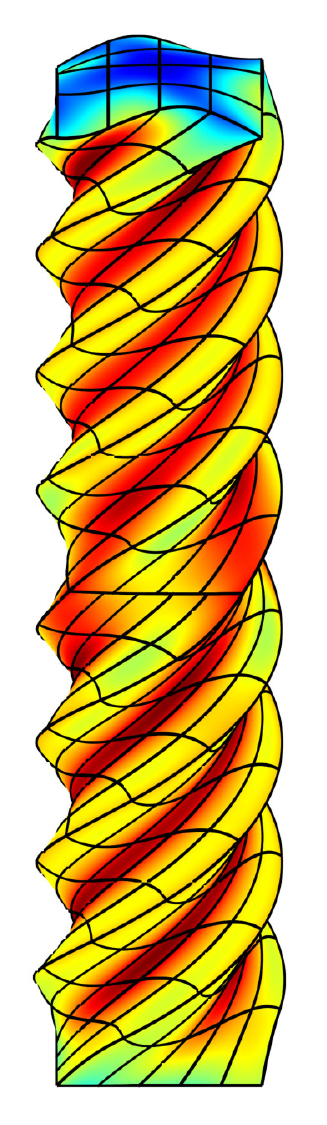}\hspace{7mm}
\includegraphics[height=0.3\textheight]{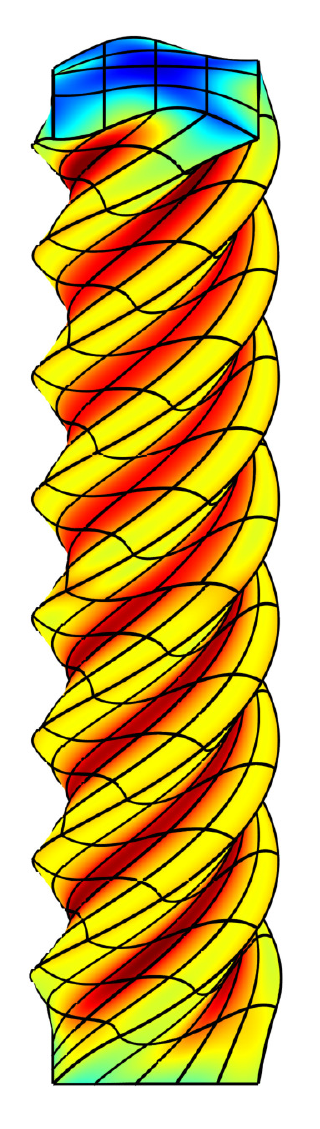}\hspace{7mm}
\includegraphics[height=0.3\textheight]{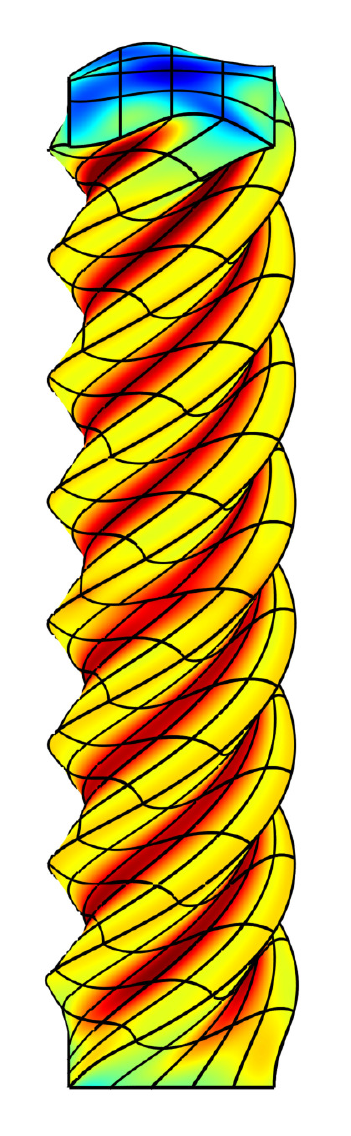}\hspace{7mm}
\includegraphics[height=0.3\textheight]{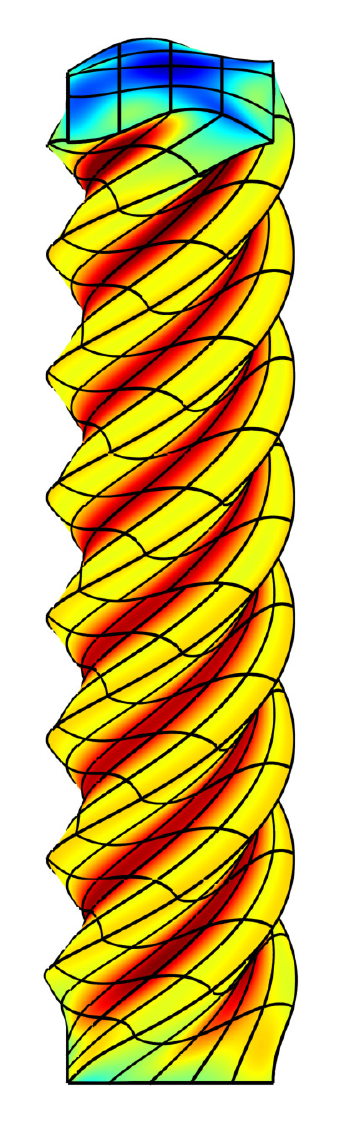}\hspace{7mm}
\includegraphics[height=0.3\textheight]{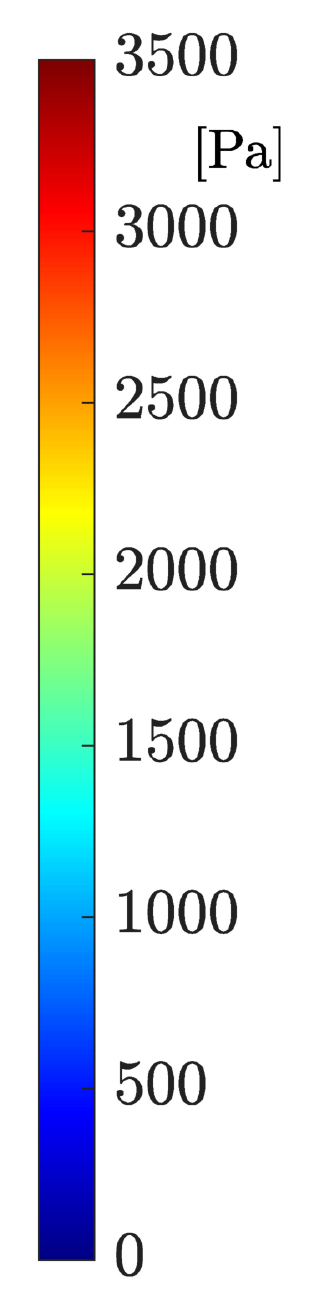}
\end{tabular}
\end{center}
\caption{\textbf{Twisted block.} Von Mises stress result for the quadratic discretization with full and reduced modification and the cubic 
discretization with full and reduced modification (from left to right).}
\label{fig:twistedStress}
\end{figure}

To demonstrate the applicability on large deformations, we introduce an example composed of two parts, which are bonded together via the basis modification approach. The lower surface is fixed in space and the upper surface is rotated by an angle of \(\phi=720^{\circ}\). For both parts, we apply quadratic as well as cubic B-spline based discretization, where the lower part consists of \(4\times 4\times 10\) elements and the upper part of \(5\times 5\times 10\) elements. In order to consider locking effects, we apply the full set of dependent degrees of freedom (dof), referred to as full modification, as well as a reduced set of dependent dofs, referred to as reduced modification. The von Mises stress distribution is depicted in Figure \ref{fig:twistedStress}. Concerning the result for the quadratic discretization with full modification, we observe a locking behavior at the interface since only \(54\) degrees of freedom are applied at the interface, reducing the approximation quality significantly. In contrast, the deformation is not suppressed for the reduced modification, where \(216\) degrees of freedom remain for the approximation of the interface. Moreover, the result of the cubic discretization with full modification shows a non significant locking behavior, visible only as a slightly reduced maximum value of the von Mises stress at the interface. Here, \(96\) degrees of freedom remain for the approximation at the interface. Eventually, for the cubic discretization with reduced modification, \(294\) degrees of freedom are used for the approximation at the interface such that we obtain a nearly perfect stress distribution across the interface without any disorders due to the basis modification. 
\subsection{Crosspoint modification }\label{sec:crosspoints}

In a last step concerning higher-order coupling conditions using an extended mortar method, we consider crosspoint modifications at the crosspoints between sub-patches of a multipatch geometry. This subsection is based on \cite{dittmann2020}, where a modification of the Lagrange multipliers is shown to decouple the interfaces, avoid overconstraint situations and resume the best approximation property (BA), as discussed in section \ref{sub:biorth}. 

\begin{figure}[t]
\begin{center}
\footnotesize
\begin{tabular}{ccc}
\includegraphics[width=0.3\textwidth]{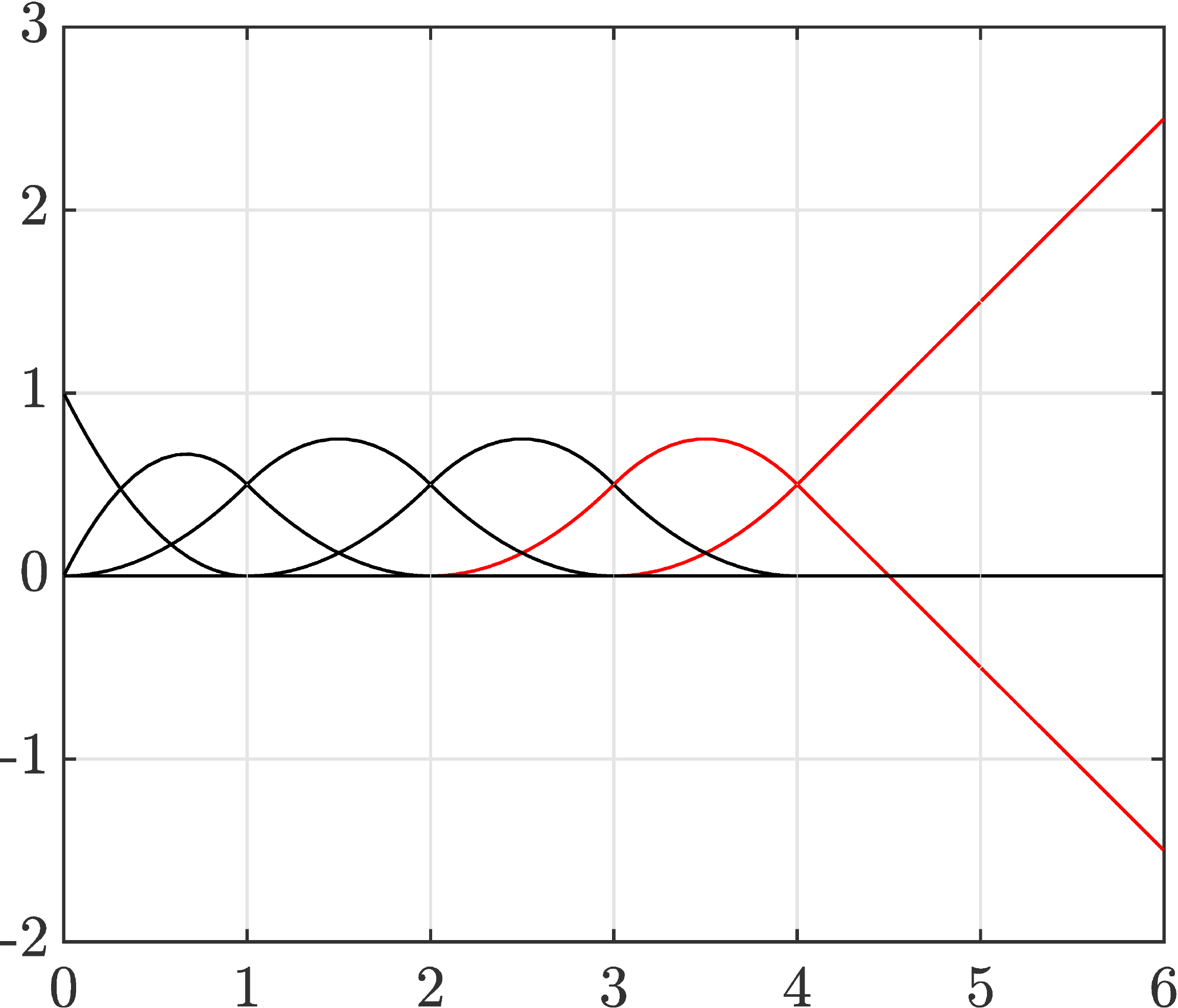}\hspace{2mm}
\includegraphics[width=0.3\textwidth]{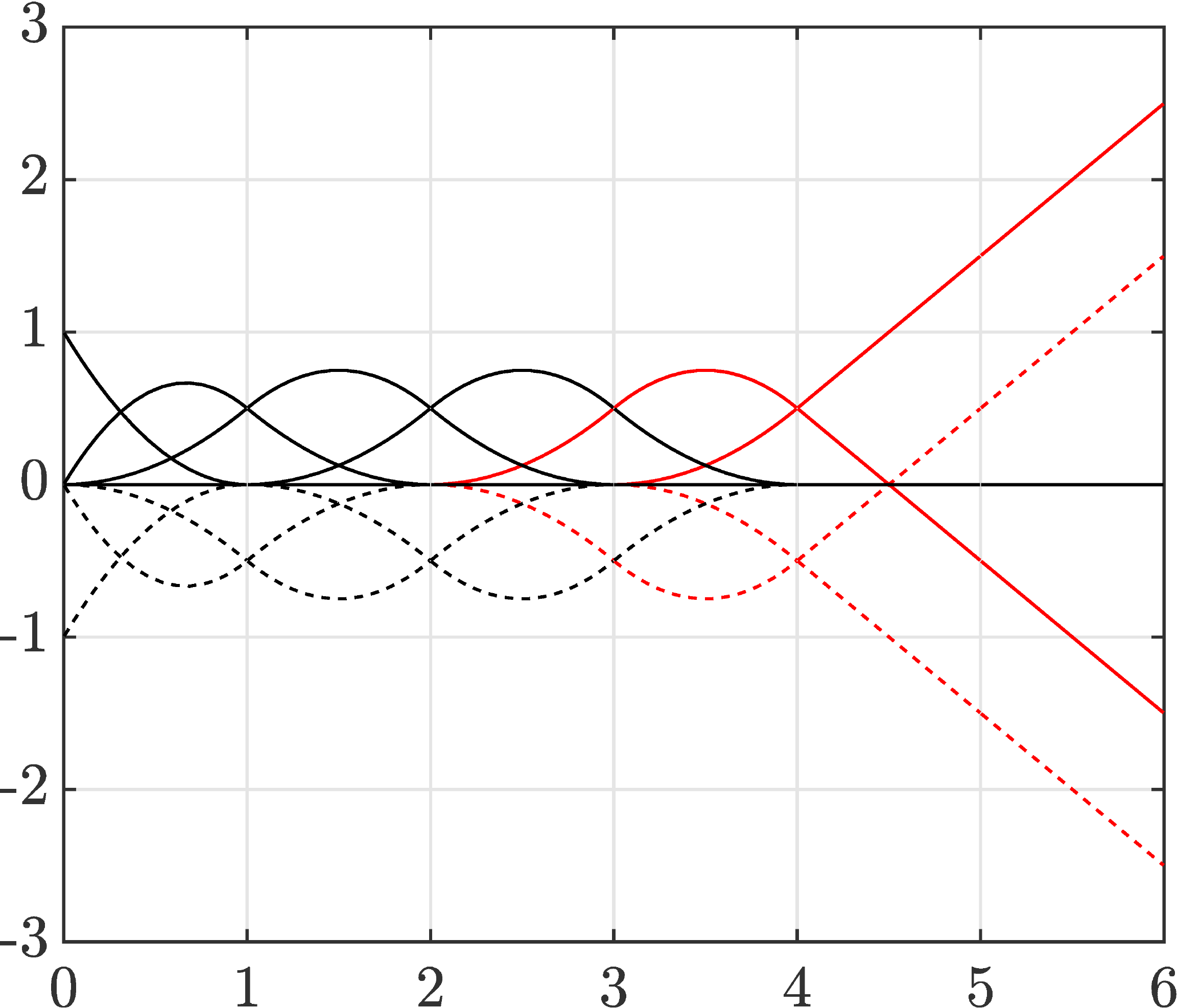}\hspace{2mm}
\includegraphics[width=0.31\textwidth]{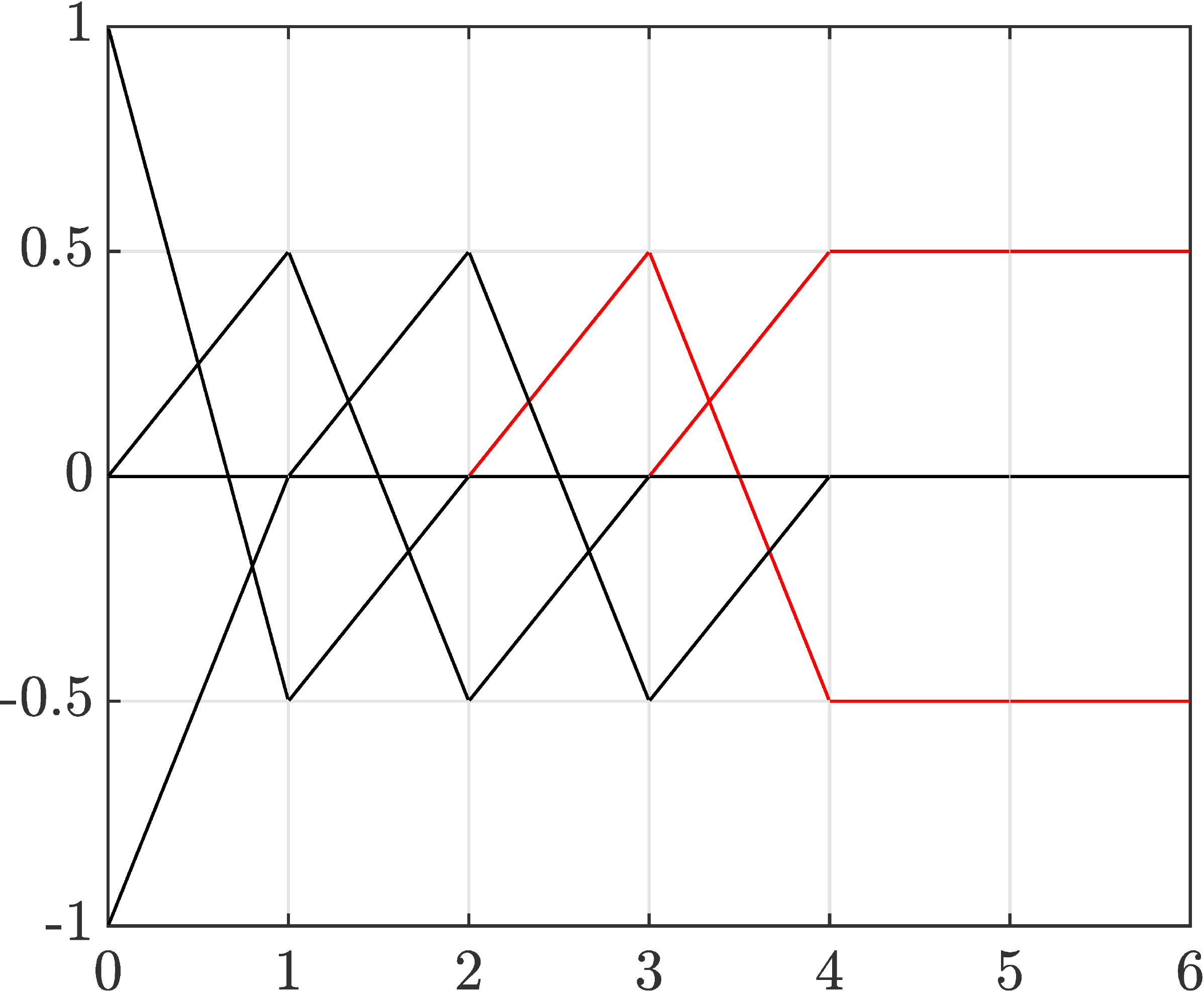}
\end{tabular}
\end{center}
\caption{\textbf{Crosspoint modification.} Evaluation of quadratic B-spline basis functions and their first derivatives in normal and tangential direction at the interface (from left to right). Modified functions and derivatives at the crosspoint are colored in red. The dashed curves denote derivatives associated with the interior of the slave patch.}
\label{fig:cpMod1}
\end{figure}
Our modification will be carried out on the parametric space of the crosspoint for the slave side of each interface separately. Thus, without restriction it is sufficient to consider one interface and one crosspoint at a time. In case of a weak $C^{l-1}$, $ 1 \leq l \leq p$  coupling we have to remove the first $l$ basis functions on the interface but also functions associated with the interior of the subdomain. While the interior basis functions are affected by normal derivatives, the ones on the interface by tangential derivatives. Now we want to modify the following next $p$ such that the new reduced basis functions are given as
\begin{align}
  R_i^m &= \sum_{j=1}^l c_{ij}\, R_j  + R_{i+l}, \qquad i = 1, \ldots,p, \label{eq:defmod} \\
  R_i^m &= R_{i+l}, \qquad i> p, %\nonumber
\end{align}
with coefficients matrix $\vec{C}  \in \mathbb{R}^{p \times l}$, \([\vec{C}]_{ij} = c_{ij}\), see \cite{dittmann2020}, Section 2.2 for details on the definition of the coefficient matrix. Obviously, the new $R_i^m$ are linearly independent, and in case of maximal continuity $\vec{C}$ forms a square matrix.  Let $\vec{A}_1 \in \mathbb{R}^{p\times l}$ and $\vec{A}_2 \in \mathbb{R}^{p \times p}$ with components $[\vec{A}_1]_{ij} = a_{ij} $ and $[\vec{A}_2]_{ij} = a_{i j+l}$. These matrices can be obtained from computing the $L^2$ scalar products $[\vec{Q}]_{ij} := (q_i, R_j )$ on the corresponding boundary with $ i=1, \ldots , p$, $j=1 , \ldots , p+l$, and $[\vec{M}]_{ij} :=(R_i , R_j)$, $i,j = 1 , \ldots , p+l$ and setting \((\vec{A}_1,\vec{A}_2) = \vec{Q}\vec{M}^{-1}\). Now, the matrix $\vec{C}$ can be formally computed from $\vec{A}_1 $ and $\vec{A}_2$ by
\begin{equation} \label{condmod}
\vec{C}:= \vec{A}_2^{-1} \vec{A}_1.
\end{equation}
 For $ p=2$ it reads as
\begin{equation}\label{modMat1}
[\vec{C}] = 
\begin{bmatrix}
\frac 52 & 2 \\ -\frac 32 & -1
\end{bmatrix}
\;\text{for}\;l=2,\quad
[\vec{C}] = 
\begin{bmatrix} \frac{3}{2} \\ -\frac{1}{2}
\end{bmatrix}
\;\text{for}\;l=1.
\end{equation}
In case of $p=3$ we can impose up to weak $C^2 $ interface continuity and find
\begin{equation}\label{modMat2}
[\vec{C}] =
\begin{bmatrix}
\frac{37}{6} & 5 & 3 \\
- \frac{25}{3} & - \frac{19}{3} & -3 \\
\frac{19}{6} & \frac 73 & 1
\end{bmatrix} 
\;\text{for}\;l=3.
\end{equation}
Note that we have considered open knot vectors with \(p+1\) repeated knots at the crosspoint. Further matrices can be found in \cite{dittmann2020}. Figure \ref{fig:cpMod1} illustrates the crosspoint modification of a quadratic B-spline basis. Therein, the bases are evaluated at the interface where modified functions and their derivatives are colored in red. Note that basis functions associated with the interior of the slave patch only contribute to derivatives normal to the interface due to the assumed construction of the parametric domain.

\begin{figure}[t]
\begin{center}
\tiny
\psfrag{Crosspoint}{Crosspoint}
\psfrag{CrosspointMod}{Crosspoint mod.}
\psfrag{Masterside}{Master side}
\psfrag{Slaveside}{Slave side}
\psfrag{PeriodicCoup}{Periodic conditions}
\psfrag{M}{M}
\psfrag{S}{S}
\psfrag{BxA}{\([205\times 200]\)}
\psfrag{AxB}{\([200\times 205]\)}
\begin{tabular}{cc}
\includegraphics[width=0.52\textwidth]{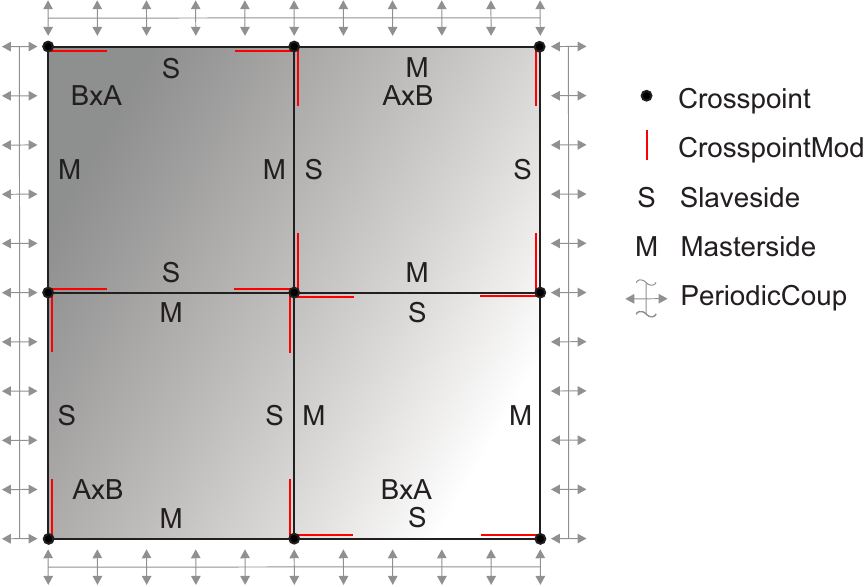}\hspace{7mm}
\includegraphics[width=0.375\textwidth]{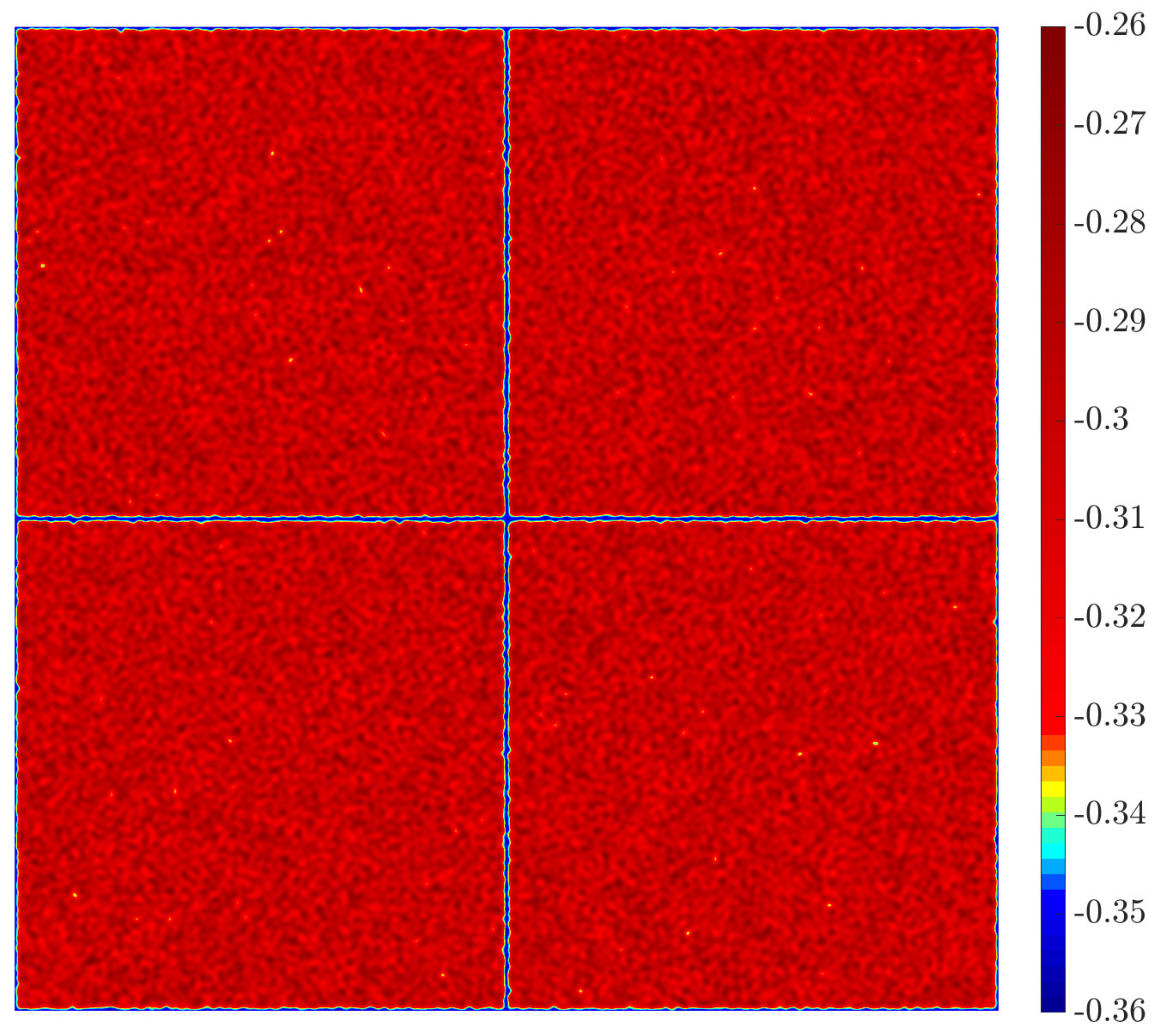}
\end{tabular}
\end{center}
\caption{\textbf{Grain growth in crystalline materials.} Setting of multi-patch problem (left) and initial field (right).}
\label{fig:pfcRef}
\end{figure}

\begin{figure}[t]
\begin{center}
\footnotesize
\begin{tabular}{ccc}
\includegraphics[width=0.3\textwidth]{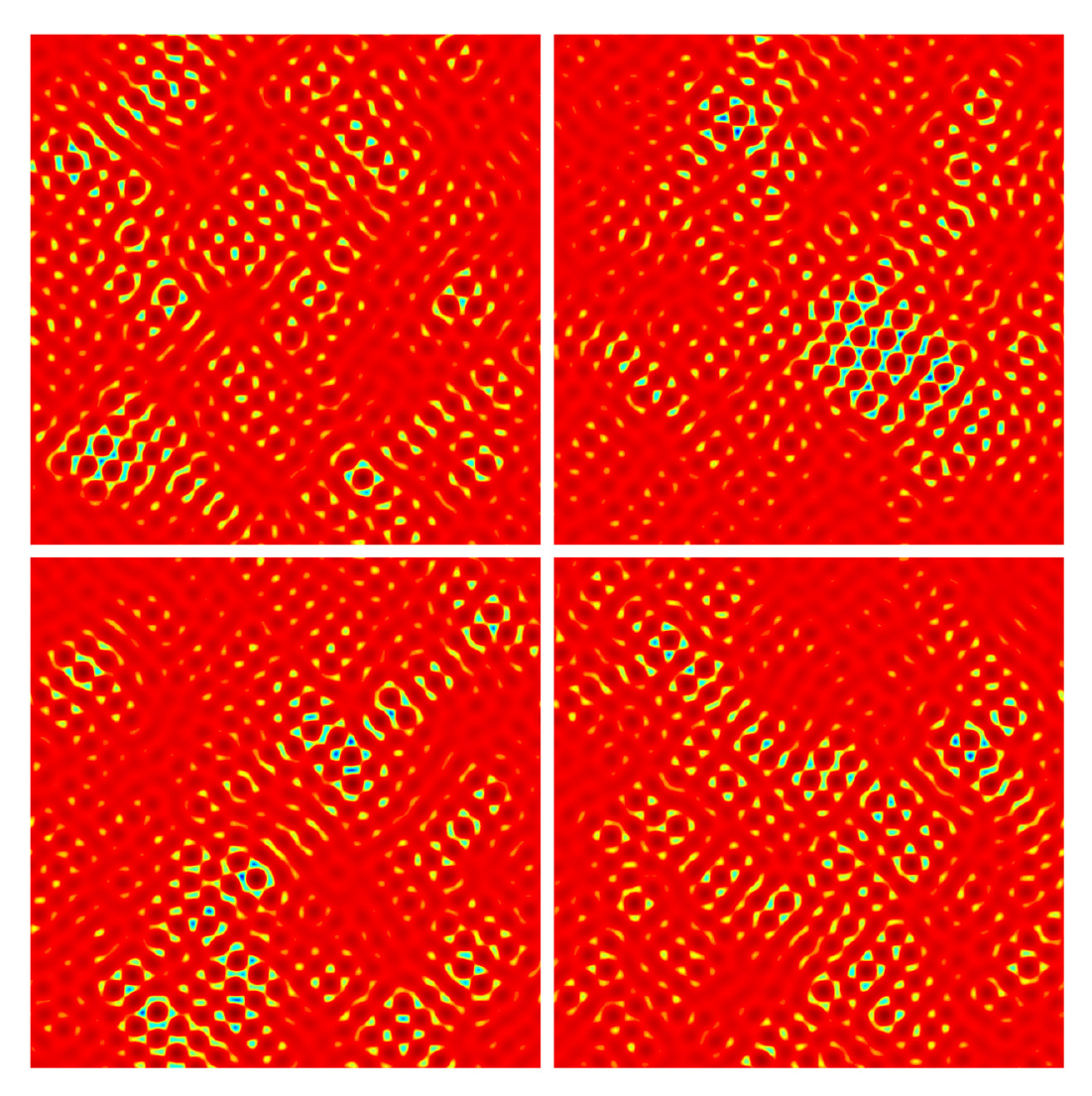}\hspace{4mm}
\includegraphics[width=0.3\textwidth]{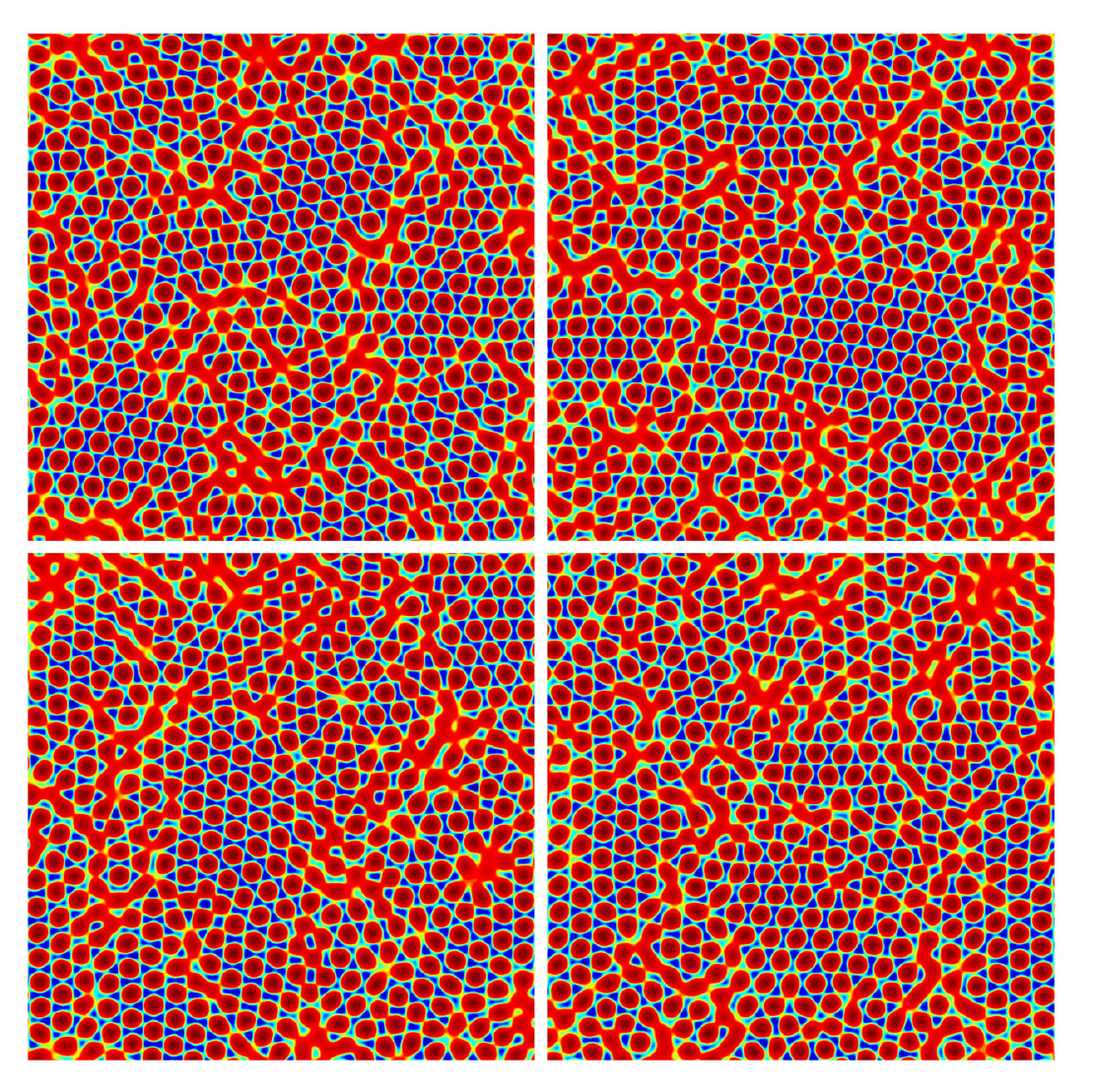}\hspace{4mm}
\includegraphics[width=0.3\textwidth]{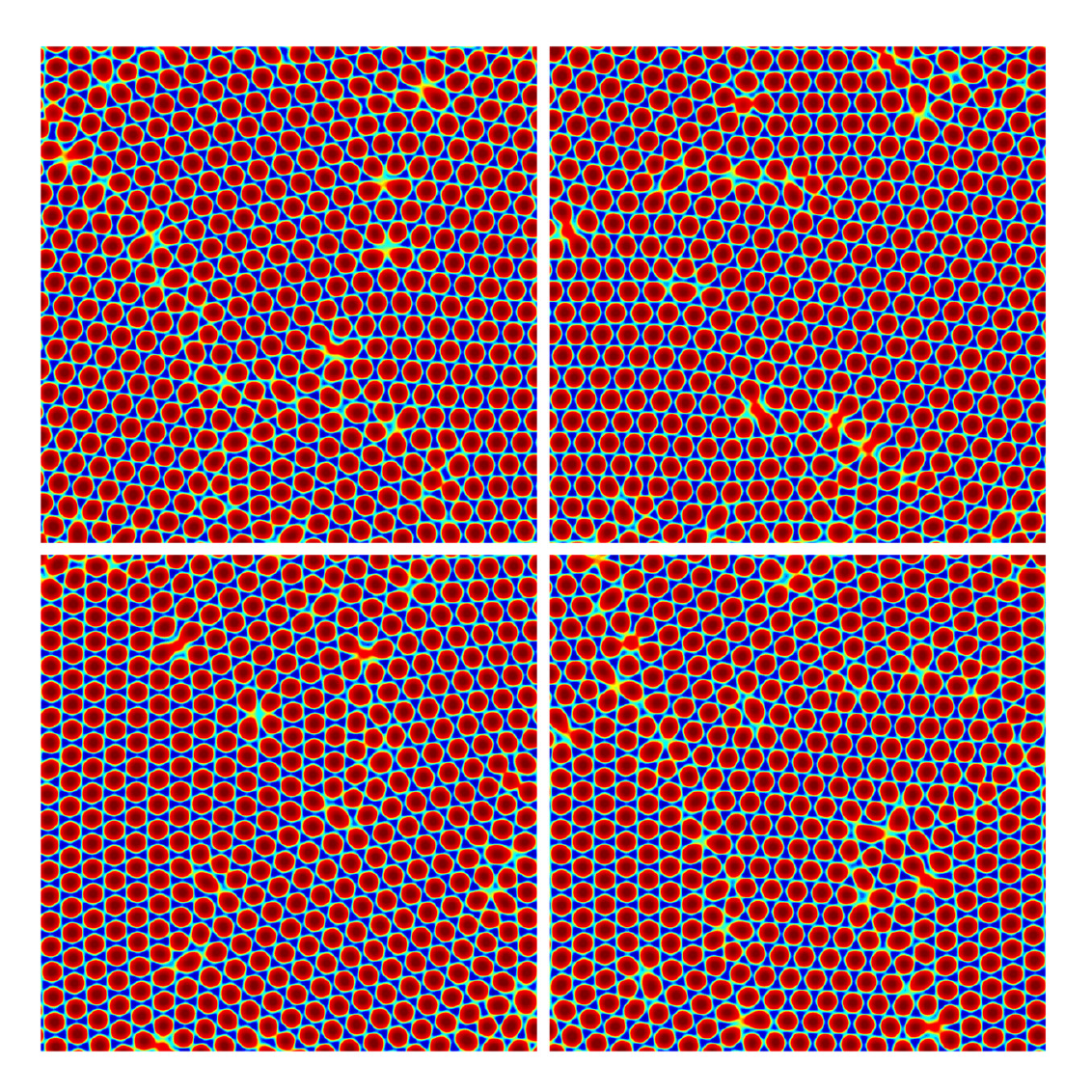}\vspace{4mm}
\end{tabular}
\includegraphics[width=0.5\textwidth]{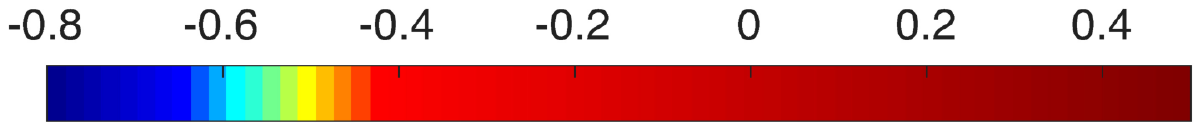}
\end{center}
\caption{\textbf{Grain growth in crystalline materials.} Solution of multi-patch simulation at different times \(t=[50,65,100].\)}
\label{fig:pfcSolution}
\end{figure}

To demonstrate the applicability on systems with $C^2$ continuity requirements, we investigate a phase-field crystal equation defined in terms of an order-parameter \(\psi(\vec{X},t):\mathcal{B}\times\mathcal{I}\rightarrow\mathbb{R}\) which describes a local deviation from a reference mass density. Therefore, we introduce a Swift-Hohenberg energy function as
\begin{equation}
% F(\psi)=\int\limits_{\mathcal{B}}\frac{1}{4}\,\psi^4+\frac{1}{3}\alpha\psi^3+\frac{1}{2}(r+1)\psi^2+\psi\Delta\psi+\frac{1}{2}\psi\Delta\Delta\psi\d V,
F(\psi)=\int\limits_{\mathcal{B}}\frac{1}{4}\,\psi^4+\frac{1}{2}\,(r+1)\,\psi^2+\psi\,\Delta\psi+\frac{1}{2}\,\psi\,\Delta\Delta\psi\d V,
\end{equation}
where the parameter \(r\) represents an undercooling of the system. The phase-field crystal model is derived as a Wasserstein gradient flow of the Swift-Hohenberg energy 
\begin{equation}\label{eq:pfc}
% \partial_{t}\psi=\nabla\cdot\left[\psi^{+}\,\nabla\frac{\delta F]}{\delta\psi}\right]\quad\text{with}\quad\frac{\delta F}{\delta\psi}=\psi^3+\alpha\psi^2+(1+r)\psi+2\Delta\psi+\Delta\Delta\psi.
\dot{\psi}=\nabla\cdot\left(\psi^{+}\,\nabla\frac{\delta F}{\delta\psi}\right),\quad\forall \, (\vec{X},t)\in\mathcal{B}\times\mathcal{I},
\end{equation}
where the variational derivative reads
\begin{equation}
\frac{\delta F}{\delta\psi}=\psi^3+(1+r)\,\psi+2\,\Delta\psi+\Delta\Delta\psi.
\end{equation}
Here, \(\psi^{+}=\psi-\psi^{\mathrm{min}}\geq 0\) denotes a mobility parameter with lower bound \(\psi^{\mathrm{min}}\). 
Moreover, assuming again a quadratic domain $\mathcal{B}=[a_1,a_2]$, the strong form (\ref{eq:pfc}) is supplemented 
by periodic boundary conditions and initial conditions given by
\begin{equation}
\psi(\vec{X},0) = \psi_{0}(\vec{X}).
\end{equation}
Therein, \(\psi_{0}\) is a prescribed initial deviation of the mass density. A multi-patch setting is now introduced, where each patch is of size \(145\times 145\) and we apply \(205\times 200\) and \(200\times 205\) cubic B-spline based elements per patch such that each interface is non-conform as shown in Figure \ref{fig:pfcRef}. The simulation parameters are specified as follows. For the lower bound we set \(\psi^{\mathrm{min}}=-1.5\) and for the parameter regarding the undercooling of the system we set \(r=-0.35\). In addition, for the initial field we apply the configuration illustrated in Figure \ref{fig:pfcRef}. Results of the simulation are shown in Figure \ref{fig:pfcSolution} at different times. Note that no disturbances at the interface are observed.

\subsection{Hybrid approaches for higher-order continuity constraints}
\begin{figure}[ht]
\raisebox{1cm}{\includegraphics[width=0.25\textwidth]{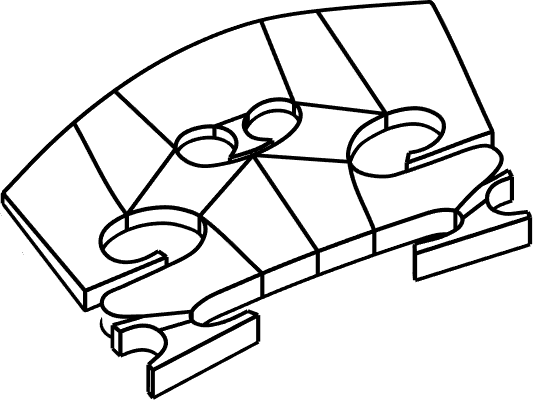}}
\includegraphics[trim = 0mm 0mm 0mm 100mm, clip,width=0.35\textwidth]{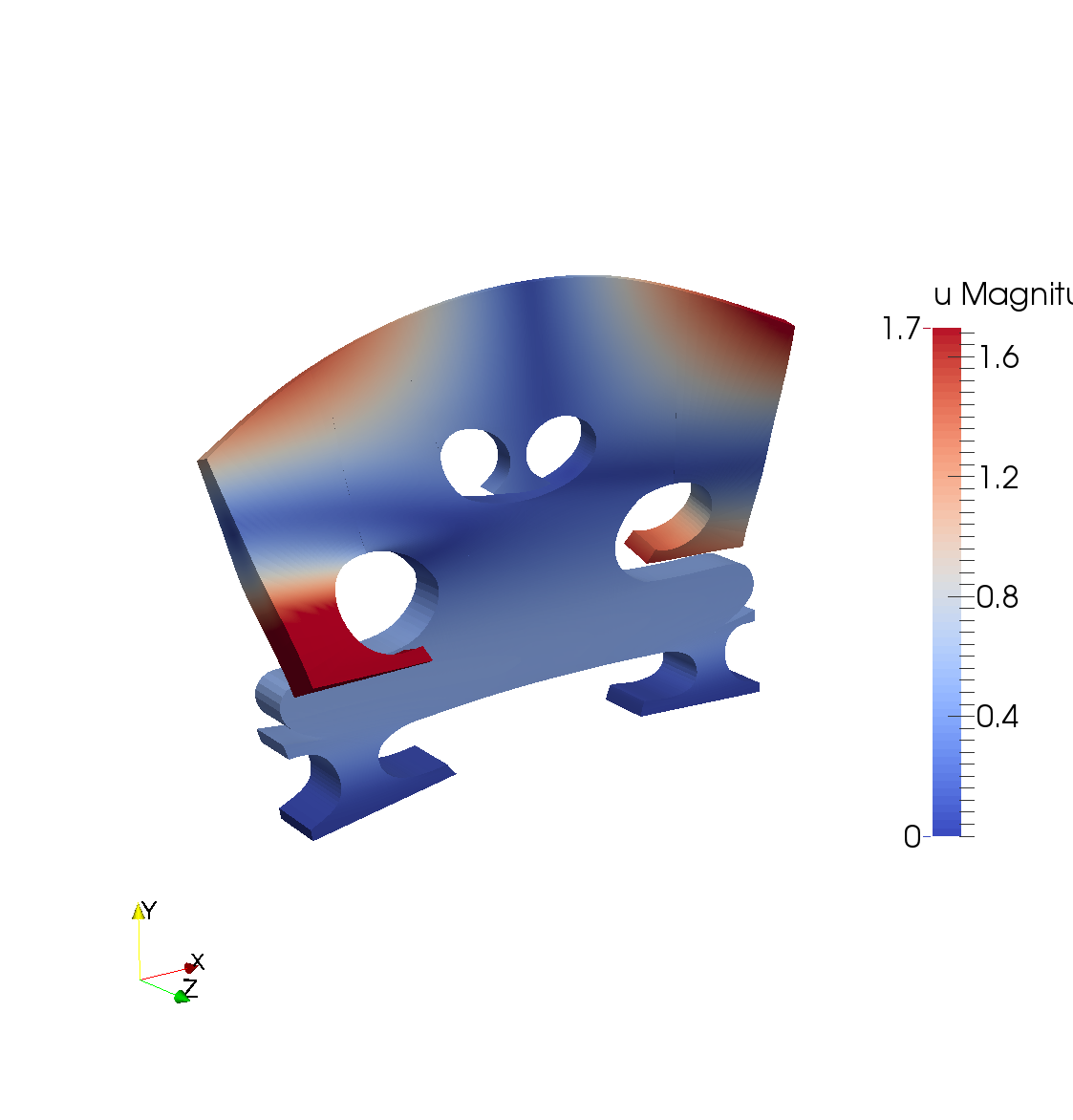}
\includegraphics[trim= 0mm 0mm 0mm 100mm, clip,width=0.35\textwidth]{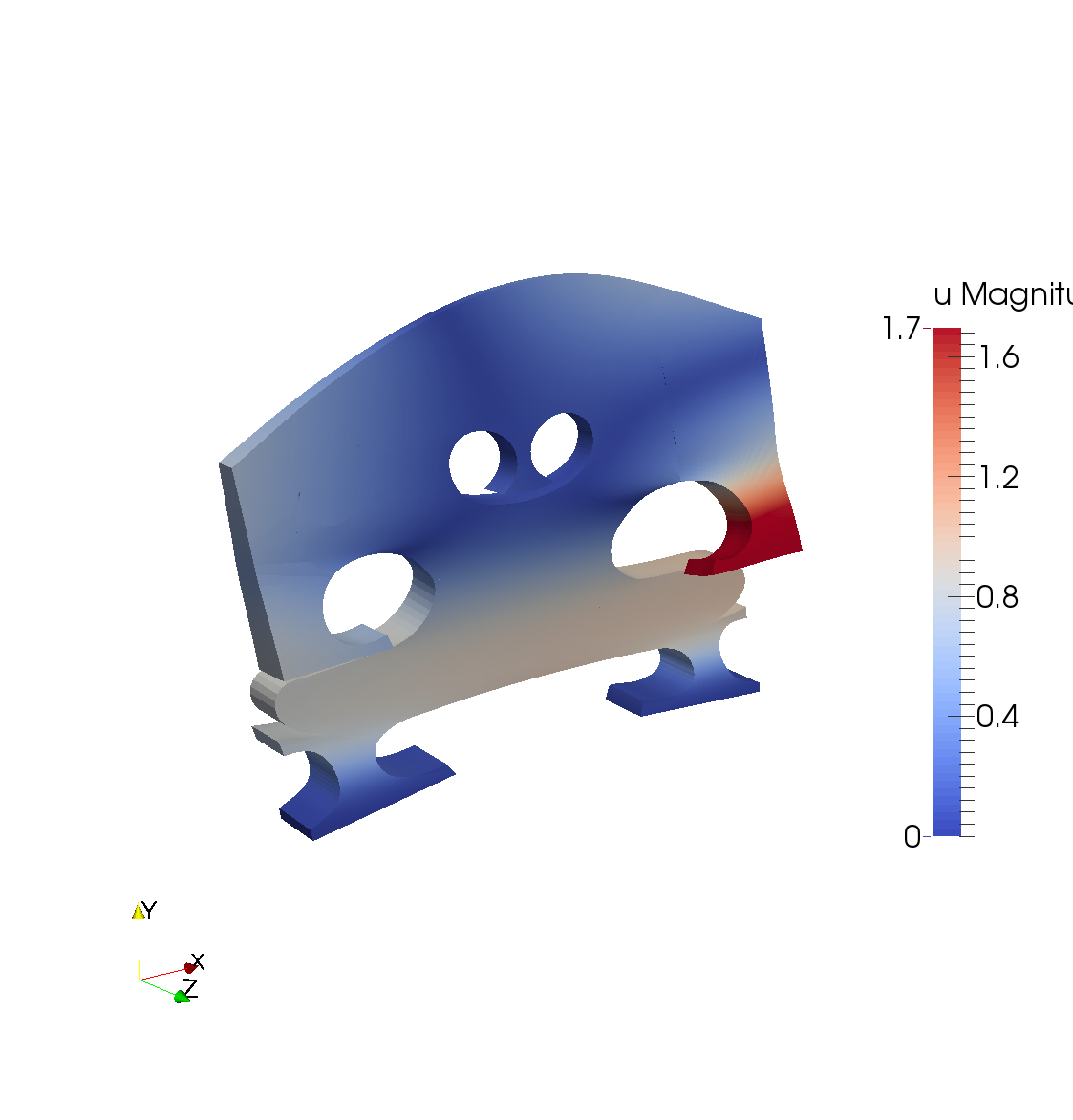}
\caption{Decomposition of a 3D violin bridge into sub-patches (left), the fourth eigenmode: homogeneous (middle) and inlay (right)}
\label{fig:eig}
\end{figure}

This subsection is based on \cite{horger2019}. In contrast to the previous two subsections, we do not enforce higher-order continuity conditions weakly in terms of Lagrange multipliers. Here we combine ideas from $C^0$-mortar based coupling techniques with DG methods. The resulting scheme yields a hybrid approach in the sense that we use discrete Lagrange multipliers only for the handling of the $C^0$ continuity condition but all higher regularity constraints are included by terms resulting from DG. As already mentioned IgA approaches are natural candidates for the approximation of fourth- and sixth-order partial differential equations. But they also may yield excellent numerical approximation results for second-order elliptic eigenvalue problems. Figure \ref{fig:eig} shows the fourth eigenmodes of a maple wood violin bridge having nine orthotropic material parameters in the elasticity tensor. The difference between the middle and right picture results from the fact that in the middle picture the material parameters are constant, whereas in the picture on the right an inlay of a harder wood is inserted, and thus only piecewise constant material parameters are assumed. On the left the decomposition into the sub-patches is given for the homogeneous case. We recall that crosspoint/wirebasket modifications of the Lagrange multiplier basis functions have to be worked out.

In comparison to classical conforming finite elements, IgA approaches provide much better results for higher frequency modes. However in case of non-matching meshes and weak $C^0$ continuity constraints across interfaces between sub-patches, one can observe severe outliers. To overcome these shortcomings and to improve quantitatively the  error decay, a simple strategy is to enforce higher continuity by penalty terms. While for second-order elliptic problems, it is not mandatory to impose higher-order constraints on the regularity of the discrete solution, it is for higher-order PDEs.  

For simplicity of presentation, we discuss here only the model problem of a biharmonic equation and refer to \cite{horger2019} for more general situations and numerical results. The approach can be easily adapted to a sixth-order PDE using cubic splines with maximal regularity in the sub-patches. Also for second-order PDEs, we can penalize jumps in the derivatives up to the maximal regularity, i.e., for a cubic spline approach we can introduce a suitable penalization for jumps in the first and second derivatives. However it is important to note that all jump terms have to be scaled properly for not loosing the continuity of the bilinear form. The scaling is dictated by the order of the PDE and the order of the derivative. Having the abstract framework of defining Lagrange multipliers up to maximal weak regularity of the two previous subsections at hand and knowing how to penalize jumps in the derivatives and adding consistency terms in a DG setting allow us in a flexible way to combine both approaches. Of special interest is to keep the first layer of degrees of freedom in the Lagrange multiplier space associated with the nodes which sit physically on the interface between sub-patches but remove the layers which are associated with nodes in the interior of the sub-patches. This simplifies the required data-structure for handling higher-order continuity constraints.

Following \cite{MR2142191}, the weak formulation of a $C^0$ symmetric interior penalty formulation for biharmonic equations reads as:
\begin{equation}
\begin{aligned} 
a (w,v) := \,&  \label{eq:consistency}
\sum_{T \in {\mathcal T}_h} \int_T D^2u : D^2 v \d x + \sum_{f \in {\mathcal F}_h} \int_f \left( \{ w_{n,n} \} \,[v_n] + \{v_{n,n}\}\, [ w_n ] \right) \d s \\
& + \sum_{f \in {\mathcal F}_h} \int_f \frac{\tau}{h_f}\, [ v_n] \, [ w_n ] \d s .
\end{aligned}
\end{equation}
Here $D^2$ stands for the Hessian, ${\mathcal T}_h $ denotes the set of elements,  ${\mathcal F}_h$ the set of faces, and $\cdot _n $ and $\cdot _{n,n}$ denote the first and second normal derivatives on the faces, respectively. The second term on the right of \eqref{eq:consistency} reflects consistency terms involving the first and second derivatives across the faces in normal direction. As it is standard, the  parenthesis $[ \cdot  ] $ stands for the jump and $\{ \cdot \} $ for the average. On a boundary face, these terms have to be defined such that the boundary conditions are reflected properly. The last term in \eqref{eq:consistency} depends on the penalty parameter $\tau \geq \tau_0 > 0$ and the diameter $h_f$ of the face $f$. If $\tau_0 $ is large enough well-posedness is guaranteed, and the bilinear form is uniformly elliptic on a suitable space.

Using a IgA approach, which is locally on each sub-patch $C^k$, $ k \geq 1$, the jump terms with respect to faces associated with the interior of the sub-patches vanish and do not have to be taken into account. However for non-matching meshes at the sub-patch interfaces, it is, in general, not possible to preserve the strong $C^1$ continuity within the IgA framework.

Thus we adapt the bilinear form defined by  \eqref{eq:consistency} in two steps. The first step follows directly from the approach above and reduces the sum  over all faces to a sum over all interfaces. It reads as
\begin{align} 
 \sum_{l=1}^K \left( \int_{\Gamma_l} \left( \{ w_{n,n} \}\, [ v_n] + \{ v_{n,n} \}\, [ w_n ] \right) \d s +
\sum_{f \in {\mathcal F_l^s}_h}  \frac{\tau}{h_f} \,\int_f [ v_n] \, [ w_n ]  \d s \right), 
\end{align}
where  ${\mathcal F_l^s}_h$ stands for all faces on the slave side of the interface $\Gamma_l$.

To avoid locking of the approach, we relax in a second step the penalty term and only consider the jumps projected onto piecewise constants with respect to the mesh associated with the slave side, i.e., the modified bilinear form reads as 
\begin{equation}
\begin{aligned} 
a_{IgA} (w,v) := & 
\sum_{m= 1 }^M \int_{\Omega_m} D^2u : D^2 v \d x + \sum_{l=1}^K \int_{\Gamma_l}  \left( \{ w_{n,n} \}\, [ v_n] + \{ v_{n,n} \}\, [ w_n ] \right) \d s \\
& + \sum_{l=1}^K \sum_{f \in {\mathcal F_l^s}_h} \frac{\tau}{h_f}\,  \int_f \pi_0^s[ v_n]\,  \pi_0^s[ w_n ] \d s ,
\end{aligned}
\end{equation}
where $\pi_0^s$ stands for the projection operator onto piecewise constants with respect to the slave side mesh.

Using standard inverse and trace estimates in combination with DG techniques it is easy to show that the resulting bilinear form is uniformly continuous and elliptic on a suitable space provided that $C^0$ continuity across the interfaces is given.  However, even this is, in general in a mortar context, not possible. In contrast to the normal derivative where a discontinuity is penalized, we impose the $C^0$ continuity of the solution weakly. As it is standard for mortar techniques this is realized in terms of a Lagrange multiplier space. Here we can use all options which are known for the standard case of a second-order elliptic operator. To summarize, the hybrid formulation for the biharmonic equation guarantees that the solution is in the constraint mortar IgA-space, i.e., it satisfies a weak $C^0$ continuity but no weak higher-order continuity. The discontinuity in the normal derivative is penalized by the bilinearform $a_{IgA} (\cdot,\cdot)$. This is the big difference between the hybrid approach discussed in this subsection and the weak $C^n$ continuity of subsections \ref{sec:higher}-\ref{sec:crosspoints}. We point out that due to the need of a uniform inf-sup condition, crosspoints in 2D and the wirebasket in 3D have to be very carefully handled within the Lagrange multiplier approach. As it is typical for this situation, the number of degrees of freedom in the Lagrange multiplier space has to be reduced without compromising the approximation property. In the case of the hybrid approach one can also add terms which only penalize jumps at the crosspoints and wirebasket, respectively, see \cite{horger2019}. Due to the scaling different weights have to be used in the penalty formulation. In 2D a crosspoint is a geometrical object of dimension zero while the interface is of dimension one. This difference in the dimension has to be balanced by different mesh-size depending weight factors.

\section{Mortar contact formulations for Isogeometric Analysis}\label{sec:contact}
Mortar low-order finite element methods are widely used in contact mechanics. In contrast to penalty methods they allow for a variational consistent formulation 
of the non-penetration condition and  a friction law. An optimal a priori estimate can be derived, for both the displacement and the surface stress being approximated by the Lagrange multiplier, \cite{hueber2005a}. The performance can be increased by applying adaptive mesh refinement techniques based on a posterior error indicators 
\cite{W07} and by specially designed energy preserving time integration schemes \cite{hager2008}. For an overview of variationally consistent formulations of inequality constrained problems, we refer to \cite{wohlmuth2011}.
Of special interest are formulations which allow for local static condensation such as biorthogonal based Lagrange multiplier techniques. By this one can easily apply all-at once semi-smooth Newton techniques which can be implemented in form of primal-dual active set strategies, \cite{hueber2008,hueber2005b}. In each Newton iteration, one has to decide for each node on the slave side of the contact interface the type of boundary condition. In case of a thermo-mechanical contact problem typically non-linear Robin type conditions occur, and the heat flux can be eliminated locally, \cite{hueber2009}. Most theoretical results and algorithmic approaches can be easily adapted to the IgA framework. In the following
subsections, we report on  recent results for contact mechanics and IgA approaches.
\subsection{Biorthogonal basis functions applied to contact mechanics} \label{sub:biorthcontact}
This subsection is related to \cite{MSWW18,Seitz2016}. While the condition (BA) in subsection \ref{sub:biorth} is of crucial importance to obtain optimal order best approximation properties for the constrained IgA space, this condition can be considerably relaxed in case of contact mechanics. Here the solution is typically not of high global regularity. Thus we cannot expect convergence rates of order $p$, $p \geq 2$, in the energy norm if uniform refinement is used. Numerically one observes typically a sub-optimal convergence rate of $\approx 3/2$ if quadratic or even cubic basis functions are used. If the numerically convergence rate is bounded not by the best approximation order of the involved discrete spaces but by the regularity of the solution, then the best approximation order might be reduced without loosing the observed convergence order. More precisely, if the solution is in $H^{5/2} (\Omega)$ but not in $H^s (\Omega)$ with $s > 5/2$, then we cannot expect a better order than $3/2$ for the error decay in the energy norm. Typically due to the inequality constraints of contact problems such as the non-penetration condition or  a friction law which determines about sliding,  the solution of a mechanical contact problem is in $H^s (\Omega)$ with $s < 5/2$.

Therefore the required best approximation property for the discrete Lagrange multiplier space $M_h$ reads as
\begin{align}
\inf_{\psi_h \in M_h} \| \psi - \psi_h \|_{H^{-1/2} (\Gamma)} \leq C\, h^{3/2}\, \| \psi \|_{H^1(\Gamma)},
\end{align}
where $H^{-1/2}(\Gamma) $ stands here for the dual of the trace space on the possibly contact boundary $\Gamma$. This property holds on mild assumptions on the shape of the basis functions $\psi_i$ of $M_h$ and the following two conditions:
\begin{itemize}
\item Each $\psi_i $ is locally supported, in the sense that the support of $\psi_i $ contains at most $K_1$ elements  and each element is at most contained in the support
of $K_2$ basis functions. Both $K_1$ and $K_2$ are supposed to be meshsize independent.
\item The constant function equal one is an element of $M_h$. (RE)
\end{itemize}
Let us now consider the dual basis obtained in subsection \ref{sub:biorth} after Step II (see also Figure \ref{fig:localdual}). As mentioned it does not satisfy an order $p$ best approximation property for $p \geq 2$ but it satisfies (RE). Recalling that $M_h \subset \text{span } \{ \phi_{i,e} \}$ and $1 \in \text{span } \{\phi_{i,e}\}$, it is easy to show that $\psi_i$ form a partition of unity. Let 
$\Psi := \sum_i \psi_i$, then we get for all $\phi_{j,e}$ that
\begin{equation}
\begin{aligned}
\int_{\Gamma} \Psi \,\phi_{j,e} \d s &= \sum_{i, \text{supp } \phi_i \supset e}  \int_e \psi_{g(i,e),e}\, \phi_{j,e} \d s 
=\sum_{i, \text{supp } \phi_i \supset e}  \int_e \phi_{j,e} \d s \,\delta_{g(i,e),j} \\ & =  \int_e \phi_{j,e} \d s  = \int_\Gamma \phi_{j,e} \d s.
\end{aligned}
\end{equation}
Thus, we found that $\Psi = 1 \in M_h$. In other words: although the order $p$ best approximation property is lost for $p \geq 2$, this dual basis may still preserve the observed convergence rate for problems where the convergence is bounded by the regularity of the solution as discussed above. While sub-optimal convergence results are to be expected in IgA patch coupling situations (as has been confirmed in section \ref{sub:biorth}, in particular with Figure \ref{fig:platewithhole_results}), the element-wise dual basis is still an attractive candidate for contact problems in IgA.

In \cite{MSWW18}, low-order dual Lagrange multipliers have been applied to a dynamic viscoelastic contact problem with short memory. Existence and uniqueness results have been shown for the associated  mixed formulation. For the discretization of the primal space,  low-order conforming finite elements have been applied as a special case of IgA. Numerical results for higher-order NURBS-IgA in the mesh tying case but also for finite deformation contact problems can be found in \cite{Seitz2016}. As can be expected from the lack of a higher-order reproduction property of $M_h$, the error decay in the mesh tying examples is asymptotically not optimal for general non-matching meshes. 

\begin{figure}[ht]
\begin{center}
\includegraphics[width=0.88\textwidth]{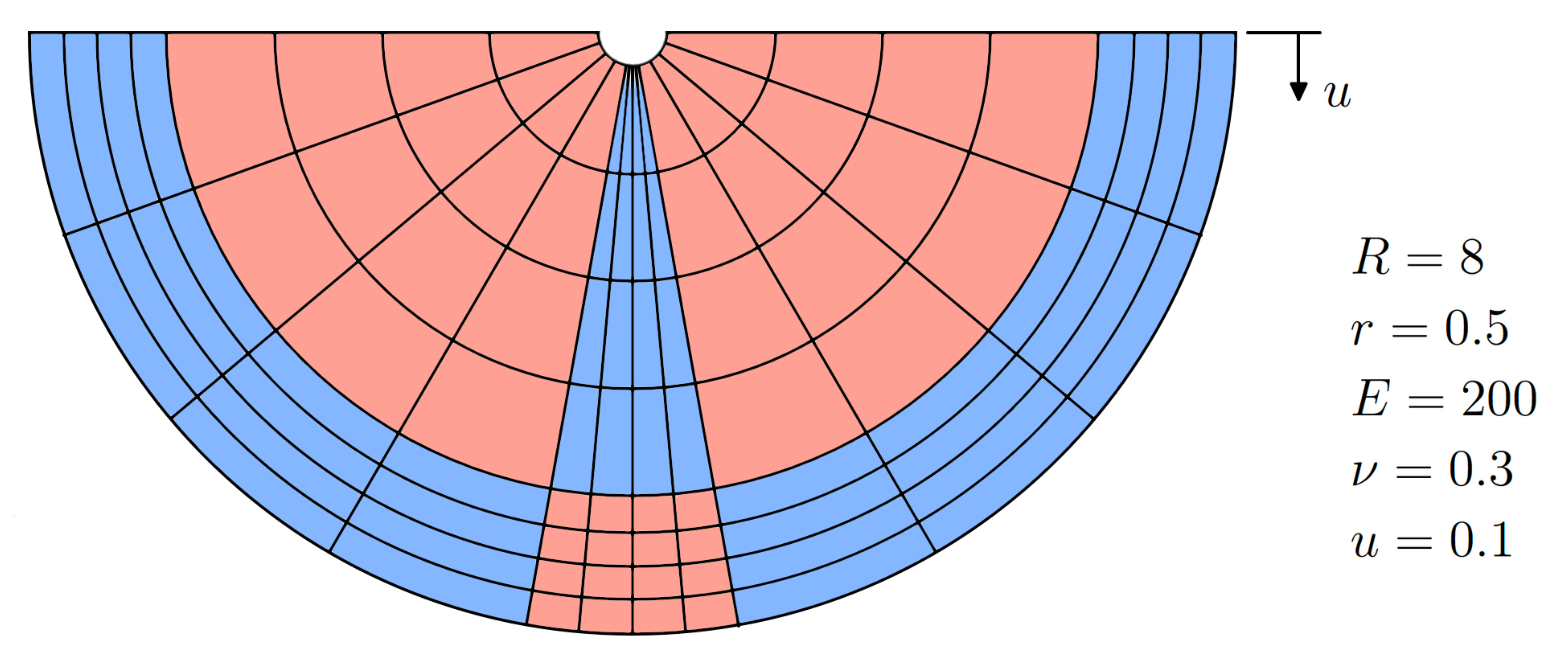}
\end{center}
\caption{\textbf{2D Hertzian contact.} Geometry setup, material parameters and the chosen patch parametrization with an exemplary coarse NURBS mesh (level 2). Reproduced and slightly modified from \cite{Seitz2016}.}
\label{hertz_setup}
\end{figure}

To better illustrate the theoretical findings, a two-dimensional Hertzian-type contact example of a cylindrical body (radius $R$) with a rigid planar surface under plane strain conditions is reproduced and slightly modified from \cite{Seitz2016}. To avoid singularities in the isogeometric mapping, a small inner radius (radius $r$) is introduced, see Figure \ref{hertz_setup} for the geometric setting, the material parameters and the parametrization (different IgA patches are marked with different shading). The two horizontal upper boundaries undergo a prescribed vertical displacement. Meshes using second-order and third-order NURBS basis functions are employed, which is also illustrated in Figure \ref{hertz_setup} for a very coarse mesh (level 2). In this setup, half of the elements on the potential contact surface are located within one ninth of the circumferential length and $C^{p-1}$ continuity is ensured over the entire active contact surface.

\begin{figure}[ht]
\begin{center}
\includegraphics[width=0.88\textwidth]{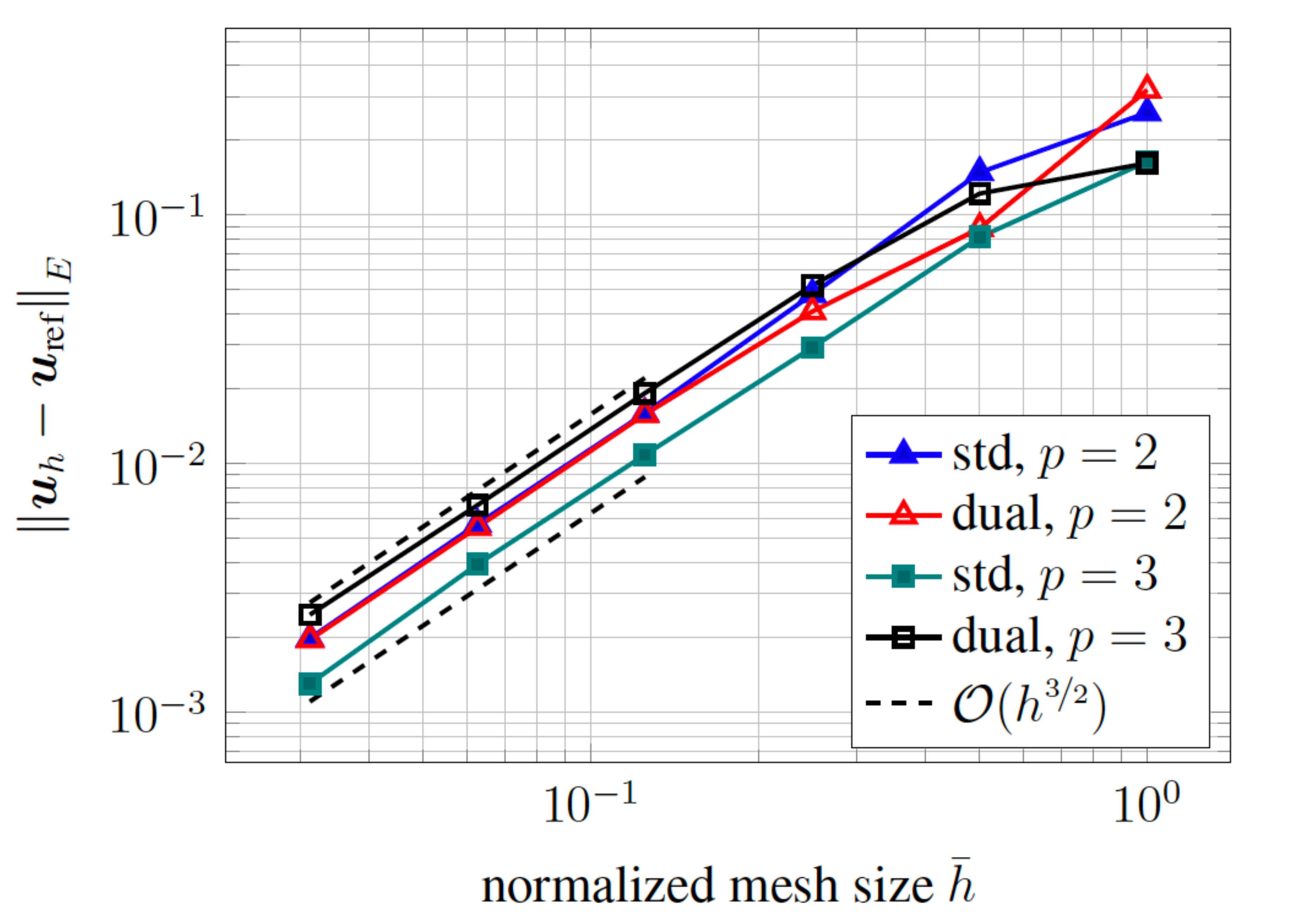}
\end{center}
\caption{\textbf{2D Hertzian contact.} Convergence results for standard and dual Lagrange multiplier bases for $p=2$ and $p=3$. The biorthogonality construction for the dual case is based on Step I and II in Section \ref{sub:biorth}. Reproduced and slightly modified from \cite{Seitz2016}.}
\label{hertz_results}
\end{figure}

In the convergence study, uniform mesh refinement via knot insertion is performed on each of the patches resulting in a constant local element
aspect ratio. Although only relatively small deformations are to be expected, a fully nonlinear description of the continuum using nonlinear kinematics and a Saint-Venant-Kirchhoff material under plane strain condition is assumed. Figure \ref{hertz_results} depicts the convergence behavior in terms of the energy norm. Since no analytical solution is available, the finest mesh (level 7) with standard third-order NURBS is used as a numerical reference solution. In the limit, all methods converge with the expected order of $\mathcal{O}(h^{3/2})$ in the energy norm and also the absolute error values are quantitatively very similar. In the second-order case ($p = 2$) the standard and dual Lagrange multiplier bases yield the same error asymptotically, whereas for third-order NURBS, a slightly elevated error of the dual variant as compared to the standard one can be observed. In view of Figure \ref{hertz_results}, the use of a simple (i.e. element-wise) biorthogonal basis for the Lagrange multiplier (as obtained in section \ref{sub:biorth} after Step II) instead of primal ones does not come at the expense of a reduced accuracy for contact problems, but yields equally accurate results while reducing the total system size to the number of displacement degrees of freedom only. In contrast to the IgA patch coupling case in section \ref{sub:biorth}, the convergence is now limited by the regularity of the solution, such that both the standard and biorthogonal Lagrange multiplier variants converge with the same order. The use of higher-order NURBS for contact problems with reduced regularity, i.e. third-order in Figure \ref{hertz_results} or even higher seems questionable from this viewpoint, since no faster convergence is gained from the higher-order interpolation.

%\todoB{Alexander willst du hier auch noch eine Numerik einfuegen - ist wahrscheinlich gar nicht so schlecht dann ist das Kapitel nicht so trocken - mehr theorie aus der anderen arbeit einzufuegen ist fast unmoeglich ohnen einen riesen formalismus aufzubauen und die Numerik ist da eher aus einem anderen projekt heraus entstanden}

%\todoA{Erledigt!}

\subsection{Thermomechanical contact problems}

The isogeometric mortar methods for isothermal contact derived in the previous section can also be extended to include thermal coupling effects consisting of heat conduction across the contact interface, frictional heating and a temperature-dependent coefficient of friction. The following remarks are based on \cite{Seitz2018,Seitz2019}, and the interested reader is referred to the original publications for further details. From the continuum mechanical perspective, the first two coupling effects are included in the contact interface heat fluxes, while the last one enters in Coulomb's law of friction via a temperature-dependent coefficient of friction.

The thermomechanical coupling in the bulk continuum (e.g. thermo-elasticity or thermo-plasticity) is not revisited here, but the focus is clearly set on the thermomechanical interface and the choice of a discrete Lagrange multiplier basis in IgA. As in the isothermal case, a Lagrange multiplier field $\vec{\lambda}$ is introduced to enforce the mechanical contact constraints and can be identified as the negative slave-side contact traction, i.e. $\vec{\lambda} = - \vec{t}_c^{(1)}$. In a similar fashion, a thermal Lagrange multiplier field ${\lambda}_T$ is now introduced to enforce the thermal interface constraint and will be chosen as the slave side heat flux ${\lambda}_T = q_c^{(1)}$. Specifically, the variational formulation of the thermal interface constraint is as follows:
\begin{equation}
\int_{\Gamma} \left( \lambda_T - \beta_c \,\lambda_n\, (T^{(1)}-T^{(2)}) - \delta_c \,\vec{\lambda} \cdot \vec{v}_{\tau} \right)\, \delta \lambda_T \d\gamma = 0,
\end{equation}
where $\vec{v}_{\tau}$ represents the relative tangential velocity (sliding velocity), $\lambda_n$ is the normal part of the mechanical Lagrange multiplier (contact pressure), $\beta_c$ is the contact heat conductivity and $\delta_c$ is the distribution parameter for frictional heat dissipation. In the limit cases $\delta_c = 0$ or $\delta_c = 1$ the entire frictional dissipation is converted to heat on the master or slave side, respectively. Interestingly, the choice of a discrete Lagrange multiplier basis follows the exact same steps for the thermal contact part as for the mechanical contact part described in section \ref{sub:biorthcontact}. The main complexity in terms of mortar discretization and algebraic system representation lies in the third and last part of the thermal interface constraint described above, which represents the frictional heat dissipation at the contact interface. Firstly, an objective kinematic measure has to be defined for the relative tangential velocity $\vec{v}_{\tau}$. Secondly, and most importantly,  the term involves a so-called 'triple' integral, i.e.~an integral over a product of three shape functions at the contact interface, since $\vec{v}_{\tau}$ as constraint interface shape functions besides $\vec{\lambda}$ and $\delta \lambda_T$.
This poses very high demands on the quadrature accuracy at the contact interface, especially when dealing with higher-order approximations using Lagrange polynomials or NURBS, see e.g.~\cite{dittmann2014}. Following the work in~\cite{hueber2009}, an appropriate lumping technique is applied to reduce the computational cost without compromising on accuracy.

The fully coupled nonlinear system to be solved in each time step is comprised of the structural and thermal equilibrium equations, the nonlinear complementarity (NCP) function of normal and tangential contact and, finally, the thermal contact interface condition. The global vector of discrete unknowns consists of four groups of degrees of freedom: vectors containing all nodal values of the displacements $\vec{D}$ and temperatures $\vec{T}$ as well as the discrete Lagrange multipliers $\bar{\vec{\lambda}}$ and $\bar{\vec{\lambda}}_T$. As in the isothermal case, the system is non-smooth due to the involved NCP functions, but still amenable to semi-smooth versions of Newton`s method as discussed in \cite{popp2010,popp2013,Seitz2015}. If biorthogonal basis functions as introduced in section \ref{sub:biorth} and section \ref{sub:biorthcontact} are used for the Lagrange multiplier fields $\vec{\lambda}$ and $\lambda_T$, the local support (LS) property from section \ref{sub:biorthcontact} is again satisfied by construction due to to the similar structure of the variational formulations including $\vec{\lambda}$ and $\lambda_T$. Hence, both the usual Lagrange multiplier increments and the thermal Lagrange multiplier increments can be trivially condensed, and therefore the saddle point structure of the system matrix is successfully removed. The condensed linear system to be solved consists of displacement and temperature degrees of freedom only. In an abstract notation, it reads
\begin{equation}
\begin{bmatrix}
\mathcal{K}_{DD} & \mathcal{K}_{DT} \\ \mathcal{K}_{TD} & \mathcal{K}_{TT}
\end{bmatrix}
\begin{bmatrix}
\Delta \vec{D} \\ \Delta \vec{T}
\end{bmatrix}
= -
\begin{bmatrix}
\mathcal{R}_D \\ \mathcal{R}_T
\end{bmatrix}
\end{equation}

\begin{figure}[ht]
\begin{center}
\includegraphics[width=1.0\textwidth]{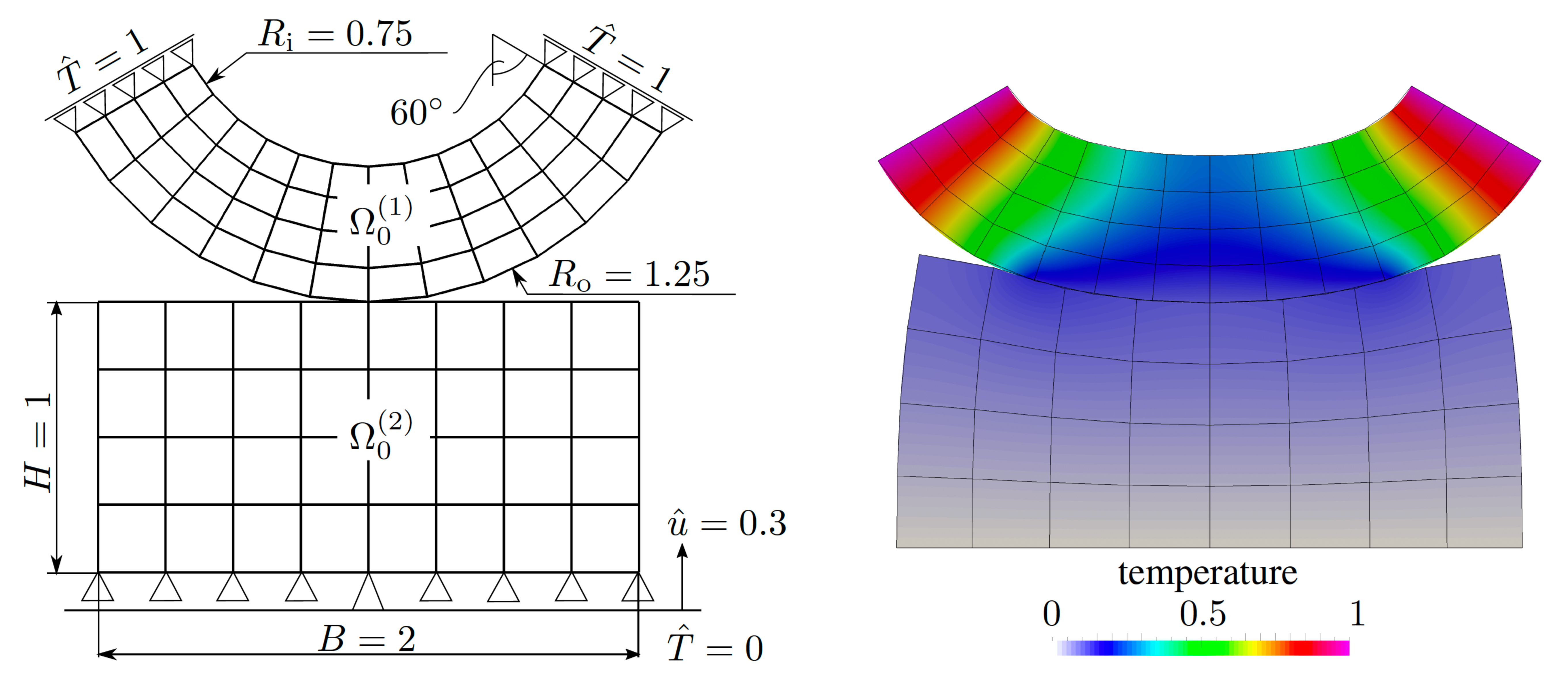}
\end{center}
\caption{\textbf{Thermomechanical contact.} Geometry setup (left) and exemplary temperature solution (right). Reproduced and slightly modified from \cite{Seitz2019}.}
\label{thermoconv_setup}
\end{figure}

Only one representative numerical example is presented in the following to highlight the most important features of isogeometric mortar methods for thermomechanical contact, see also \cite{Seitz2019} for further details and results. First, convergence under uniform mesh refinement is analyzed with a two-body contact setup as given in Figure \ref{thermoconv_setup}. Both bodies are modeled with a Neo-Hookean material law with $E^{(1)} = 5$, $E^{(2)} = 1$ and $\nu^{(1)} = \nu{(2)} = 0.2$. Furthermore, thermal expansion is included with the coefficient of thermal expansion being $\alpha^{(1)}_T = \alpha^{(2)}_T = 0.01$ and thermal conductivities are set to $k^{(1)}_0 = 1$ and $k^{(2)}_0 = 5$. At the contact interface, frictionless contact is assumed with a contact heat conductivity $\beta_c = 10^3$. The final configuration and temperature distribution are also illustrated in Figure \ref{thermoconv_setup}.

\begin{figure}[ht]
\begin{center}
\includegraphics[width=1.0\textwidth]{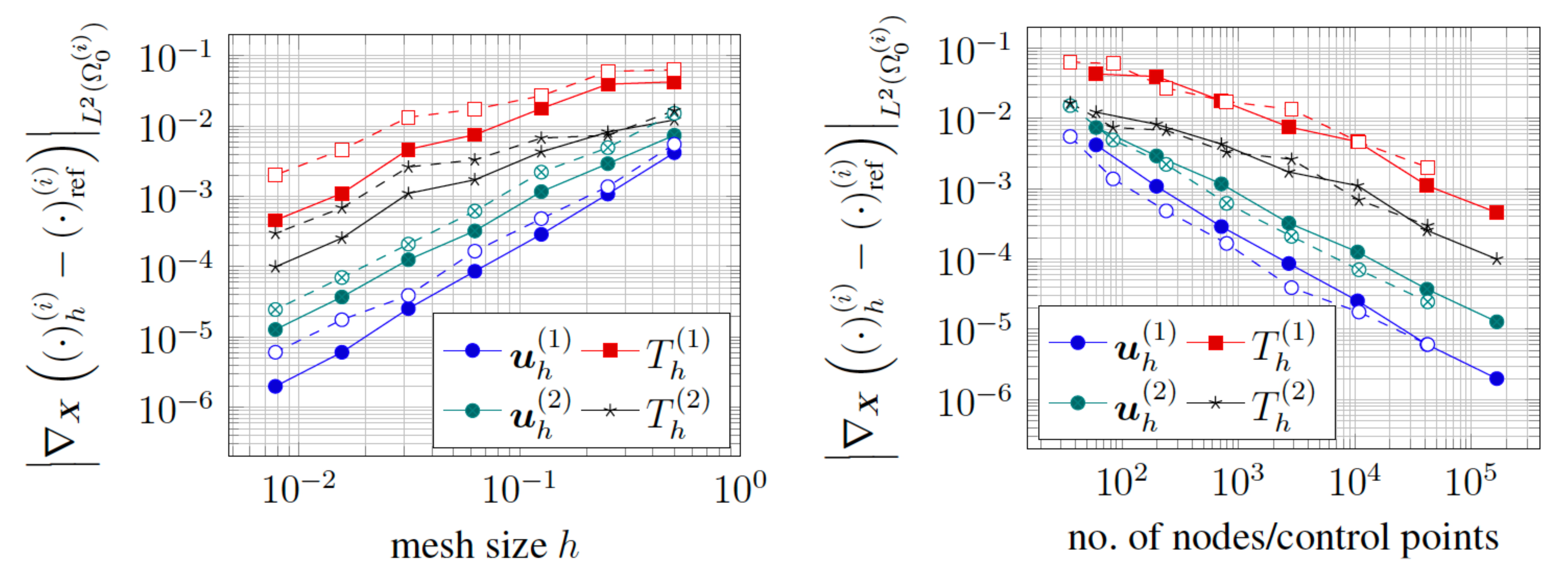}
\end{center}
\caption{\textbf{Thermomechanical contact.} Convergence results for second-order finite elements (solid lines) and second-order NURBS (dashed lines). For both approximations, dual Lagrange multipliers are employed. Comparison by mesh size (left) and by number of control points/nodes (right). Reproduced and slightly modified from \cite{Seitz2019}.}
\label{thermoconv_results}
\end{figure}

Figure \ref{thermoconv_results} exemplarily depicts the convergence behavior in the $H^1$ semi-norms of the discrete displacement
and temperature fields within the two bodies for mesh sizes $h \in [2^{-7} \, , \, 2^{-1}]$. In particular, classical finite elements with Lagrange multiplier bases according to~\cite{popp2012} and IgA with biorthogonal Lagrange multiplier bases according to \cite{wunderlich2018} are compared for the quadratic case ($p=2$). All variants converge with the optimal order to be expected based on the regularity of the solution, i.e.~$\mathcal{O}(h^{3/2})$. For second-order NURBS approximation, the absolute error values are slightly larger than for quadratic finite elements when the same mesh size is analyzed. This is not surprising since, at the same mesh size $h$, the isogeometric approximation has a smaller function space. More specifically, the B-spline basis used for the discretization at a certain mesh size is included entirely in the corresponding quadratic finite element discretization. If, however, the errors are
analyzed with respect to the number of nodes or control points, respectively, the isogeometric case is slightly more accurate in the displacement solution, whereas the error in the discrete temperature field is of similar accuracy as compared to finite elements.

To analyze the effects of frictional heating, thermoplasticity and nonlinear dynamics including mechanical and thermal energy conservation over an isogeometric contact interface, several more examples have been collected in \cite{Seitz2019}. The interested reader is referred to the original publication for details on problem geometry, loading and material parameters.

\section{Multi-dimensional coupling}\label{sec:multi}
In this last chapter, we consider a dimension reduced model for a fiber-matrix coupling. The fiber is modeled by a one dimensional beam theory and is embedded
into a three dimensional body. This approach follows fundamental ideas as introduced in \cite{peskin1972,peskin1989}. For more recent contributions, we refer to \cite{hesch2014b} in the context of immersed finite element methods for fluid-structure interaction problems and to \cite{vidotto2019hybrid,koeppl2020coupled}, in the context of 3D-1D transport models in microvascular networks. Working with a 1D-3D model has clear advantages with respect to meshing  in cases of stochastic fiber distributions.

We start with a classical continuum degenerated beam model, assuming that the motion of the  beam is given by the restricted position field
\begin{equation}\label{eq:fib:position}
\tilde{\vec{x}}(\theta^{\alpha},s) = \tilde{\vec{\varphi}}(s) + \theta^{\alpha}\,\vec{d}_{\alpha}(s).
\end{equation}
As usual, Greek indices are ranging from $\alpha = 1,2$ and latin from $i=1,2,3$. Here, the orthonormal triad $\vec{d}_i$ is related to the reference triad via the rotation tensor $\tilde{\vec{R}}\in SO(3)$, i.e. $\tilde{\vec{R}} = \vec{d}_i\otimes\vec{D}_i$,  orthogonal to the beam cross-section.

As strain energy, we use a simple form
\begin{equation}
\tilde{\Psi}(\vec{\Gamma},\vec{K}) = \frac{1}{2}\,\vec{\Gamma}\cdot\mathbb{K}_1\,\vec{\Gamma} + \frac{1}{2}\,\vec{K}\cdot\mathbb{K}_2\,\vec{K},
\end{equation}
where $\mathbb{K}_1 = \op{Diag}[G\,A_1,\,G\,A_2,\,E\,A]$ and $\mathbb{K}_2 = \op{Diag}[E\,I_1,\,E\,I_2,\,G\,J]$.
The standard beam strain and curvature measures, $\vec{\Gamma}$ and $\vec{K}$, are respectively given as
\begin{equation}
\vec{\Gamma} = \tilde{\vec{R}}^T\,\tilde{\vec{\varphi}}^{\prime}-\vec{D}_3,\quad\vec{K} = \operatorname{axl}\left(\tilde{\vec{R}}^T\,\tilde{\vec{R}}^{\prime}\right).
\end{equation}
The rotation tensor $\tilde{\vec{R}}$ is parameterized  by quaternions $\mathfrak{q}\in\mathbb{R}^4$, where $|\mathfrak{q}|=\sqrt{\mathfrak{q}\cdot\mathfrak{q}}=1$.

%The deformation gradient then reads
%\begin{equation}
%\tilde{\vec{F}} = \tilde{\vec{R}}\left(\vec{\Gamma}\otimes\vec{D}_3 + \hat{\vec{K}}\theta^{\alpha}\vec{D}_{\alpha}\otimes\vec{D}_3+\vec{I}\right),
%\end{equation}

Let us introduce the beam load and moment $\tilde{\vec{n}} = \tilde{\vec{R}}\,\mathbb{K}_1\,\vec{\Gamma}$ and $\vec{m} = \tilde{\vec{R}}\,\mathbb{K}_2\,\vec{K}$, respectively.
Postulating that the principle of virtual work holds, we obtain the strong form
\begin{equation}
\begin{aligned}
\tilde{\vec{n}}^{\prime} + \vec{n}_{ext} &= \vec{0},\\
\tilde{\vec{m}}^{\prime} + \tilde{\vec{\varphi}}^{\prime}\times\tilde{\vec{n}} + \vec{m}_{ext} &= \vec{0}.
\end{aligned}
\end{equation}
 Here, we made use of the resultant contact force $\tilde{\vec{n}}$ and resultant contact torque $\tilde{\vec{m}}$, and $\vec{n}_{ext}$, $\vec{m}_{ext}$ are external contributions. 
Without loss of generality, we assume~$\vec{m}_{ext} = 0$.

The strong form can be discretized using a method of weighted residuals, also known as collocation type method.
Here we use the isogeometric collocation method proposed in~\cite{weeger2017} for the beam discretization. For the matrix, we apply a standard variational IgA approach.
%As we have to couple both balance equations, i.e. the displacements of the collocation points as well as the Euler angles to the three-dimensional continuum material, we consider two types of constraints.
%We establish the coupling of both position field

We enforce the continuity condition 
\begin{equation}\label{eq:phi1}
%\vec{\Phi}_1(\vec{\varphi},\tilde{\vec{\varphi}}) := \vec{\varphi} - \tilde{\vec{\varphi}}\;\text{in}\;\Omega\cap\tilde{\Omega}.
\vec{\varphi} = \tilde{\vec{\varphi}}\;\text{in}\;\tilde{\Omega},
\end{equation}
%and second the coupling of the rotation matrix with the deformation gradient
%\begin{equation}\label{eq:phi2}
%\vec{\Phi}_2(\tilde{\vec{R}},\vec{\varphi}) := \op{skew}(\tilde{\vec{R}}^T\vec{F}(\vec{\varphi})).
%\end{equation}
%Both constraints can be enforced in different ways. For the position field, 
for matrix-fiber deformation fields weakly in terms of Lagrange multipliers.
% and apply a null-space reduction scheme, following the approach as proposed in the context of immersed finite element methods (IFEM).
%For the latter rotation field, it is reasonable to consider a penalty method to allow small deformations between the rotation of the beam and the matrix material.
In absence of other external forces for the beam, $\vec{n}_{ext}$ contains only the matrix-fiber interaction force, which corresponds to our Lagrange multiplier.
After condensation of the Lagrange multiplier along with the boundary conditions for~$\tilde{\vec{n}}$, we obtain the following matrix-fiber system:
\begin{align}
	\int_{\Omega} \frac{\partial \Psi}{\partial \vec{F}}:\nabla\vec{\delta\varphi}\d V
	+ \int_{\tilde{\Omega}} \tilde{\vec{n}}\cdot\frac{\partial}{\partial s}\,\vec{\delta\varphi} \d s &= 0,
	\intertext{for all $\vec{\delta\varphi}$ from an appropriate functional space, and}
	\begin{aligned}
	\tilde{\vec{\varphi}} &=\vec{\varphi}, \\
	\tilde{\vec{m}}^{\prime} + \tilde{\vec{\varphi}}^{\prime}\times\tilde{\vec{n}} &= \vec{0}, \\
	\tilde{\vec{n}} &= \tilde{\vec{n}}(\tilde{\vec{\varphi}}, \mathfrak{q}), \\
	\tilde{\vec{m}} &= \tilde{\vec{m}}(\mathfrak{q}), \\
	\mathfrak{q}\cdot\mathfrak{q} &= 1,
	\end{aligned}
\end{align}
in the collocation points.
Here, $\Psi(\vec{F})$ and $\vec{F}=\nabla\vec{\varphi}$ denote the strain energy and the deformation gradient of the matrix, respectively. 

%\subsection{Simulation}

%For a numerical example, we consider a unit cube domain of Mooney-Rivlin material with coefficients $c_1 = 10^3$, $c_2 = 2\cdot 10^3$, $c = 10^4$ (Pa), coupled with a fiber of radius~$r = 10^{-1}$~m, centerline $(0, 0, s)$, $s\in[-0.75,1.75]$ (lying on a diagonal), Young modulus~$10^6$~Pa and Poisson ratio~$0.3$.
%To the upper bound plane $Z = 1$, we apply Dirichlet boundary condition $\vec{x}(\vec{X}) = [X+0.1, Y, Z]$ (shear).
%We descritize the matrix using NURBS space of order $3$ (see Section 2 for the details) with $10$ elements in each direction.

\begin{figure}[ht]
%	\centering
	\begin{minipage}[top]{0.48\textwidth}
		\vspace{-5ex}
		\includegraphics[width=\textwidth]{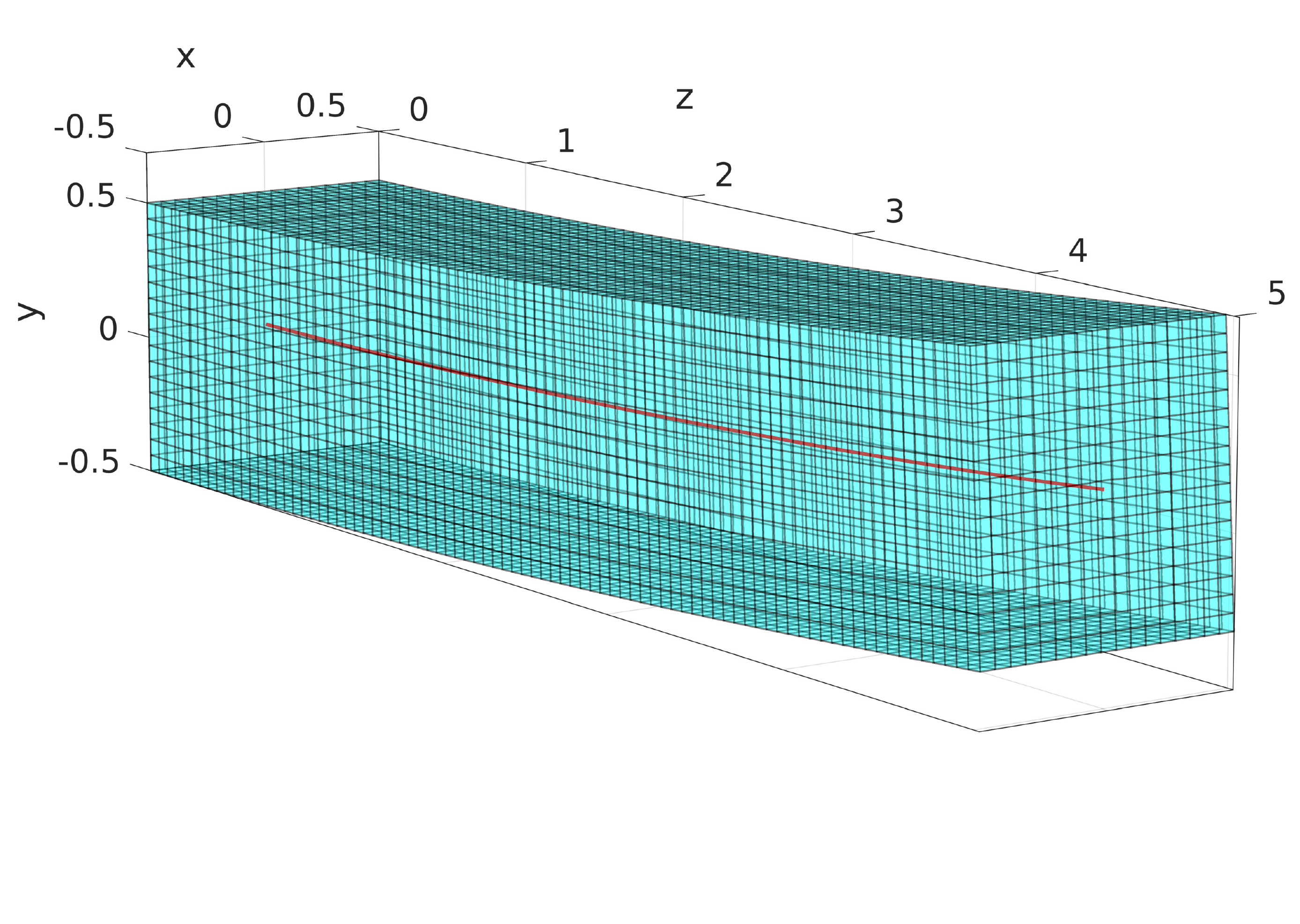}
	\end{minipage}
	\hfill
	\begin{minipage}[top]{0.50\textwidth}
		\includegraphics[width=\textwidth]{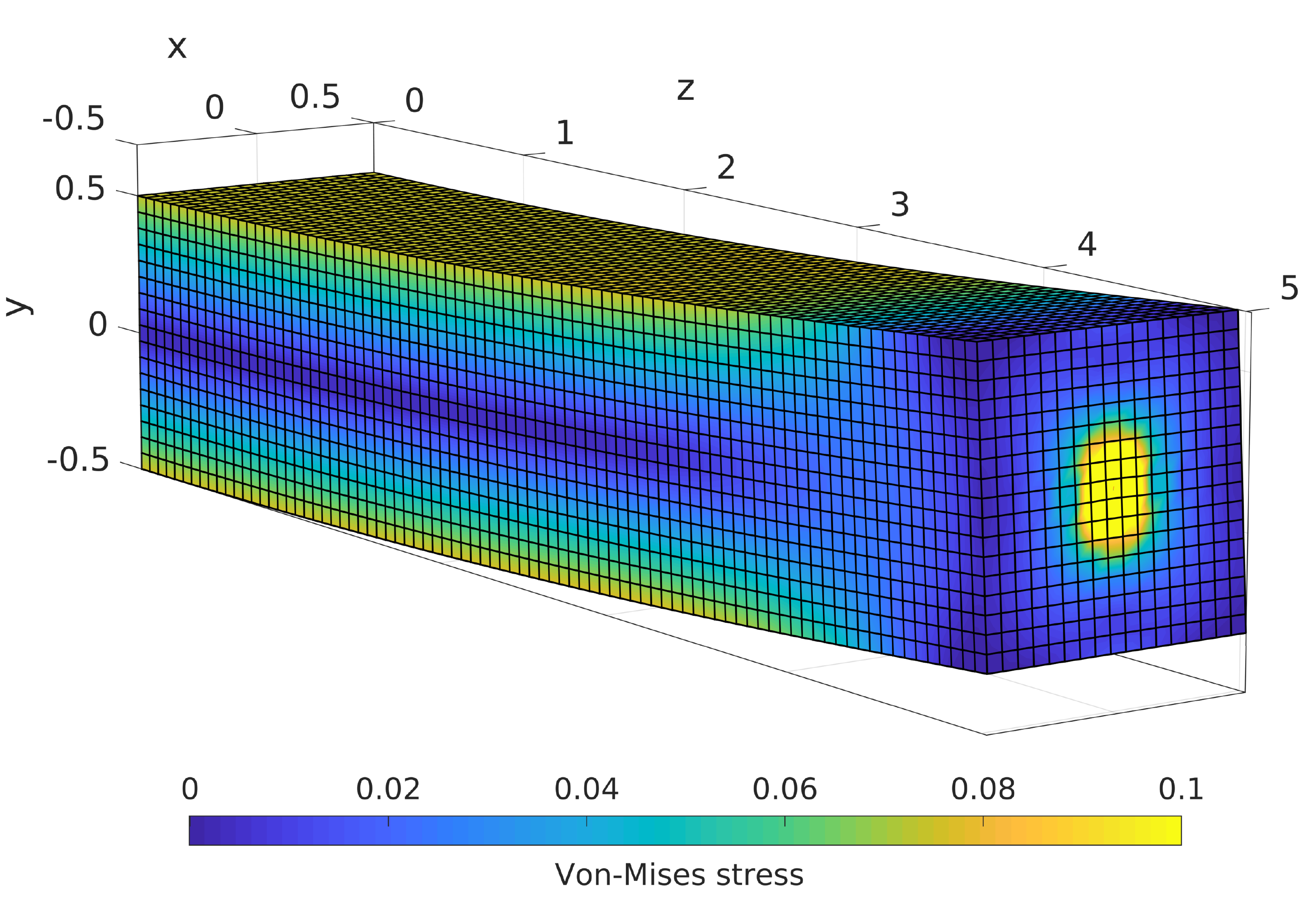}
	\end{minipage}
	\hfill
	\caption{Deformed fiber-matrix system with $p=[2,2,4]$ and $n=17$ (left) and the associated von Mises stress of the matrix (right).}
	\label{fig:BeamMatrixBending}
\end{figure}

\begin{figure}[ht]
	\centering
	\includegraphics[width=0.49\textwidth]{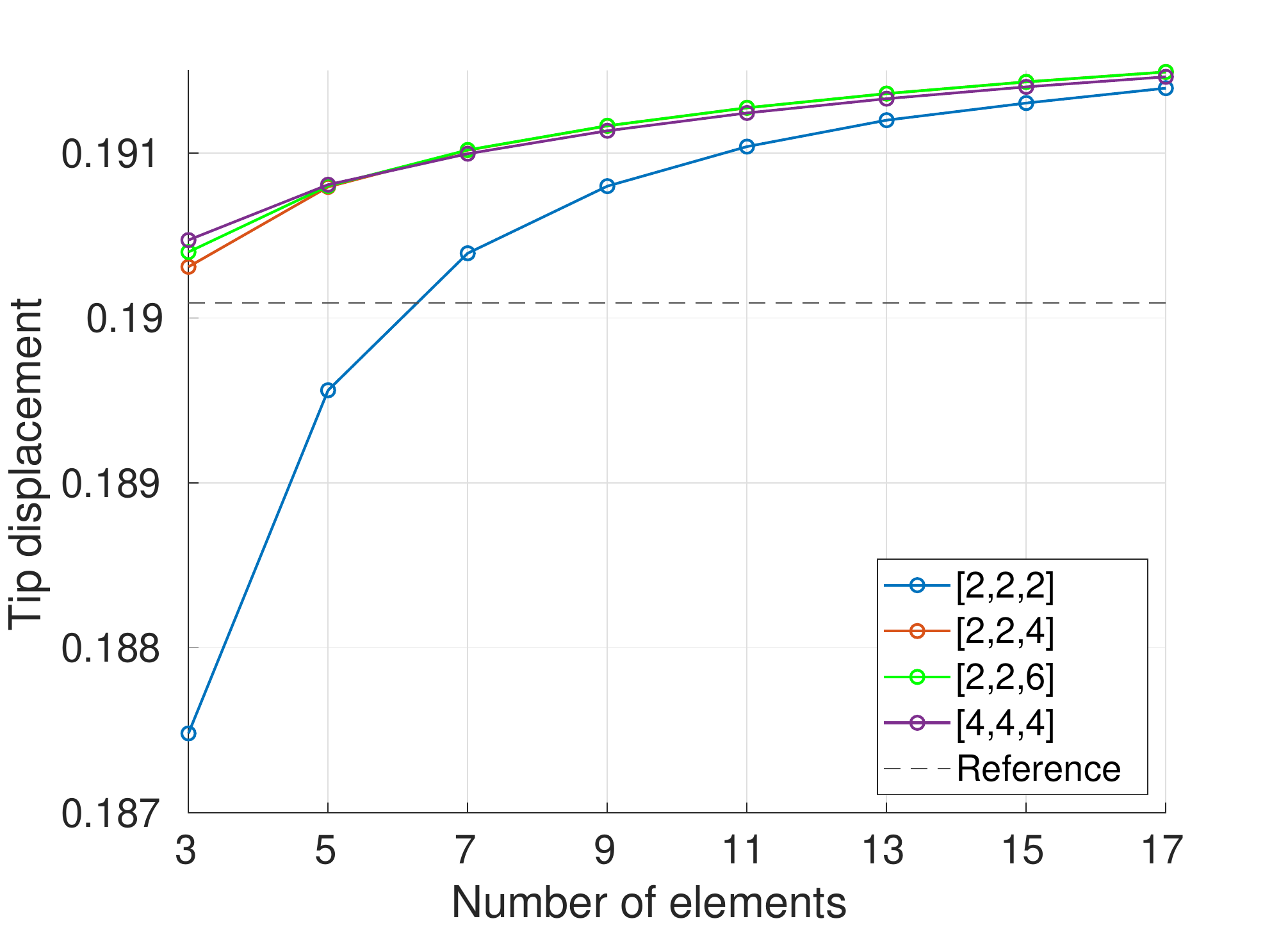}
	\hfill
	\includegraphics[width=0.49\textwidth]{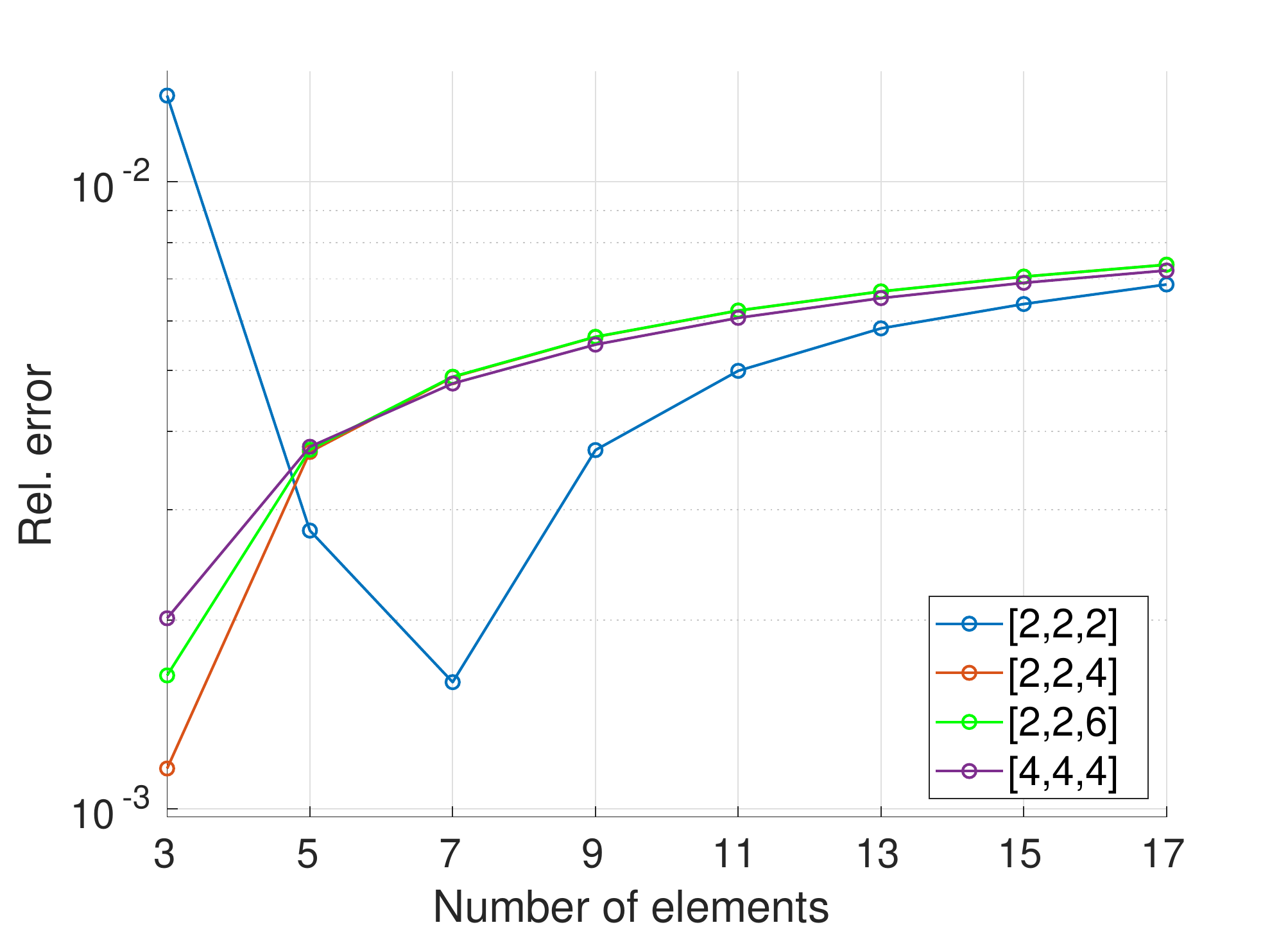}
	\caption{Convergence of the tip displacement $u(tip)$, left, and relative error $|u(tip)-u_{ref}(tip)|/ |u_{ref} (tip)|$, right.}
	\label{fig:TipDispAbsErr}
\end{figure}

For a numerical example, we consider a simple model problem from~\cite[section 4.2]{steinbrecher2019c}: a beam of length~$5\,\mathrm{m}$ and radius~$r=0.125\,\mathrm{m}$ with Young modulus~$4346\,\mathrm{N/m^2}$ is embedded into the $1\,\mathrm{m}\times1\,\mathrm{m}\times5\,\mathrm{m}$ matrix block of Saint-Venant--Kirchhoff material with Young modulus~$10\,\mathrm{N/m^2}$.
Poisson ratio is zero for both materials.
The matrix and the beam are both fixed at $z=0$, and a moment~$-0.025\,\mathrm{Nm}$ in~$x$-direction is applied to the beam tip~$z=5$.
This simple test allows us to illustrate the stability of the proposed approach, its convergence and the model error.
Simulation is performed using \textit{Esra} code developed in Siegen university by C. Hesch group.
The matrix is discretized with NURBS of degree~$p=[p_x,p_y,p_z]$.
We consider the same degree in $x$ and $y$ directions, $p_x=p_y$, and the degree in $z$ direction is the same as for the beam.
The number~$n$ of elements in $x$ and $y$ directions is the same and by a factor of five smaller than in $z$ direction.
The deformations and von Mises stress for $p=[2,2,4]$ and $n=17$ are depicted in Figure~\ref{fig:BeamMatrixBending}.
 The left picture in Figure~\ref{fig:TipDispAbsErr} shows convergence of the tip displacement~$u(tip)$ with the number~$n$ of elements in $x$ direction increasing for different spline degrees.
On the right, we also compare it with the reference finite element solution~$u_{ref}(tip)=0.19009\,\mathrm{m}$ from~\cite{steinbrecher2019c} obtained with 2D-3D (surface-to-volume) coupling. As it can be clearly seen, the numerical solution does not converge to the reference one. This results
from the reduced model approach and asymptotically we obtain the model error between a 1D-3D and a computationally more expensive full 3D-3D coupling. For a detailed analysis and a more sophisticated framework with a reduced model error for the 1D-3D coupling, we refer to \cite{hesch2020}.

%\todoB{please use larger fonts in all the plots see the style of Fig. 19 or 21}
%\todoB{in the last figure you show absolute differences in the displacements - this is meaningless
%you have to report on the relative difference}

%
%\begin{figure}
%	\centering
%	\includegraphics[width=\textwidth]{pictures/Coupling/convergence.eps}
%%	\caption{Average Von-Mises stress convergence with respect to the fiber discretization.}
%\end{figure}

\section{Conclusions}\label{sec:conclusions}
We summarized modern mortar based IgA methodologies and their application to a large variety of problems in structural mechanics ranging from non-linear contact, thermo-mechanical friction and fracture to fiber reinforced material simulations. While some of the proposed methods are a mere combination of well-established techniques out of the
mortar finite element community with IgA approaches, the  nature of IgA brings also new challenges. IgA approaches are typically used in the higher-order context and allow higher regularity of the discrete solution. To preserve this higher smoothness in the case of non-matching meshes is not as simple as in the finite element context.
A variationally consistent approach requires a careful design of suitable discrete Lagrange multiplier spaces and for all Lagrange multipliers a suitable modification at crosspoints. Also the construction of biorthogonal basis function is not as local as in the low-order finite element context. However it can be achieved on the prices of enlarging the local support by at most $p$ elements. Mortar based IgA methodologies provide flexible and robust discretization schemes for approximating a large class of partial differential equations including higher-order equations such as Kirchhoff Love shells which require a $G^1$-continuity and non-smooth problems such as contact mechanics. Traditional mortar methods are typically based on a non-overlapping domain decomposition of the physical $d$-dimensional domain and implement a weak coupling in terms of Lagrange multipliers defined on a $(d-1)$-dimensional interface. However, the concept is not restricted to such situations and can be also generalized to a multi-dimensional setting which opens the possibility for many more applications.

\vspace*{-3mm}\section*{Acknowledgments} 
We would like to thank T. Horger and A. Reali as co-authors of  \cite{horger2019}, L. Wunderlich as co-author of \cite{brivadis2015,wunderlich2018,horger2019}, M.D.
 Alaydin as co-author of \cite{wunderlich2018}, A. Matei, S. Sitzmann and K. Willner as co-authors of \cite{MSWW18} as well as P. Farah and J. Kremheller  as co-authors of \cite{Seitz2016}.
Funds provided by the Deutsche  Forschungsgemeinschaft  under  the  contract/grant  num\-bers:  PO1883/3-1, WO671/11-1  as  well  as  PO1883/1-1,  WA1521/15-1  and  WO671/15-1,
WO671/15-2 (within  the  Priority Program SPP 1748, ”Reliable Simulation Techniques in Solid Mechanics.  Development of Non-standard Discretisation Methods, Mechanical and Mathematical Analysis”) are gratefully acknowledged. Moreover, we would like to thank S. Schu{\ss} and M. Dittmann as co-authors of \cite{schuss2018,dittmann2018c,dittmann2020}, S. Klinkel as co-author of \cite{schuss2018} and $\dot{\text{I}}$. Temizer and M. Franke as co-authors of \cite{hesch2016a}. Funds provided by  the  Deutsche  Forschungsgemeinschaft  under  the  contract/grant  numbers: HE5943/3-2, HE5943/6-1, HE5943/8-1, DI2306/1-1 and HE5943/5-1 (within  the  Priority Program SPP 1748, ”Reliable Simulation Techniques in Solid Mechanics. Development of Non-standard Discretisation Methods, Mechanical and Mathematical Analysis”) are gratefully acknowledged. R. Krause wants  to thank the Swiss National Science Foundation for their support through project 154090 and the Platform for Advanced Scientific Computing for support through the projects FASTER and AV-FLOW.

\bibliographystyle{plain}
\bibliography{literature,bibliographyMS}
\end{document}